\documentclass{article}
\usepackage{graphicx} 

\title{Joint Modelling of Line and Point Data on Metric Graphs}
\author{Karina Lilleborge$^a$, Sara Martino$^a$, Geir-Arne Fuglstad$^a$, \\Finn Lindgren$^b$ and Rikke Ingebrigtsen$^c$} 

\date{\small$^a$ Department of Mathematical Sciences, Norwegian University of Science and Technology, \\$^b$ School of Mathematics and Maxwell Institute for Mathematical Sciences, University of Edinburgh, \\$^c$ Institute of Transport Economics (Norway)\normalsize}

\usepackage{amssymb}
\usepackage{natbib}
\usepackage{amsmath}
\usepackage{url}
\usepackage[margin=1.2in]{geometry}
\usepackage{txfonts}
\usepackage{enumitem}
\usepackage{subcaption}
\usepackage{multirow}
\usepackage{siunitx}
\usepackage{listings}
\usepackage{array}
\usepackage{placeins}
\usepackage{appendix}

\usepackage{xcolor}

\newcommand{\vect}[1]{\ensuremath{\boldsymbol{\mathbf{#1}}}}  
\newcommand{\matr}[1]{\ensuremath{\boldsymbol{\mathbf{#1}}}}  
\newcommand*\dif{\mathop{}\!\mathrm{d}} 
\newcommand{\yp}{\ensuremath{y^\mathrm{P}}}
\newcommand{\yl}{\ensuremath{y^\mathrm{L}}}

\newcommand{\SMs}{\texttt{WSM} } 
\newcommand{\IMs}{\texttt{CSM} } 
\newcommand{\SM}{\texttt{WSM}} 
\newcommand{\IM}{\texttt{CSM}} 

\definecolor{codegreen}{rgb}{0,0.6,0}
\definecolor{codegray}{rgb}{0.5,0.5,0.5}
\definecolor{codepurple}{rgb}{0.58,0,0.82}
\definecolor{backcolour}{rgb}{0.95,0.95,0.92}

\lstdefinestyle{mystyle}{
    backgroundcolor=\color{backcolour},   
    commentstyle=\color{codegreen},
    keywordstyle=\color{magenta},
    numberstyle=\tiny\color{codegray},
    stringstyle=\color{codepurple},
    basicstyle=\ttfamily\footnotesize,
    breakatwhitespace=false,         
    breaklines=true,                 
    captionpos=b,                    
    keepspaces=true,                 
    numbers=left,                    
    numbersep=5pt,                  
    showspaces=false,                
    showstringspaces=false,
    showtabs=false,                  
    tabsize=2
}

\begin{document}
\maketitle
\begin{abstract}
Metric graphs are useful tools for describing spatial domains like road and river networks, where spatial dependence act along the network. We take advantage of recent developments for such Gaussian Random Fields (GRFs), and consider joint spatial modelling of observations with different spatial supports. Motivated by an application to traffic state modelling in Trondheim, Norway, we consider line-referenced data, which can be described by an integral of the GRF along a line segment on the metric graph, and point-referenced data. Through a simulation study inspired by the application, we investigate the number of replicates that are needed to estimate parameters and to predict unobserved locations. The former is assessed using bias and variability, and the latter is assessed through root mean square error (RMSE), continuous rank probability scores (CRPSs), and coverage. Joint modelling is contrasted with a simplified approach that treat line-referenced observations as point-referenced observations. The results suggest joint modelling leads to strong improvements. The application to Trondheim, Norway, combines point-referenced induction loop data and line-referenced public transportation data. To ensure positive speeds, we use a non-linear link function, which requires integrals of non-linear combinations of the linear predictor. This is made computationally feasible by a combination of the R packages \texttt{inlabru} and \texttt{MetricGraph}, and new code for processing geographical line data to work with existing graph representations and \texttt{fmesher} methods for dealing with line support in \texttt{inlabru} on objects from \texttt{MetricGraph}. We fit the model to two datasets where we expect different spatial dependency and compare the results.
\end{abstract}


\section{Introduction}\label{sec:introduction}
Gaussian random fields (GRFs) are a central tool for modelling spatial and spatio-temporal dependence \citep{ diggle,cressie2011statistics}. They are used in a broad range of fields from environmental sciences \citep{gelfand} to global health \citep{ribeiro2019mapping}. GRFs are popular because they can be specified in an interpretable way through a mean function and a covariance function, and computations involve multivariate Gaussian distributions. 

However, defining valid covariance functions that exhibit diverse, useful behaviors while remaining interpretable is challenging. As a result, it is common to use established families of covariance functions. In Euclidean domains, such as $\mathbb{R}$ and $\mathbb{R}^2$, the Mat\'ern covariance function is the most commonly used choice \citep{stein},
\begin{equation}\label{eq:matern}
	c(h) = \sigma^2\frac{2^{1-\nu}}{\Gamma(\nu)}\left(\sqrt{8\nu} \frac{h}{\rho}\right) K_\nu \left(\sqrt{8\nu} \frac{h}{\rho}\right), \quad h \geq 0.
\end{equation}
Here $h$ is the Euclidean distance between two spatial locations, $\sigma^2>0$ is the marginal variance, $\rho>0$ the spatial range, $\nu>0$ the smoothness parameter, and $ K_\nu(\cdot)$ is the modified Bessel function of second kind of order $\nu$. In this parametrization, $\rho$ is the distance at which correlation is approximately $0.135$.

Several phenomena, for example traffic flow or road accidents on a road network or water temperature along a river network, are not inherently suited for representation in $\mathbb{R}^2$. Instead, a metric graph --- a structure consisting of curves joined at vertices, as illustrated in Figure \ref{fig:studyarea} --- provides a more intuitive and appropriate framework. 
There is considerable interest in statistical modeling of data on metric graphs based on GRFs. E.g., there has been much development for point processes on networks \citep{Baddeley2021, Møller2024},  
and there is interest in developing complex spatio-temporal dependence structures \citep{porcu2023}. 
Substantial work has also been done on modeling stream flows on river network where direction may be important as described by up-stream and down-stream models \citep{cressie2006spatial,hoef2006spatial, ver2010moving}. 
Such models consider a directed graph, or a tree, where the direction is fixed apriori, and covariance functions are not invariant to changing the direction of edges. In such models loops are not allowed, something that makes their use in traffic modeling on road networks problematic.
  
A more direct way to define spatial dependence on metric graphs would be to consider the Mat\'ern covariance with a distance measure that is appropriate for metric graphs. 
Unfortunately, defining distances on metric graphs is not straightforward. 
Alternative distance measures, such as geodesic distance (i.e., the shortest path) and the resistance metric inspired by electrical network theory, have been proposed. However, it has been proven that for $\nu > 0.5$, it is possible to build Euclidean tree graphs such that the Mat\'ern covariance function is not valid \citep{anderes2020isotropic}.

A promising way to specify GRFs on metric graphs is to extend the SPDE approach introduced by \citet{SPDE} to metric graphs, as proposed in \citet{metric_graph}. This method automatically accounts for the graph’s geometry while ensuring a valid covariance structure that follows the metric graph’s topology.
On $\mathbb{R}^2$ , the SPDE approach \citep{SPDE} exploits the fact that Mat\'ern GRFs can be expressed as the stationary solutions of 
\begin{equation}\label{eq:SPDE_Rd}
    (\kappa^2-\Delta)^{\alpha/2}(\tau u(\vect s)) = \mathcal W(\vect s), \quad \vect s\in  \mathbb R^{d},
\end{equation}
where $\Delta$ is the Laplacian, $\mathcal W(\cdot)$ is a Gaussian white noise process, 
and $\alpha$, $\kappa$ and $\tau$ are linked to the standard parametrization by 
$\nu=\alpha-d/2$, 
$\rho={\sqrt{8\nu}}/{\kappa}$, and $\sigma^2=\Gamma(\nu)/[\Gamma(\nu+1/2)(4\pi)^{d/2}\kappa^{2\nu}\tau^2]
$.
This connection forms the basis for a computationally efficient approach based on sparse matrices that has seen large developments over the last decade \citep{SPDE_10y}, and combined with the INLA approach \citep{INLA}, it is a popular tool for fast spatial inference \citep{bakka2018spatial}.

\citet{matern_metric_graph} have extended Equation \eqref{eq:SPDE_Rd} to metric graphs. The resulting field does not have a Mat\'ern covariance structure, but the covariance resembles the Mat\'ern if one is far enough from any vertices of the graph. 
There has been great effort in defining and validating theoretical properties related to this approach
\citep{metric_graph, matern_metric_graph,nonstationary_gaussian_metric_graph} and providing software \citep{MetricGraph}.
However, less attention has been given thorough application-focused studies involving diverse observation models. Additionally, there is a lack of studies addressing the non-trivial practical challenges associated with implementing and using these models in real-world scenarios.

The aim of this paper is to provide an accessible presentation of the SPDE approach on metric graphs, provide new general code necessary for handling spatial data on the graphs, and demonstrate a complex application to joint modeling of data with different spatial supports in traffic modeling. 
In particular, we are interested in jointly modeling standard point-referenced observations of speeds with measurements of average speed across segments of the metric graph.
Representing metric graphs is easily done with \texttt{MetricGraph} \citep{MetricGraph} and standard point-referenced data is handled within the package and has interface with \texttt{inlabru} \citep{inlabru}, which is a wrapper of \texttt{R-INLA} \citep{INLA} (available through: \citet{INLA_web}) that is software to perform fast Bayesian inference on latent Gaussian models (LGMs), and \texttt{inlabru} allows one to have non-linear predictors \citep{inlabru}. However, line-referenced data is not available in \texttt{MetricGraph}, and representing valid paths and practical implementation of line-referenced data are key contributions of this work. The new code allows spatial data to be defined separately from the metric graph object, enabling more flexible models within the \texttt{R-INLA}-framework. It bridges the metric graph representation in \texttt{MetricGraph} and the \texttt{rSPDE}-package \citep{rSPDE,rspde_article} with \texttt{inlabru} in a similar way to how \texttt{fmesher} \citep{fmesher} is providing mesh (discretization) representation for $\mathbb R$ and $\mathbb R^2$ domains. We provide methods for defining line segments on metric graphs from geometric line objects in \texttt{sf} into a format which \texttt{MetricGraph} and \texttt{rSPDE} can handle. Additionally, we expand existing code for general one-, two- and three-dimensional meshes to meshes defined on metric graphs. Code and a reproducible example are available from \citet{code}.

In this study, we analyze traffic flow — a phenomenon naturally represented on metric graphs—in the city of Trondheim, Norway. 
Automated vehicle location (AVL) data from multiple bus lines was provided by the public transport authority in Trondheim, AtB. Data was processed to average bus velocities between consecutive bus stops along multiple bus lines.
These line-referenced observations differ from the point-referenced observations —typically the focus of existing literature—  as they are defined over road segments rather than specific locations. While line-referenced data is less informative than point-referenced data, it is far more abundant in our case study and can provide valuable insights. In fact, Trondheim has only six stations with point-referenced traffic data (see Figure \ref{fig:studyarea}), making the additional availability of line-referenced data particularly useful for assessing the traffic state at various parts of the road network.

To assess whether the available data is sufficient for meaningful parameter estimation and spatial predictions, we conduct a simulation study that mimics the conditions of the application. 
First, we evaluate parameter estimation by examining the bias and variability of the parameter estimates. Next, we assess spatial prediction performance using root mean square error (RMSE), continuous ranked probability score (CRPS) \citep{gneiting2007strictly}, and coverage.

The rest of the paper is organized as follows: In Section \ref{sec:motivation}, we motivate our application to traffic flow and discuss the available data. Section \ref{sec:preliminaries} provides an overview of metric graphs and the SPDE approach on metric graphs. In Section \ref{sec:model}, we present the hierarchical joint model and detail the computational aspects. Section \ref{sec:sim_study} focuses on a simulation study to evaluate parameter estimation and spatial prediction. This is followed by the application to traffic modeling in Section \ref{sec:case}. Finally, Section \ref{sec:discussion} offers a discussion of the approach, highlighting its strengths and limitations, and outlines potential directions for future work with metric graphs.

\FloatBarrier

\section{Motivating application: Traffic modelling}\label{sec:motivation}
Efficient and smooth traffic systems have been a central part of traffic engineering for years and are topics of interest to this day with increasing urban areas and a global growth in transportation \citep{traffic_model_overview, urban-traffic-planning}. 
The state of the traffic is typically well-monitored on main roads with, e.g., loop detectors registering vehicle speed or cameras monitoring traffic flow. However, on smaller roads there is less data about traffic and there has been an interest in using information from moving vehicles (fleeting car data) to assess speed and flow at these roads \citep{floating-car}.
 
However, fleeting car data (e.g.\ from private cars or taxies) are often not easy to access for various reasons (e.g.\ privacy etc.). In contrast, data from buses are usually administrated by public transport authorities and more available to researchers and transport planners. Since buses operate at fixed routes and schedules in mixed traffic conditions, they provide a stable and reliable source of traffic information. This study leverages recent advances in Gaussian random fields (GRFs) on metric graphs and efficient Bayesian inference methods to gain insights into traffic states and flow by integrating traffic speed data with public transport data.

Models for describing traffic are of interest to road authorities, bus companies, and city planners, and common approaches include visual representations of data \citep{spat_AVL_visual}, regression models \citep{delay_AVL_linreg, spatial-regression}, and recently more advanced deep learning networks have been introduced \citep{transit_data_analytics}. To our knowledge, there has been limited use of spatial statistics for this purpose, and our goal is to close this gap. We believe that considering the spatial domain on which the random field of interest is taking place will improve the model's ability to capture the true trend of traffic flow in the road network. 

\begin{figure}[htb]
    \centering
    \begin{subfigure}[c]{0.487\textwidth}
        \centering
        \includegraphics[width=\linewidth]{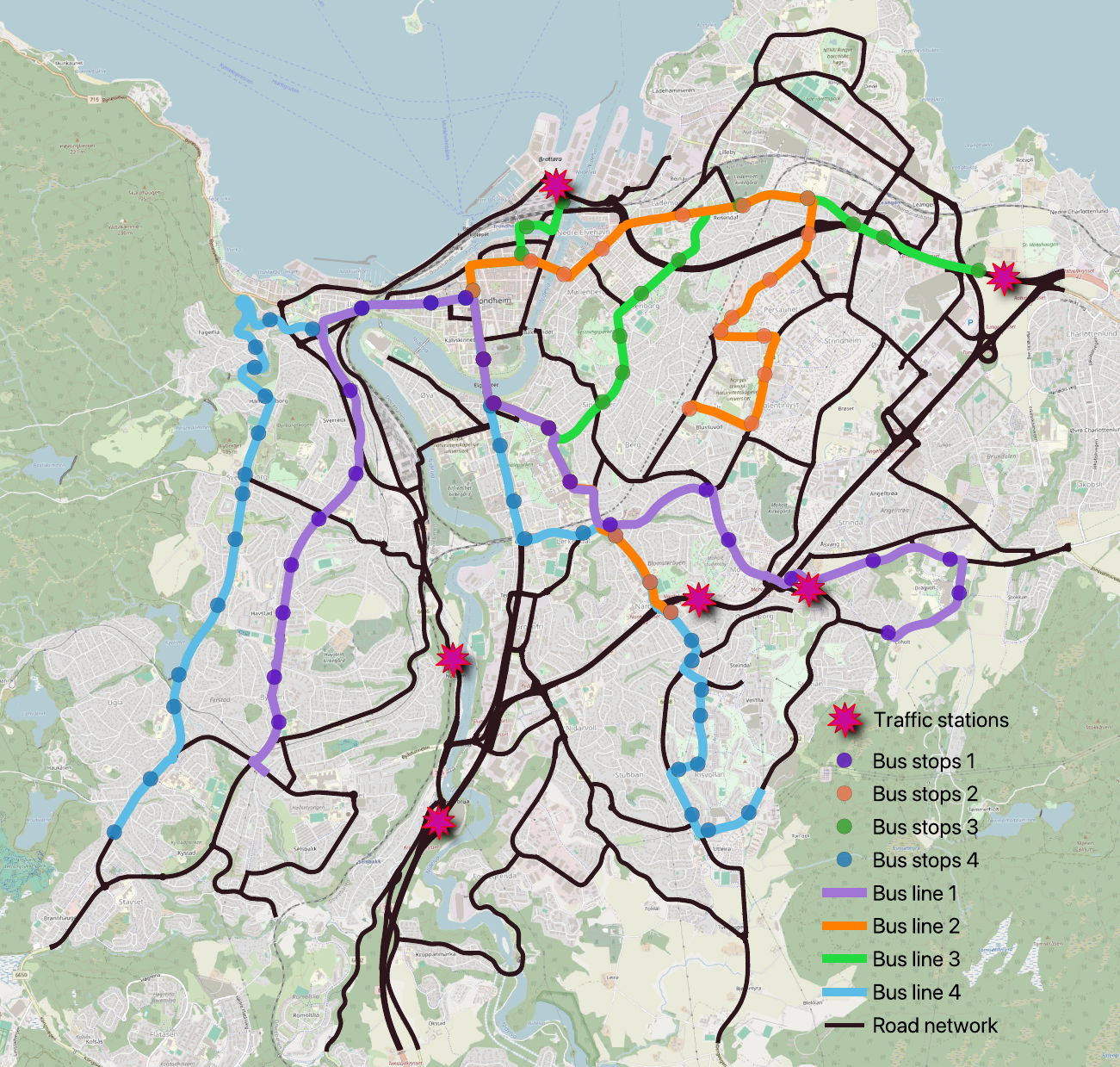}
        \caption{Graph and data.}
        \label{fig:studyarea}
    \end{subfigure}%
    ~
    \begin{subfigure}[c]{0.48\textwidth}
        \centering
        \includegraphics[width=\linewidth]{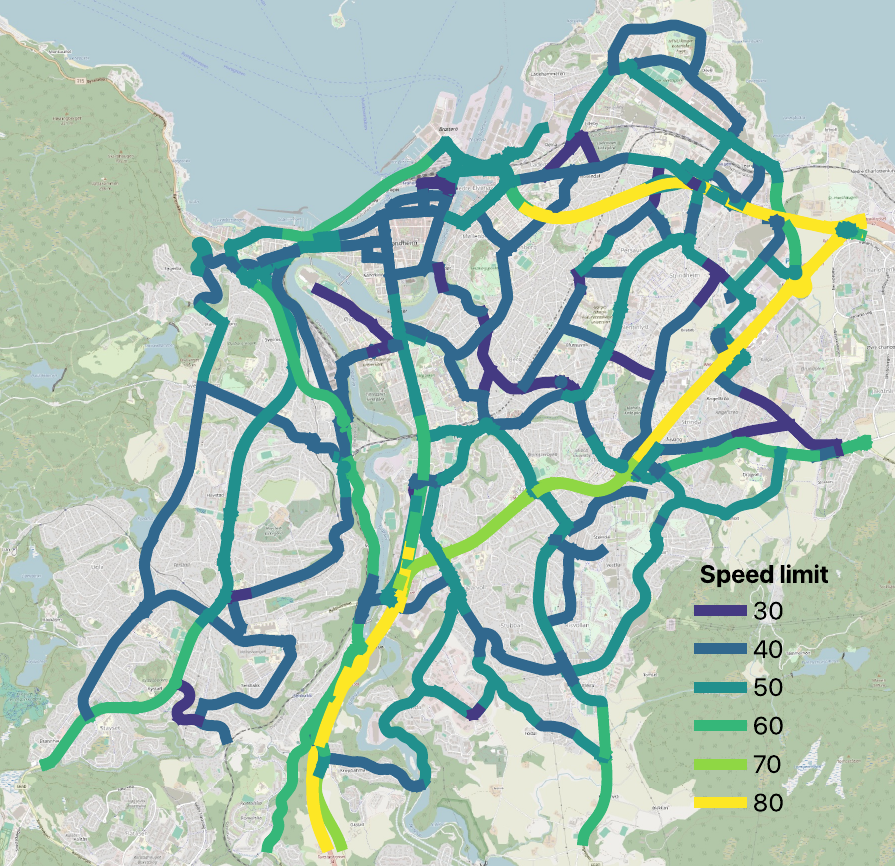}
        \caption{Speed limit.}
        \label{fig:case_covariate}
    \end{subfigure}%
    \caption{The study area with (a) roads (black lines), bus lines (blue, purple, orange and green lines), bus stop (blue, purple, orange and green dots) and stations (pink stars) constructed from \citet{osm}, and (b) the spatial covariate considered in the case study shown on the road network, where the color represents the speed limit on that road segment (minimum 30\si{\kilo\meter/\hour} and maximum 80\si{\kilo\meter/\hour}) and missing values are filled in as described. This is obtained from \citet{osm}, and missing values are filled in.}
\end{figure}

The study area is the city center and adjacent neighborhoods in Trondheim, the third largest city in Norway. We construct the road network using Open Street Map (\texttt{OSM}) \citep{osm}. Figure \ref{fig:studyarea} shows the network consisting of 713 intersections, 1064 roads connecting the intersections. There is large variation in road segment lengths from 1 \si{\meter} to 2.8 \si{\kilo\meter}, and the entire network consists of 177 \si{\kilo\meter} of roads. The diameter of the graph --- longest shortest path between two intersections/end points --- is 13.2 \si{\kilo\meter}. Note that this is only a medium-sized graph, and that \citet{bolin_point} demonstrate that \texttt{MetricGraph} can handle much larger, more complex graphs.
To explain part of the variability in traffic we use 
the speed limit as shown in Figure \ref{fig:case_covariate}, as a covariate. The speed limit is obtained from \texttt{OSM} and varies from 30 \si{\kilo\meter/\hour} up to 80 \si{\kilo\meter/\hour} with average  speed limit across the network of 53.5 \si{\kilo\meter/\hour}. Information about the speed limit is not available for all road segments in the network. We have therefore reconstructed the missing information  assuming  that roads with missing information have a speed limit of 40 \si{\kilo\meter/\hour}. The resulting covariate is shown in Figure \ref{fig:case_covariate}.

Traffic exhibits strong spatio-temporal variation, 
with rush hours characterized by high vehicle density and slow-moving queues. During these periods, traffic load is higher than in off-peak hours, typically reflecting commuter flows from rural areas to the city center in the morning and in opposite direction in the afternoon. 
To analyze traffic patterns, we define departure time windows (DTWs), following \citet{DTW-travel-time}, where journeys occurring in the same DTW on a given weekday are analysed jointly while journeys in different DTWs are analysed separately. \citet{DTW-travel-time} show that for narrow DTWs, travel time can be well described by the normal distribution. In this study we consider 1-hour DTWs and focus on two specific periods: Mondays 07:00a.m.--08:00a.m.\ (rush hour) and Wednesdays 07:00p.m.-08:00p.m.\ (outside rush hour).
We conduct separate spatial analyses for these two times, to investigate differences in behavior.

We consider two sources of traffic data: the first consists of six vehicle loop detectors that measure the speed of passing vehicles, providing point-reference data similar to that used in \citet{metric_graph}. Loop detectors are induction loops placed under the road surface with a short distance between them, recording the time between inductions in the two loops and converting this to momentary speed. Such data are typically collected and owned by road authorities, in this case The Norwegian Public Roads Administration (Statens vegvesen). Raw data from these detectors are not openly available for General Data Protection Regulation (GDPR) reasons, and access to summary statistics is available upon request. We have average speeds from each station within the DTW considered. 
While data from the six loop detectors are precise, this source is very sparse (see station locations in Figure \ref{fig:studyarea}) and insufficient for robust inference.
The second data source is provided by the local public transport company (AtB), and includes recorded arrival and departure times for four bus lines at designated bus stops. These bus lines and the bus stops are represented in Figure \ref{fig:studyarea} by different colors.
These data are line-referenced as they are informative on the average buses' speeds over specific segments of the road system, namely between two consecutive bus stops.
Such information cannot be summarized by the speed at a single location. 

In the application we consider two key aspects of the data: 
the geometry, represented by the road network, and the data collection mechanisms, which include both point and line data. To address the first aspect, we consider metric graphs, described in Section \ref{sec:preliminaries}. The second aspect is discussed in Section \ref{sec:model}. Here we consider two alternative models: one where the correct spatial support is explicitly formulated, and a simpler one where line data are treated as point observations of (average) speed at the midpoints between bus stops.

\section{Gaussian random fields on metric graphs}\label{sec:preliminaries}
\subsection{Metric graphs}\label{ssec:graph_field}
\begin{figure}[htb]
	\centering
	\includegraphics[width=0.65\linewidth]{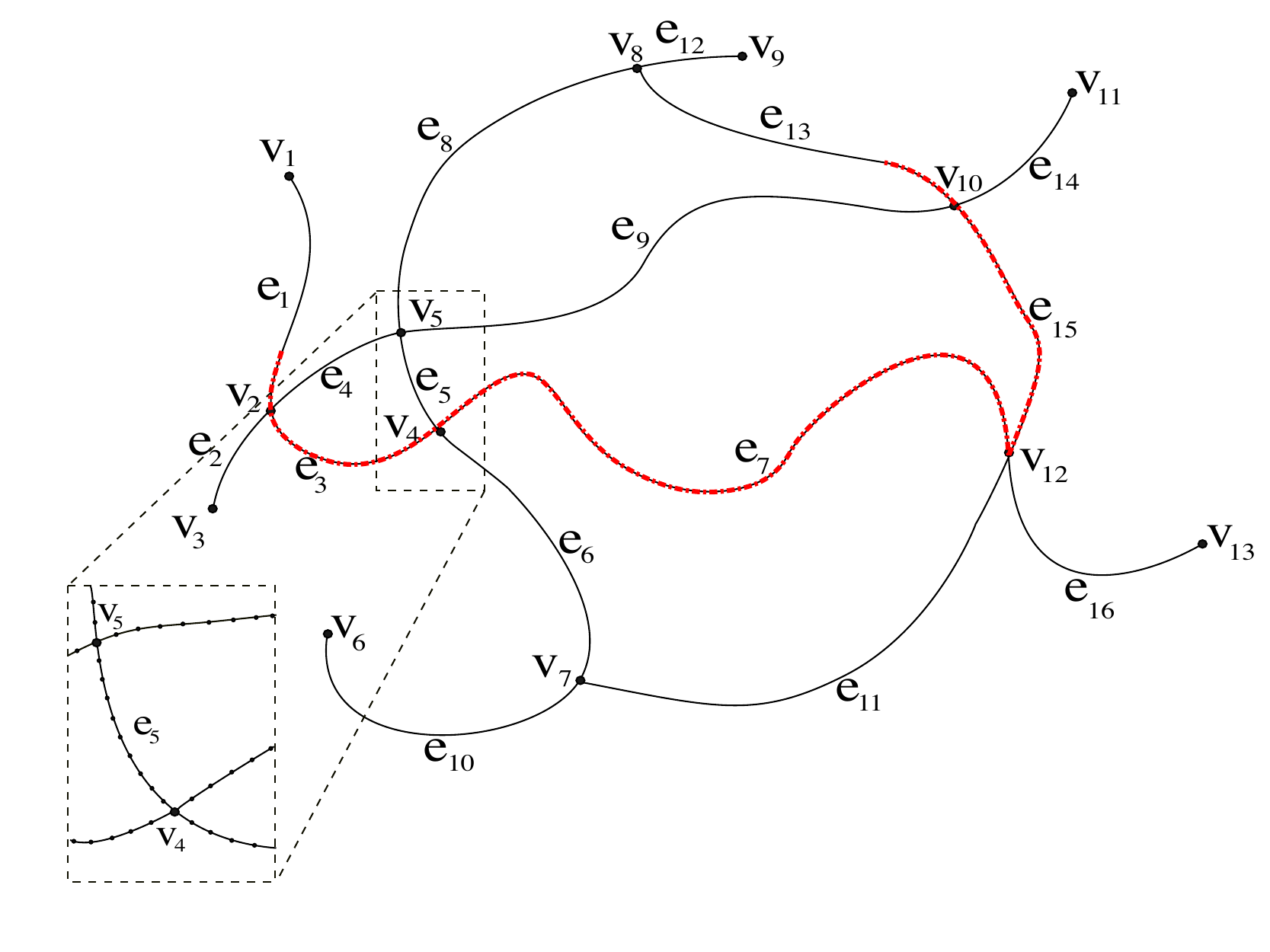}
	\caption{Illustration of a graph with 13 vertices, $\mathcal V = \{v_i\}_{i=1}^{13}$, and 16 edges, $\mathcal E = \{e_i\}_{i=1}^{16}$. In the lower left panel, the additional vertices that constitute the mesh are visible. The red dotted line indicates a valid simple path through the graph.}
	\label{fig:graphmeshillustration}
\end{figure}

A metric graph is a collection of one-dimensional curves, connected to form a network (for example a road or river network) as illustrated in Figure \ref{fig:studyarea}. In this section, we summarize the key points about metric graphs and link them to the motivating example in Section \ref{sec:motivation}. We encourage readers interested in more details to see \citet{metric_graph}. Assume there are $m$ vertices (e.g., intersections), we can then describe them as a set of $m$ vertices given by their spatial coordinates $\mathcal{V} = \{\vect v_1, \ldots, \vect v_m\}\subset\mathbb{R}^2$. An edge, $e_i$, is a continuous curve in $\mathbb{R}^2$ that connects two vertices (e.g., road), and $\mathcal{E} = \{e_1, \ldots, e_M\}$ is the set of $M$ such curves (e.g., roads). For $i = 1, \ldots, M$, each edge $e_i$ can be parameterized as an interval $[0, l_{e_i}]$ where $l_{e_i}$ is the length of the curve $e_i$. Thus, we can think of a metric graph $\Gamma$, as a collection of locations $\boldsymbol{s}\in\Gamma$ described by a tuple $\boldsymbol{s}=(e,t)$ where $e \in \mathcal{E}$ and $t\in[0, l_{e}]$. Note that for every edge $e\in\mathcal{E}$, the starting point $(e,0)\in\mathcal{V}$ and the ending point $(e, l_e)\in\mathcal{V}$. The representation as tuples is useful for describing the geometry of $\Gamma$, but $\Gamma$ can also be viewed as a subset of $\mathbb{R}^2$ for visualization. See Figure \ref{fig:graphmeshillustration} for an example of a simple metric graph. 
We assume that $\Gamma$ is connected so that it is possible to find a path between any pair of locations. An example of such path is visualized in Figure \ref{fig:graphmeshillustration} through a red dotted line.

The goal is to define GRFs with useful properties on metric graphs. We first discuss how a GRF can be defined on a single edge, $e$, which can be viewed just as a one-dimensional interval $[0, l_e]\subset\mathbb{R}$, and then discuss how to extend it to the metric graph, $\Gamma$, using the finite element method (FEM) approach described in \citet{metric_graph_FEM}.

\subsection{SPDE approach on a single edge}\label{ssec:spde1d}
Consider an interval $[0, l] \subset \mathbb R$, and describe a GRF $u(\cdot)$ as the solution of
\begin{equation}
    (\kappa^2-\Delta)^{\alpha/2}(\tau u(t)) = \mathcal W(t), \quad t\in [0,l],\label{eq:singleEdge:spde}
\end{equation}
where $\alpha > 1/2$, $\kappa, \tau > 0$, $\mathcal{W}(\cdot)$ is Gaussian white noise, and zero Neumann boundary conditions are enforced. Note that $\mathcal{W}(\cdot)$ is an abuse of notation, as point-wise evaluation of the noise process is not defined. The behavior of the solution is then approximately described by a Mat\'ern covariance function up to boundary effects. General $\alpha$ requires combining FEM with fractional approximations \citep{rspde_article}. To avoid this, we fix $\alpha = 1$ so that we have approximately the exponential covariance function,
$$
    c(|t_1-t_2|) = \frac{1}{2\tau^2\kappa}\exp(-\kappa(|t_1-t_2|)), \quad t_1, t_2\in [0, l], 
$$
up to boundary effects. Fixing $\alpha=2$, or other integer values, is straight-forward when following the same approach that is described in this section. To reduce the influence of boundary effects, the domain is often extended to a larger interval. 
See the general discussion in \citet{SPDE} for domains in $\mathbb{R}^d$.

To apply FEM, we start by discretizing the interval $[0,l]$ into a regular grid of smaller intervals,
$[t_k,t_{k+1}]_{k=0,..,K}$, where $t_k=l\cdot k/(K+1)$, $k=0,1,\ldots,K+1$. We define a finite set of piece-wise linear pyramidal basis functions $\varphi_k(\cdot)$, $k=0,\ldots,K+1$, each defined by $\varphi_k(t_i)=1$ when $i=k$ and $\varphi_k(t_i)=0$ otherwise. The FEM representation of $u(\cdot)$ is then given by
$u^\text{FEM}(\cdot) = \sum_{k=0}^{K+1}\text{w}_k\varphi_k(\cdot)$, 
where $\vect w = [\text{w}_0,\ldots,\text{w}_{K+1}]^\mathrm{T}$ contains the weights for the basis functions. We use the notation $\langle f(\cdot), g(\cdot)\rangle = \int_{[0,l]} f(t)g(t) \:\dif t,$ for the inner product on $L^2([0, l])$. An approximate solution of SPDE in \eqref{eq:singleEdge:spde} is then found by enforcing
\begin{equation*}
    \begin{bmatrix}\langle (\kappa^2-\Delta)^{1/2}(\tau u(\cdot)), (\kappa^2-\Delta)^{1/2}\varphi_0(\cdot)\rangle\\
    \vdots \\
    \langle (\kappa^2-\Delta)^{1/2}(\tau u(\cdot)), (\kappa^2-\Delta)^{1/2}\varphi_{K+1}(\cdot)\rangle \end{bmatrix} \overset{\mathrm{d}}{=} \begin{bmatrix}
\langle\mathcal W(\cdot), (\kappa^2-\Delta)^{1/2}\varphi_0(\cdot)\rangle\\
    \vdots \\
   \langle\mathcal W(\cdot), (\kappa^2-\Delta)^{1/2}\varphi_{K+1}(\cdot)\rangle\end{bmatrix}.
\end{equation*}
This is a least squares FEM, where the test functions are $(\kappa^2-\Delta)^{1/2}\varphi_j(\cdot)$. See \citet{SPDE} for details.

Let $\matr C$ be the mass matrix, given by $C_{ij} = \langle \varphi_i(\cdot), \varphi_j(\cdot)\rangle$, and $\matr G$ be the stiffness matrix, given by $G_{ij}=\langle \varphi_i'(\cdot),  \varphi_j'(\cdot)\rangle$.
Then inserting the FEM representation $u^{\mathrm{FEM}}(\cdot)$ into the above linear system of equations, gives a 
linear system of equations
$$
\matr (\kappa^2 C+G)\tau^2 \vect w \overset{\mathrm{d}}{=} \vect (\kappa^2C+G)^{1/2} \vect z,
$$
where $\boldsymbol{z} \sim \mathcal{N}(\boldsymbol{0}, \matr I)$. This leads to a Gaussian distribution $\vect{w}\sim \mathcal{N}(\boldsymbol{0}, {\matr Q}^{-1})$, where 
the precision matrix is given by
\begin{equation}
    \matr Q = (\kappa^2 \matr C + \matr G)/\tau^2.
\end{equation}
A popular parameterization of the GRF with the SPDE approach for one-dimensional domain is through range $\rho= 2/\kappa$ and marginal variance $\sigma^2=1/(2\kappa\tau^2)$.

\subsection{SPDE approach on a metric graph}\label{ssec:spdegraph}
We now consider the extension from one edge to the entire graph,
\begin{equation}\label{eq:SPDE_graph}
	(\kappa^2-\Delta_\Gamma)^{\alpha/2}(\tau u(\vect s)) = \mathcal W(\vect s) \quad \vect s\in \Gamma,
\end{equation}
where $\alpha>1/2$, $\kappa,\tau>0$, $\Delta_\Gamma$ is the Kirchhoff-Laplacian \citep{metric_graph}, $\mathcal W(\cdot)$ is Gaussian white noise. Zero Neumann boundary conditions is enforced at all terminal vertices. The Kirchhoff-Laplacian, $\Delta_\Gamma$, is the standard second-order derivative (using $t\in[0,l_e]$) in the interior of each edge. To have a unique inverse for this operator, one enforces, for the inverse, continuity at the vertices and a zero-net-flow condition for the directional derivatives at each edge that meets in a vertex. We do not extend the domain to alleviate boundary effects as we consider the graph fixed by the application. Boundary effects will occur on terminal vertices, which can be classified as either dead ends or vertices where we have cut an edge due to limiting the study area. For the second type of boundary vertices, one can apply Robin boundary conditions, which is demonstrated in \citet{metric_graph}. In this paper, we consider the two types of boundary vertices equally, and set the same boundary conditions for all terminal vertices.
The Gaussian white noise can be thought of as defined independently for each edge. Again we fix $\alpha=1$ in this work, but it is possible to combine FEM with rational approximation methods \citep{rspde_article}. Note that, when looking at the whole graph, setting $\alpha=1$, does not result in an exponential covariance function.

As seen in Section \ref{ssec:spde1d}, we can solve the SPDE in a straightforward way on a single edge with zero Neumann boundary conditions, the challenge with the graph is to handle the interface conditions between edges that meet at internal vertices. The zero Neumann boundary conditions are only enforced at terminal vertices.
Note that in this case there is an exact Markov representation available \citep{metric_graph}, but we pursue a FEM representation because we aim to be able to integrate the solution over arbitrary parts of the graph. 
\citet{metric_graph_FEM} demonstrate existence and uniqueness of the FEM-solution of the SPDE. We summarize the key parts of the derivation below.

Similar to Section \ref{ssec:spde1d}, we construct a finite set of basis functions. Each edge is discretized into a regular grid of intervals, with interval widths that may vary slightly across edges. Figure \ref{fig:graphmeshillustration} illustrates how these regular grid points are added to an edge.

\begin{figure}
    \centering
    \begin{subfigure}[c]{0.48\textwidth}
        \centering
        \includegraphics[width=\linewidth]{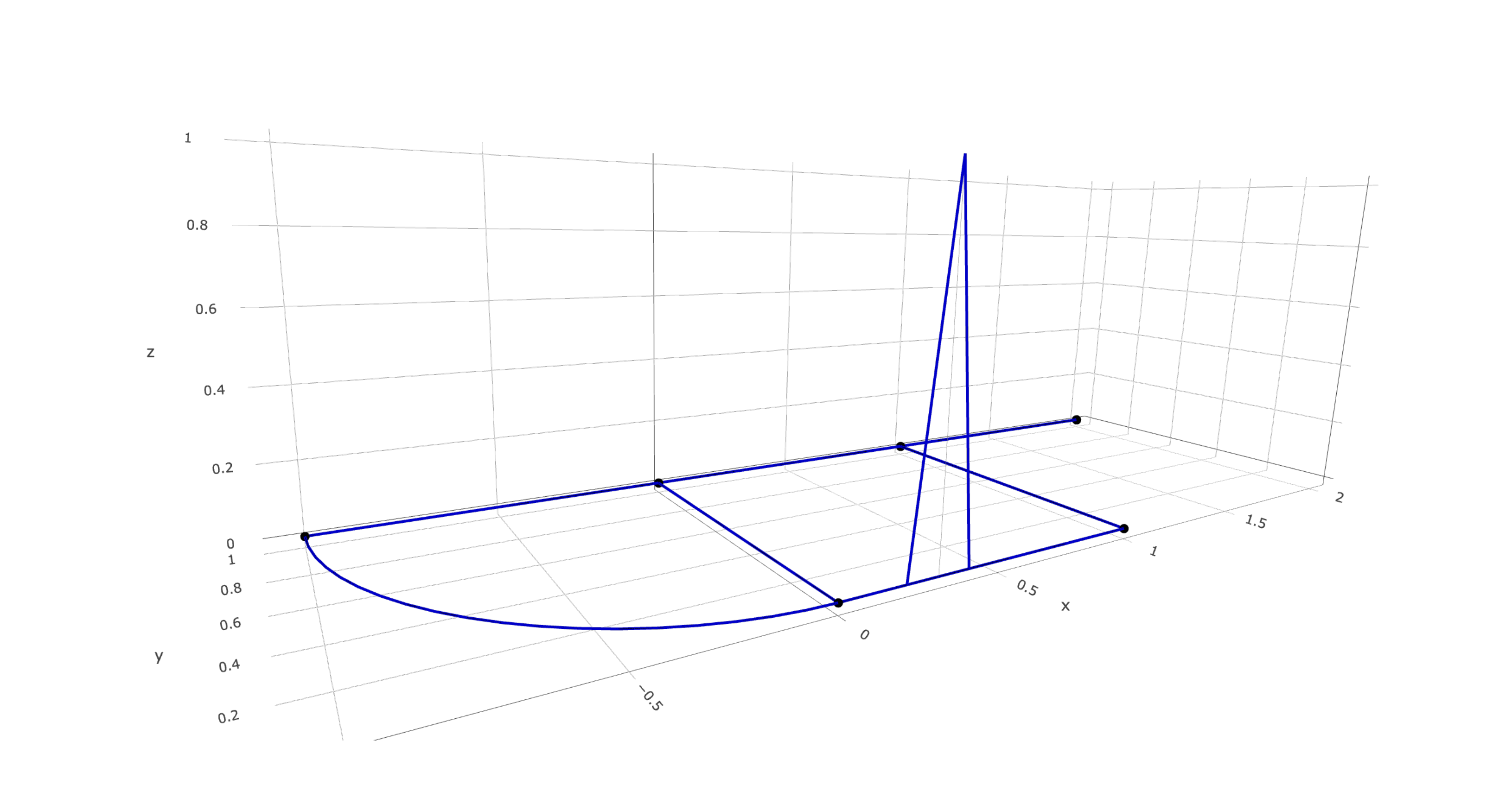}
        \caption{Non-internal basis function.}
        \label{fig:basis_func_ex2}
    \end{subfigure}
    ~
    \begin{subfigure}[c]{0.48\textwidth}
        \centering
        \includegraphics[width=\linewidth]{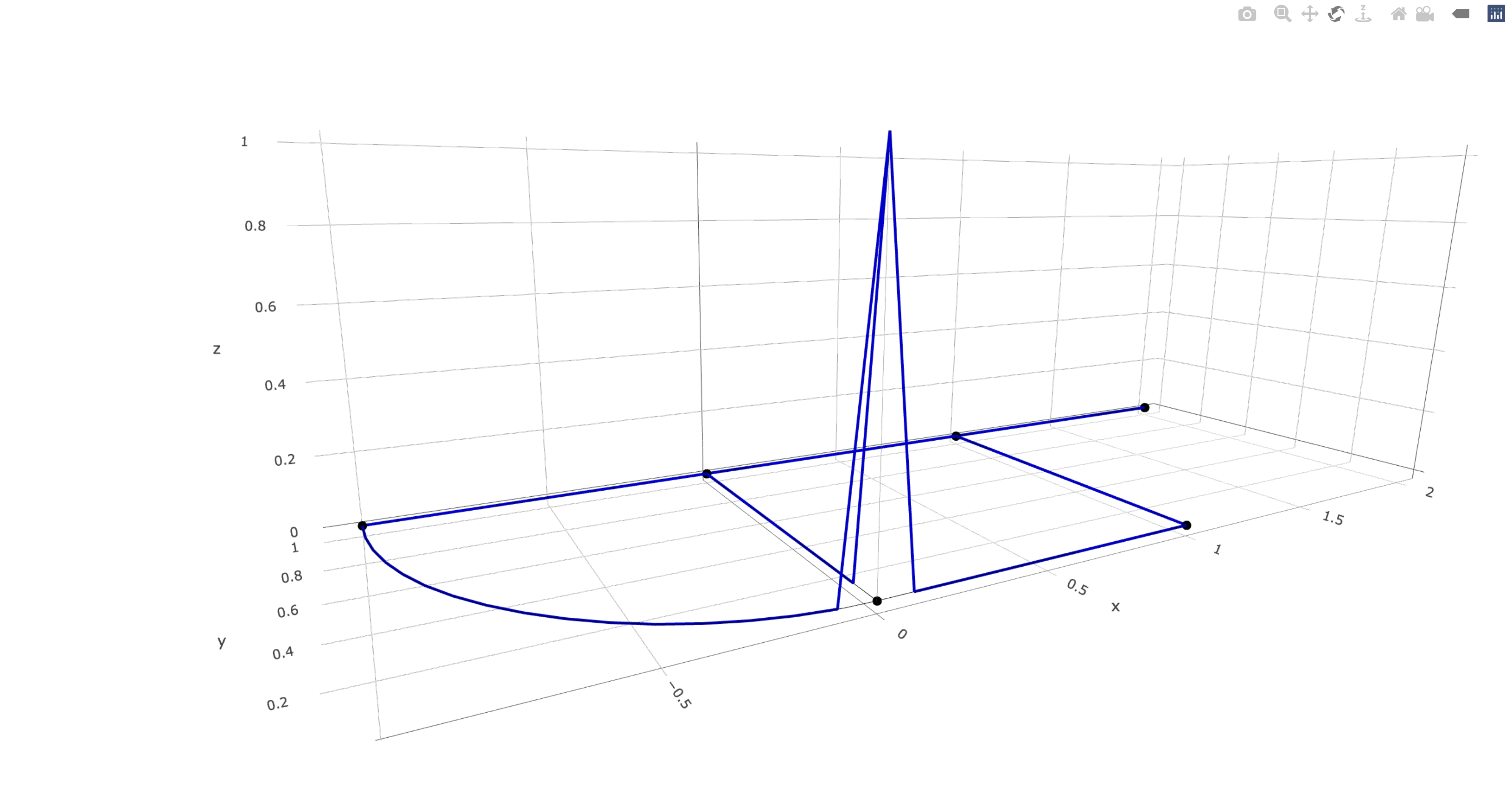}
        \caption{Internal vertex basis function.}
        \label{fig:basis_func_ex1}
    \end{subfigure}%
    \caption{Examples of (a) non-internal vertex basis function, and (b) internal vertex basis function on a simple graph with six graph vertices and seven edges. The mesh distance is approximately 0.1. Note that internal vertex basis functions linearly decrease from 1 to 0 from the graph vertex to each of the neighboring mesh vertices. For non-internal vertices, we see the similarity to a basis function for a line segment $[0,l_e]$.}
    \label{fig:basis_functions}
\end{figure}

For grid points corresponding to terminal vertices or not located at an edge endpoint, the basis functions are piecewise-linear pyramidal functions, as described in Section \ref{ssec:spde1d} (see Figure \ref{fig:basis_func_ex2}). Each edge $e$, is associated with $K_e$ such basis functions. For a grid point that is an internal vertex, the basis function remains piecewise linear but takes value of 1 at the vertex and decreases to zero at the next grid points along each connected edge. See Figures \ref{fig:basis_func_ex1} for a visualization. 
Denote the resulting set of basis functions
$\{\varphi_k(\cdot)\}_{k =1\ldots,K}$. Note that $K=m+\sum_{e\in\mathcal E}K_e$, where $m$ is the number of vertices in the graph $\Gamma$ and $K_e$ is the number discretization points (not including end points) added in each edge, as described in Section \ref{ssec:spde1d}. 

The FEM representation can then be written as $u^{\text{FEM}}(\cdot) = \sum_{k=1}^K \text{w}_k \varphi_k(\cdot)$, and the only difference from Section \ref{ssec:spde1d} is that we define the inner product as
\begin{equation*}
    \langle f(\cdot), g(\cdot)\rangle_{L_2(\Gamma)} = \sum_{e\in\mathcal E}\int_{e} f((e,t))g((e,t)) \:\dif t,
\end{equation*}
where the integral over $e$ should be understood as a standard one-dimensional integral over $[0,l_e]$ and $f((e,t))$ is the function $f(\cdot)$ evaluated at the graph location given by $(e,t)$ where $e$ is the edge index and $t\in[0,l_e]$. The corresponding system of equations is
\begin{equation*}
    \begin{bmatrix}\langle (\kappa^2-\Delta)^{1/2}(\tau u(\cdot)), (\kappa^2-\Delta)^{1/2}\varphi_1(\cdot)\rangle_{L_2(\Gamma)}\\
    \vdots \\
    \langle (\kappa^2-\Delta)^{1/2}(\tau u(\cdot)), (\kappa^2-\Delta)^{1/2}\varphi_K(\cdot)\rangle_{L_2(\Gamma)} \end{bmatrix} \overset{\mathrm{d}}{=} \begin{bmatrix}
\langle\mathcal W(\cdot), (\kappa^2-\Delta)^{1/2}\varphi_1(\cdot)\rangle_{L_2(\Gamma)}\\
    \vdots \\
   \langle\mathcal W(\cdot), (\kappa^2-\Delta)^{1/2}\varphi_K(s)\rangle_{L_2(\Gamma)}\end{bmatrix}.
\end{equation*}
Exactly as in Section \ref{ssec:spde1d}, this results in $\boldsymbol{w}\sim \mathcal{N}(\boldsymbol{0}, {\matr Q}^{-1})$, where the precision matrix is given by
\begin{equation}
    \matr Q = (\kappa^2 \matr C + \matr G)/\tau^2,
\end{equation}
and $\matr C$ is the mass matrix of the basis and $\matr G$ is the stiffness matrix of the basis.

The resulting field from this approach does not generally exhibit a covariance structure strictly following the Mat\'ern form. 
Between points on the same edge, i.e $(e, t_1)$ and $(e,t_2)$ for $t_1,t_2\in(0,l_e)$ and sufficiently distant from the edge endpoints, the range $\rho$ and marginal variance $\sigma^2$ retain the same physical interpretation as in the one-dimensional case. 
It is notable that at vertices where more than two edges intersect, the covariance structure will not be Mat\'ern. This is a direct consequence of the Kirchhoff node conditions. However, we can still refer to the practical correlation range $\rho=2/\kappa$ and marginal variance $\sigma^2=1/(2\kappa\tau^2)$, which are more interpretable than $\kappa$ and $\tau$ when constructing models with \texttt{rSPDE} and \texttt{MetricGraph}. 

\FloatBarrier

\section{Spatial modelling on metric graphs}\label{sec:model}
\subsection{Observation models}
Inspired by the application in Section \ref{sec:motivation}, we focus on observation models for two types of data: point observations and line observations. Assume that the true signal is given by $\eta(\boldsymbol{s})$, $\boldsymbol{s}\in\Gamma$, where $\Gamma$ is the graph of interest, and
let $g$ be a real function.
The observation model
for point observations $\yp_1, \yp_2, \ldots, \yp_{n_p}$ supported at points $\boldsymbol{s}_1, \ldots, \boldsymbol{s}_{n_\mathrm{P}}\in\Gamma$, respectively, is
\begin{equation*}
	\yp_j = g(\eta(\boldsymbol{s}_j)) + \varepsilon^\mathrm{P}_j \quad j=1,\ldots,n_\mathrm{P},
\end{equation*}
where $\vect \varepsilon^\mathrm{P} = (\varepsilon^\mathrm{P}_1, \ldots, \varepsilon^\mathrm{P}_{n_\mathrm{P}})^\mathrm{T}|\sigma_\mathrm{P}^2\sim \mathcal{N}_{n_\mathrm P}(\vect 0, \sigma_\mathrm{P}^2\matr I)$, 
and the point measurement noise variance is $\sigma_{\mathrm{P}}^2 >  0$.

Line observations $\yl_1, \ldots, \yl_{n_\mathrm{L}}$ are supported by lines $\mathrm L_1, \ldots, \mathrm L_{n_\mathrm{L}}\subset \Gamma$, respectively, and need to be connected to the behavior along the line. We consider line observation models of the form
\begin{equation}
	\yl_i = \sum_{e\in\mathcal{E}}\int_{\mathrm{L}_i\cap e} g(\eta((e,t)) \dif t + \varepsilon^\mathrm{L}_i\quad i=1,\ldots,n_\mathrm{L},
 \label{eq:integral}
\end{equation}
where the integral is a standard one-dimensional integral for each edge over the part of the edge intersected by the line. The errors $\vect \varepsilon^\mathrm{L} = (\varepsilon^\mathrm{L}_1, \ldots, \varepsilon^\mathrm{L}_{n_\mathrm{L}})^\mathrm{T}|\sigma_\mathrm{L}^2\sim\mathcal{N}_{n_\mathrm L}(\vect 0, \matr D_\mathrm{L})$, where the matrix $\matr D_\mathrm{L}$ is diagonal and line measurement variances are linked to the length of the line through $D_{ii} = h(|\mathrm L_i|)\sigma_\mathrm{L}^2$, $ i = 1,\ldots, n_\mathrm{L}$ for a positive real function $h$ and line variance parameter $\sigma_\mathrm{L}^2> 0$.
Such line observations are naturally arising when observing a section speed of a vehicle on a road segment, which is a result of integrating a momentary inverse speed along the line segment. This is further explained in Section \ref{sec:case}.

\subsection{Hierarchical model}\label{ssec:hierarchicalmodel}
We model the latent signal through
\begin{equation*}
	\eta(\boldsymbol{s}) = \beta_0 + \vect x(\boldsymbol{s})^\mathrm{T}\boldsymbol{\beta} + u(\boldsymbol{s}),\quad \boldsymbol{s}\in \Gamma,
\end{equation*}
where $\beta_0$ is the intercept, $\vect \beta$ is a $p$-dimensional vector of coefficients,   $\vect x(\cdot)$ is a vector of $p$ spatially varying covariates, and $u(\cdot)$ is a zero-mean GRF. We assume $u(\cdot)$ is the GRF constructed in Section \ref{sec:preliminaries} with parameters $\vect \theta=(\sigma^2, \rho)^\mathrm{T}$. Then $\eta(\cdot)|\beta_0, \vect \beta, \vect \theta$ is a GRF with mean function $\mu(\cdot)=\beta_0+ \vect x(\cdot)^\mathrm{T}\vect\beta $ and the covariance structure arising from the SPDE approach on the graph $\Gamma$. 

Assume $u(\cdot)$ is described by $K$ weights $\vect w|\vect \theta \sim \mathcal{N}_{K}(\boldsymbol{0}, \matr Q(\vect \theta)^{-1})$. Then we can write the point observations in vector form $$
    \vect \yp = g(\beta_0 \vect 1 + \matr X_\mathrm{P}\vect \beta +\matr A^\mathrm{P}\boldsymbol{w}) + \vect \varepsilon^\mathrm{P},
$$ 
where $\vect 1$ is a vector of ones, $\matr X_\mathrm{P}$ is the design matrix for the observation locations, and $\matr A^\mathrm{P}$ is $n_\mathrm{P}\times K$ matrix from evaluating the compactly supported basis at the different measurement locations. The line observation model can be written in the form
$$
    \vect \yl = \matr A^\mathrm{L}g(\beta_0 \vect 1 + \matr X_w\vect \beta +\vect w)+\vect \varepsilon^\mathrm{L},
$$
where $\vect 1$ is a vector of ones, $\matr X_w$ is the design matrix corresponding to grid locations, and $\matr{A}^\mathrm{L}$ is the $n_\mathrm{L} \times K$ matrix needed to compute the integrals in Equation \eqref{eq:integral}. Note that this makes an assumption that covariates are approximated as piece-wise linear on the discretization, and that $g(\cdot)$ is assumed to be applied element-wise to the vector.

The hierarchical model is then
\begin{align*}
	\vect \yp \mid \vect w , \vect\theta, \sigma_\mathrm{P}^2,  \beta_0, \vect \beta &\sim \mathcal{N}_{n_\mathrm{P}}(g(\beta_0 \vect 1 + \matr X_\mathrm{P}\vect \beta +\matr A^\mathrm{P}\boldsymbol{w}),\sigma_\mathrm{P}^2 \matr I), \\
    \vect \yl \mid \vect w,\vect \theta, \sigma_\mathrm{L}^2, \beta_0, \vect \beta &\sim \mathcal{N}_{n_\mathrm L}(\matr A^\mathrm{L}g(\beta_0 \vect 1 + \matr X_w\vect \beta +\vect w),\matr D_\mathrm{L}),\\
	\vect w \mid\vect\theta \sim \mathcal{N}_{K}(\vect 0, \matr Q(\vect\theta)^{-1}), \:
    \beta_0 &\sim \mathcal{N}_1(0, V),\: \vect \beta \sim \mathcal{N}_p(\vect 0,V \matr I) \\
    1/\sigma^2_{\mathrm{L},i}, \: 1/\sigma^2_\mathrm{P}&\overset{\text{iid}}{\sim} \text{Gamma}(\alpha_\sigma,\beta_\sigma),\: \log(\vect \theta) \sim \mathcal{N}_2(\vect \mu_\theta, \boldsymbol{\Sigma}_\theta),
\end{align*}
where we will choose a vague prior $V = 10^3$, and set $\alpha_\sigma = 1$ and $\beta_\sigma = 5\cdot 10^{-5}$. The latter two are the default priors in \texttt{R-INLA}. The notation means $\log(\vect \theta)=(\log(\sigma^2), \log(\rho))^\mathrm{T}$. We will set $\vect\mu_\theta = (\log(1), \log(0.700))^\mathrm{T}$, that is $\sigma^2=1$ and $\rho=\SI{0.700}{\kilo\meter}$, and $\boldsymbol{\Sigma}_\theta=\text{diag}(0.1^{-1},0.1^{-1})$ for all models considered based on the graph in Figure \ref{fig:studyarea}. 
This is the model with the correct support, from here abbreviated \IM.
We also consider a simplified model with wrong support for the line observations, abbreviated \SM, where each line observation $\yl_i$ is treated as point observations assigned to the midpoint, along the line, of $\mathrm L_i$ while we keep $\sigma_{\mathrm{L},i}^2 =h(|\mathrm{L}_i|) \sigma_\mathrm{L}^2$ for $i = 1, \ldots, n_\mathrm{L}$.

\subsection{Coordinate representation on metric graphs}
For a metric graph that lives in $\mathbb R^2$, we can represent points in the graph in Euclidean coordinates, $(x,y)$. In spatial data, these coordinates are related to some defined coordinate reference system, \textit{CRS}. One common CRS is longitude-latitude coordinates that are defined on the whole globe. When working with smaller subregions, it is often more useful to switch to some local CRS, and Universal Transverse Mercator \textit{UTM} is a commonly used reference system. In the \texttt{MetricGraph}-package one refers to such coordinate representation as \texttt{XY}, and it is a two-column matrix, where the first column is typically for longitude or Easting, and the second column is for latitude or Northing.

Typically, a graph is represented by its set of vertices, $\mathcal{V}$, and set of edges, $\mathcal E$. Any point on the graph can be represented by an edge index $e$ and a normalized distance $\tilde t\in[0,1]$ ($\tilde t=t/l_e$ for an unnormalized $t\in[0,l_e]$). This representation is graph-specific, and knowledge about the graph is necessary for extracting information about the physical location of such coordinate. The advantage of such coordinate representation is that two points with the same edge index and location on that edge which is similar will refer to points that are "close" in the sense one is interested in when doing spatial modelling. This coordinate system gives a unique representation for all locations on edges in the graph, while vertex locations are ambiguous. For a vertex of degree 2 (where two edges connect), we have two possible representations for that vertex in the coordinate system, using either of the two edges connected in that vertex and $\tilde t$ equal to zero or one. For degree 3, we have three such representations. 
In the \texttt{MetricGraph}-package such coordinate representation is called \texttt{PtE}, and it is a two-column matrix. The first column is the edge index, and the second column specified the (un)normalized distance. Typically, one specifies if the coordinate is normalized or not with the additional boolean argument \texttt{normalized}.

Conversion between Euclidean coordinates and this graph representation is done through \texttt{\$coordinates()}, which is a built-in function related to the graph object itself. The metric graph object in \texttt{MetricGraph} can optionally have a mesh (previously referred to as a grid) associated with it, where the maximal distance between mesh vertices is decided upon construction. That is, the mesh is a new graph contained within the original graph, where extra vertices are added on all edges with approximately the same distance. 

Inter-edge intervals are defined as intervals which are a subset of spatial locations on a single edge of a metric graph, $(e,t_s,t_e)$ where $e$ is the edge index, $t_s$ is the start and $t_e$ is the end of the interval. 
We consider a valid path as a collection of connected inter-edge intervals, where the end point of one inter-edge interval is spatially equivalent to the start of the following inter-edge interval.
E.g. you have a valid path with start $(\text{e}_1, 0.7)$, set of visited edges $\{\text{e}_4,\text{e}_7,\text{e}_{15}\}$ and end $(\text{e}_{13}, 0.75)$, as can be seen in Figure \ref{fig:graphmeshillustration} as a red dotted line through the graph. 

\subsection{Implementation details}
We fit our model using the \texttt{INLA} framework \citep{INLA_web} which makes use of nested Laplace approximations \citep{INLA} to perform fast approximate Bayesian inference. This approach is deterministic and works on the large class of latent Gaussian models (LGMs). For known link functions between observations, $y_i$, observed in location $\vect s_i$, and the linear predictor, $\eta_i=\eta(\vect s_i)$, we can use \texttt{R-INLA}. Whenever the observations are linked to the linear predictor in a non-linear way, that is $y_i$ is not a linear function of $\eta_i$ alone, we are not within the \texttt{INLA}-framework anymore, but the wrapper \texttt{inlabru} \citep{inlabru2, inlabru} supports these types of models with non-linear predictors. The non-linearity is handled by using a first-order Taylor approximation. 
\SMs fits directly into the standard \textsc{R-INLA} framework, but \IMs does not due to the potential non-linearity due to the function $g(\cdot)$ inside the integral. However, \IMs fits into the \texttt{inlabru} framework.

The \texttt{MetricGraph} package has an interface with \texttt{R-INLA} and \texttt{inlabru} for point support. An introduction to point support modeling is given in detail in the vignettes at \citet{MetricGraph}. 
In this work, we have extended the possible spatial support for metric graphs to line support inspired by the methods in \texttt{fmesher} \citep{fmesher}. \texttt{fmesher} is an \texttt{R}-package, that was originally part of \texttt{R-INLA}, which provides tools for handling triangle meshes and other geometries.
To implement our model to handle integration as described for the \IM, it was necessary to expand existing functionality in \texttt{fmesher} to handle \texttt{metric\_graph} objects from \texttt{MetricGraph}. 
In particular, we add classes for bary-centric coordinates (\texttt{fm\_bary}), basis functions (\texttt{fm\_basis}), methods for integration (\texttt{fm\_int}) for points and intervals on \texttt{metric\_graph} objects from \texttt{MetricGraph}, and \texttt{inlabru}-mappers for our type of model. 
With these extensions, we allow users to provide well-defined paths on the graph that are related to more complex support than point observations, which can result from integration, summation, etc. We continue with a description of the code that extends the existing methods for metric graphs.

We have created classes for two coordinate types on metric graphs, named \texttt{fm\_MGG\_bary} and \texttt{fm\_MGM\_bary}, where the first is ``Metric Graph Graph'' coordinates, $(e, \tilde t)$, (equivalent to \texttt{PtE} in \texttt{MetricGraph}) and the latter is ``Metric Graph Mesh'' coordinates. We distinguish between these two classes because both are useful for different scenarios, and these classes help with recognizing which method should be used. The \texttt{fm\_MGM\_bary} coordinates refer to mesh edge indices and distances on that mesh edge, which are useful when we construct basis functions on the mesh to perform FEM. It is also possible to have a graph object that acts as the mesh which FEM is utilizing. Conversion between the two coordinate systems is handled internally when necessary, and the user only deals with the \texttt{fm\_as\_MGG\_bary} coordinates. As \texttt{fm\_as\_MGG\_bary} is equivalent to the \texttt{PtE}-format used in \texttt{MetricGraph}, users of this package will become familiar with this coordinate type.
Such coordinates are constructed using \texttt{fm\_bary()} (see documentation for what input can be handled). One of the useful input objects \texttt{fm\_bary()} supports is \texttt{sf}-objects, mainly \texttt{st\_point}-objects. 

Additionally, we provide support for constructing valid paths on metric graphs through \texttt{fm\_MGG\_intervals}. We provide two methods for obtaining such paths; (1) through a list of three elements (start location, ordered collection of edges visited and end location) or (2) through a geometric line (\texttt{sfc\_LINESTRING}) which is a subspace of the full graph, $\mathrm{L}\subset \Gamma$. 
The first function is useful when you work with smaller graphs where $\mathcal E$ is of manageable size such that the start and end location in \texttt{PtE}-format is known, and additionally the indices of edges that the path runs through between these two locations are known. This is the method used in \ref{sup:code_example}.
The second path construction is useful when working with more complex graphs that are built on multiple spatial geometries (spatial lines in \texttt{sp} or \texttt{sf}) and the paths are lines which are subsets of the graph. The function will consider \texttt{sfc\_LINESTRING} that are collections of line segments, where each segment is treated as its own path object. The output is a \texttt{tibble} (\texttt{data.frame}-like object) from CRAN-package \texttt{tibble} \citep{tibble}. It has two columns; (1) \texttt{path}: collection inter-edge intervals and (2) an \texttt{ID} referencing what line segment it came from. 

From lists of valid paths one can use \texttt{fm\_int} to construct integration points and weights. The user can choose to modify their \texttt{sfc\_LINESTRING}-object such that each line segment represents the correct block, or modify the output from \texttt{fm\_MGG\_intervals} to have the desired list construction. The first option requires some knowledge on the \texttt{sf}-package and how to work with \texttt{sf\_sfc} objects, while the latter requires manipulation of \texttt{tibble}s. 

One needs to choose a suitable numerical approximation scheme to approximate the integral in \eqref{eq:integral} and here we propose the Simpsons' rule for a function $f(\cdot)$ defined on the graph, 
\begin{align*}
    \int_{\mathrm L_i} f(\vect s) \dif \vect s = \sum_{e\in\mathcal E} \int_{L_i\cap e} f((e,t))) \dif t \approx \sum_{e\in\mathcal E}\frac{h_e}{3} &\bigg[ f(t^{\mathrm L_i\cap e}_0) + 4 \sum^{\tilde K/2}_{k=1} f(t^{\mathrm L_i\cap e}_{2k-1}) \\
    &+2 \sum^{\tilde K/2-1}_{k=1} f(t^{\mathrm L_i\cap e}_{2k})+f(t^{\mathrm L_i\cap e}_{\tilde K}) \bigg]
\end{align*}
where the integral over $\mathrm L_i$ is understood as a line integral along a subset of the full graph $\Gamma$, which is equal to the sum of integrals over all edges intersecting with $\mathrm L_i$. Let $\{t_k\}_{k=0}^{K+1}$ be a set of equidistant points on edge $e$ with distance $h_e$. Then $\{t^{\mathrm L_i\cap e}_k\}_{k=0}^{\tilde K}\subset \{t_k\}_{k=0}^{K+1}$, where the subset includes all $\tilde K+1$ grid points that are in the intersection $\mathrm L_i\cap e$. 
This is the default integration scheme for metric graph objects with aggregated observations. Note that when $g(\cdot)$ is linear, this integration scheme is exact, while for non-linear functions the scheme is approximate. 
We provide a minimal example of how the methods can be applied in \ref{sup:code_example}. 

All code is written in \texttt{R} \citep[R Version 4.4.2 (2024-10-31)]{R-language} and run with \texttt{inlabru} version 2.10.1 (\texttt{R-INLA}: 24.11.25), \texttt{MetricGraph} version 1.3.0.9000 \citep{MetricGraph} and \texttt{fmesher} version 0.2.0 \citep{fmesher}. The code which was used to obtain all results presented in this paper is available at \citet{code} along with examples of how the code works and the code used to produce the results in Section \ref{sec:sim_study}. A minimal example on how the code can be used is provided in \ref{sup:code_example}. Please note that the code is sensitive to what version of \texttt{fmesher} and \texttt{MetricGraph} that is used due to frequent updates in these two packages.
A prototype of this code, that will go into other packages, can be found on the branch \texttt{feature/MetricGraphPaper} on the inlabru-org/fmesher GitHub repository.

\section{Simulation study}\label{sec:sim_study}
\subsection{Motivation and design}
We are interested in understanding how many temporal replicates that are needed for the metric graph and spatial design introduced in Section \ref{sec:motivation} to perform well with regards to prediction and parameter estimation. This is done in the setting of medium and long range.
Additionally, we are interested in understanding the importance of using the correct observation model over the simplified observation model. 
We aim to simulate under a known latent model and simulate observations according to the correct observation model \IM, and then compare performance of \IMs and \SMs with respect to spatial prediction in terms of RMSE, CRPS and coverage, and parameter estimation in terms of bias and variability of the parameter estimates. 

We discretize the graph with maximum distance between grid vertices $h = \SI{70}{\meter}$. 
This results in $K=$ 2,825 grid points and 3,176 edges, and we consider this to be the true graph.
Line-level data paths are constructed using the true paths and bus stops of the four bus routes operating on the network.
See Figure \ref{fig:studyarea} for visualization. Each bus line provides between 16 and 34 line observations, and, in total, we have $n_\mathrm{L}=92$ line-level observations. Further, we include $n_\mathrm{P} = 6$ locations where we observe point data using the same locations as the true measurement stations.

\subsection{Scenarios}
\begin{figure}
    \centering
    \includegraphics[width=0.5\linewidth]{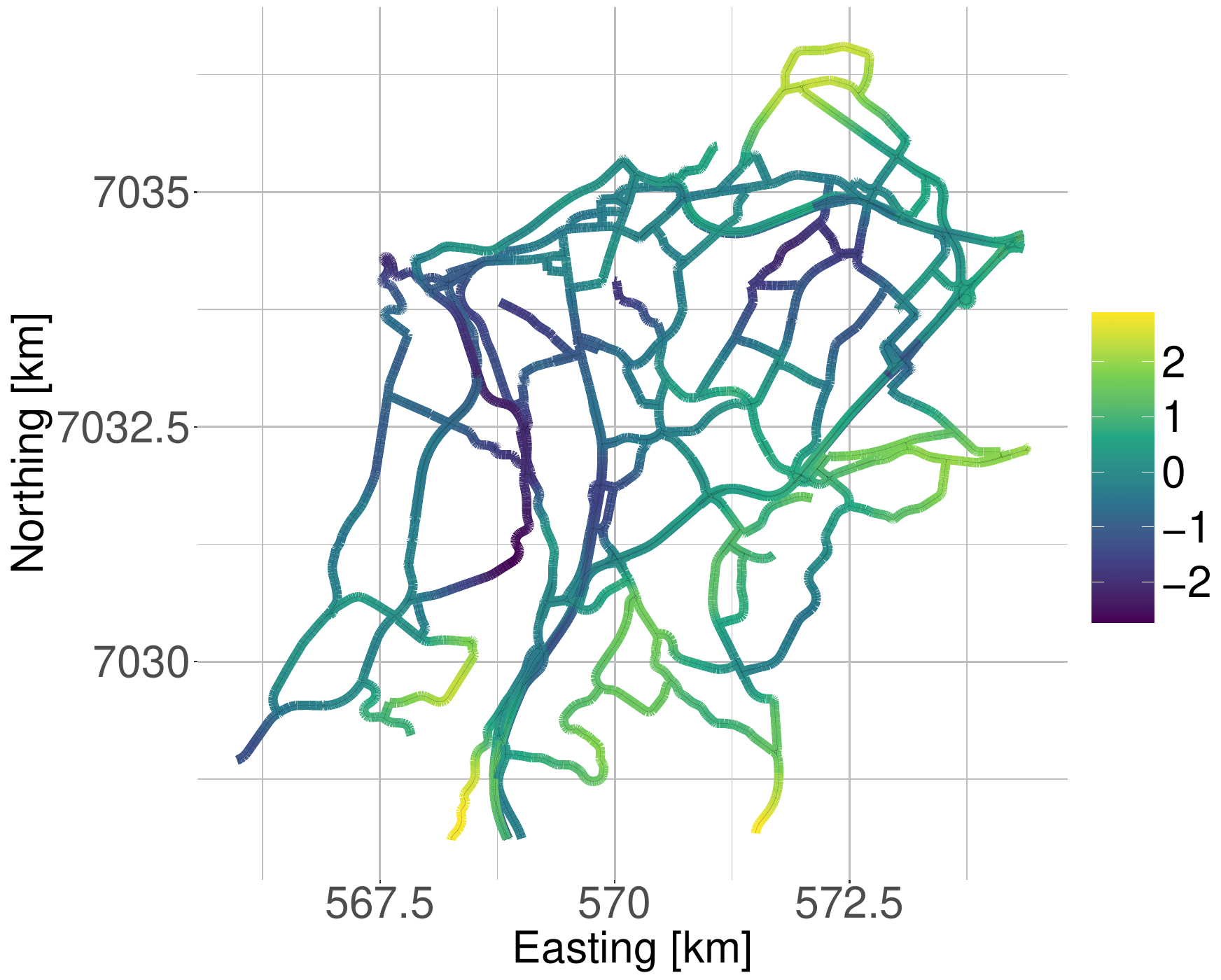}
    \caption{The standardized covariate used in the simulation study. The covariate value is known in all grid locations. The UTM zone is 32N.}
    \label{fig:sim-study-covariate}
\end{figure}

We create a spatially varying covariate $x(\cdot)$ that is fixed through all scenarios by sampling a GRF on the metric graph with parameters $\rho=\SI{6}{\kilo\meter}$ and $\sigma^2=3$. The covariate is shown in Figure \ref{fig:sim-study-covariate}.
The true latent variation is described by
\begin{equation*}
	\eta_r(\boldsymbol{s}) = 1 + x(\boldsymbol{s}) + u_r(\boldsymbol{s}),\quad \boldsymbol{s}\in \Gamma, \quad r = 1, \ldots, R,
\end{equation*}
where the GRFs $u_1(\cdot), \ldots, u_R(\cdot)$ are independent realizations of the GRF described in Section \ref{sec:preliminaries} with marginal variance $\sigma^2 = 1$, and range $\rho$ and $r=1,\ldots, R$ denotes the replicate. We consider two ranges: medium range ($\rho = 0.35\si{\kilo\meter}$) and long range ($\rho = 1\si{\kilo\meter}$). For each range, we consider $R=1$, $R=5$ and $R=25$. This gives a total of six scenarios for $\rho$ and $R$ as summarized in Table \ref{tab:scenarios}. For each scenario, we simulate 50 realizations of the true weights, $[\vect w_1(\cdot), \ldots, \vect w_R(\cdot)]$ where each realization $\vect w_r$ for $r=1,\ldots, R$ is one realization of $\vect w\mid \vect\theta$. 

From these sets of weights, we can create $R$ sets of line observations $\vect y^\mathrm{L}_r$ and point observations $\vect y^\mathrm{P}_r$ for $r=1,\ldots, R$. Inspired by the application, we set the line observation variance and point observation variance to $25\%$ and $1\%$, respectively, of the marginal variance for the spatial field, i.e., $\sigma_\mathrm{L}^2=0.25$ and $\sigma_p^2=0.01$. We use the observation model \IMs described in Section
\ref{ssec:hierarchicalmodel} with an identity link function $g(\cdot)$ and end up with $n_\mathrm{P} = 6$ point observations and $n_\mathrm{L} = 92$ line observations for each of the $R$ replicates.

\begin{table}[htb]
    \centering
    \caption{Overview of the six scenarios considered in the simulation study.}
    \begin{tabular}{c c}
       Range ($\rho$) & Number of replicates ($R$)  \\
       \hline
       Medium (0.35\si{\kilo\meter})  & 1, 5, and 25 \\
       Long (1\si{\kilo\meter}) &  1, 5, and 25 \\
    \end{tabular}
    \label{tab:scenarios}
\end{table}

\subsection{Candidate models and evaluation}
The candidate models use the latent model
\begin{equation*}
	\eta_r(\boldsymbol{s}) = \beta_0 +  x(\boldsymbol{s})\beta_1 + u_r(\boldsymbol{s}),\quad \boldsymbol{s}\in \Gamma, \quad r = 1, \ldots, R,
\end{equation*}
where the model components and priors are as described in Section \ref{ssec:hierarchicalmodel} with the exception that $u_1(\cdot), \ldots, u_r(\cdot)|\vect \theta$ are independent realizations of the GRF. Additionally we augment the observation models to handle line data $y_{r,i}^\mathrm{L}$, $i=1,\ldots, n_\mathrm{L}$, and point data $y_{r,j}^\mathrm{P}$, $j = 1,\ldots, n_\mathrm{L}$, for replicate $r = 1, \ldots, R$.
We consider both \SMs and \IM, and use $h(|\mathrm{L}_i|)=1/|\mathrm{L}_i|^2$ as the known scale for the variance of the line observations.
\IMs is fitted using the known identity link function, $g(\cdot)$. When fitting \SM, we assign the line observation to the midpoint of the line as measured along the line. To allow for a fair comparison between the models, we assign the average covariate along the line to the point. 

To evaluate the predictive performance, we compute RMSE, CRPS and coverage for the marginal posteriors of $\eta_r(\vect s_k)$, where $k=1,\ldots, K$ indexes mesh vertices and $r = 1,\ldots, R$ indexes replications. For \IMs and \SM, summary quantities are computed as the mean across all mesh vertex locations and replicates.
RMSE evaluates the point predictions, CRPS evaluates the marginal predictive distributions, and coverage assesses the accuracy of $95\%$ credible intervals. 
Lower RMSE and CRPS are better, and coverage closer to the nominal level $95\%$ is better. We use posterior means as parameter estimates for $\beta_0$ and $\beta_1$ and posterior medians as the parameter estimates for $\sigma^2$, $\rho$, $\sigma_\mathrm{P}^2$, and $\sigma_\mathrm{L}^2$. We consider bias and variability in parameter estimates.

\subsection{Results}
\begin{figure}[htb]
    \centering
    \begin{subfigure}[b]{0.45\textwidth}
        \centering
        \includegraphics[width=\linewidth]{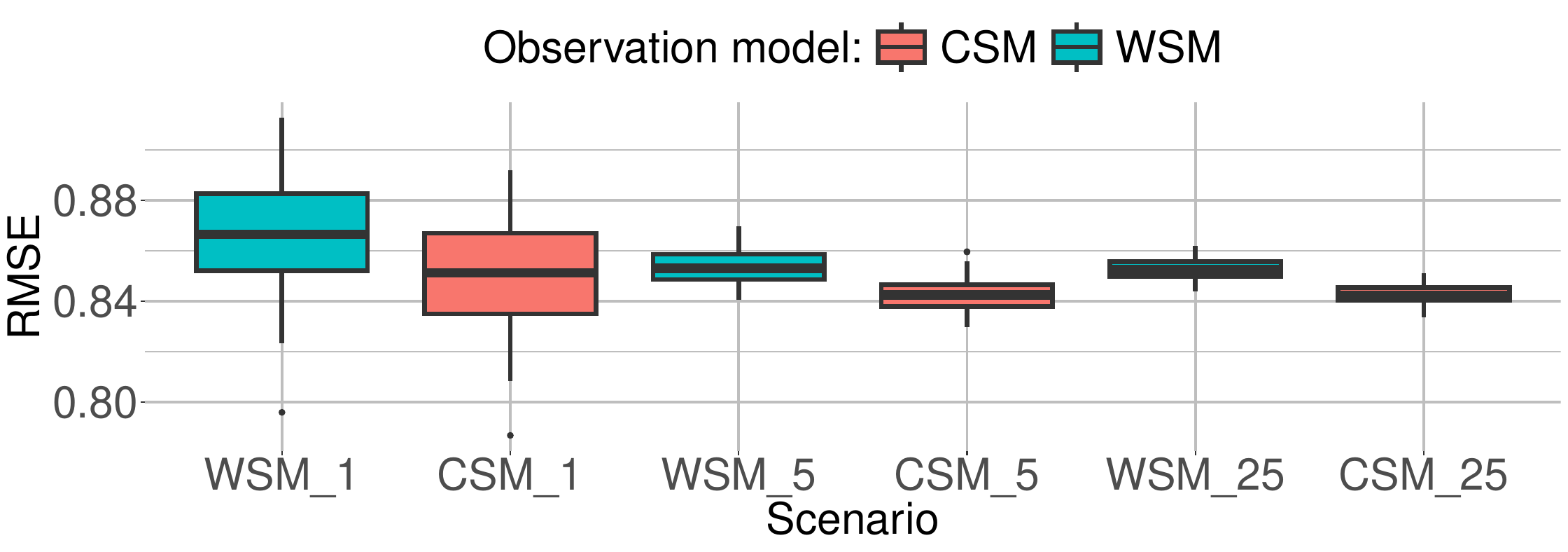}
        \label{fig:RMSE_short}
    \end{subfigure}%
    \begin{subfigure}[b]{0.45\textwidth}
        \centering
        \includegraphics[width=\linewidth]{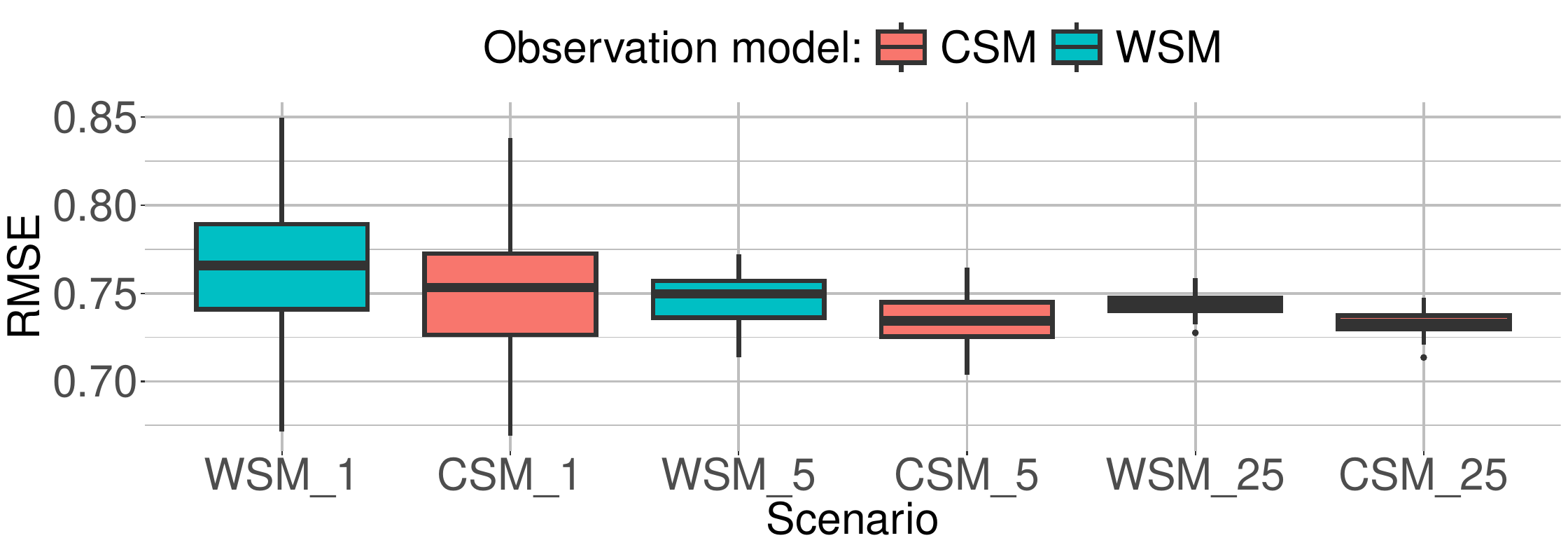}
        \label{fig:RMSE_long}
    \end{subfigure}%
    ~
    \newline
    \begin{subfigure}[b]{0.45\textwidth}
        \centering
        \includegraphics[width=\linewidth]{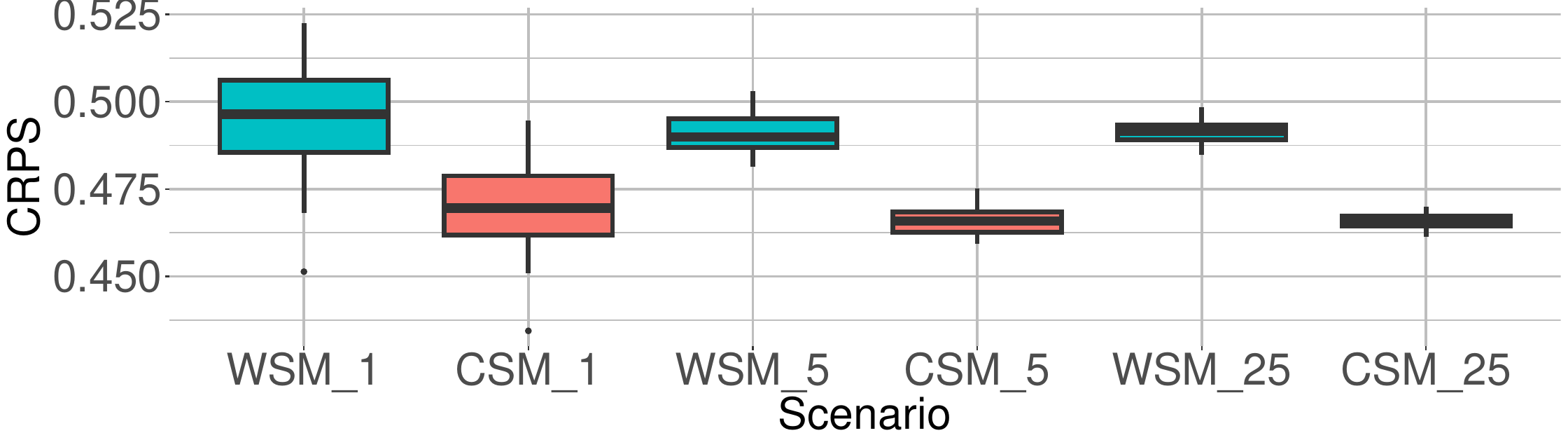}
        \label{fig:CRPS_short}
    \end{subfigure}
    \begin{subfigure}[b]{0.45\textwidth}
        \centering
        \includegraphics[width=\linewidth]{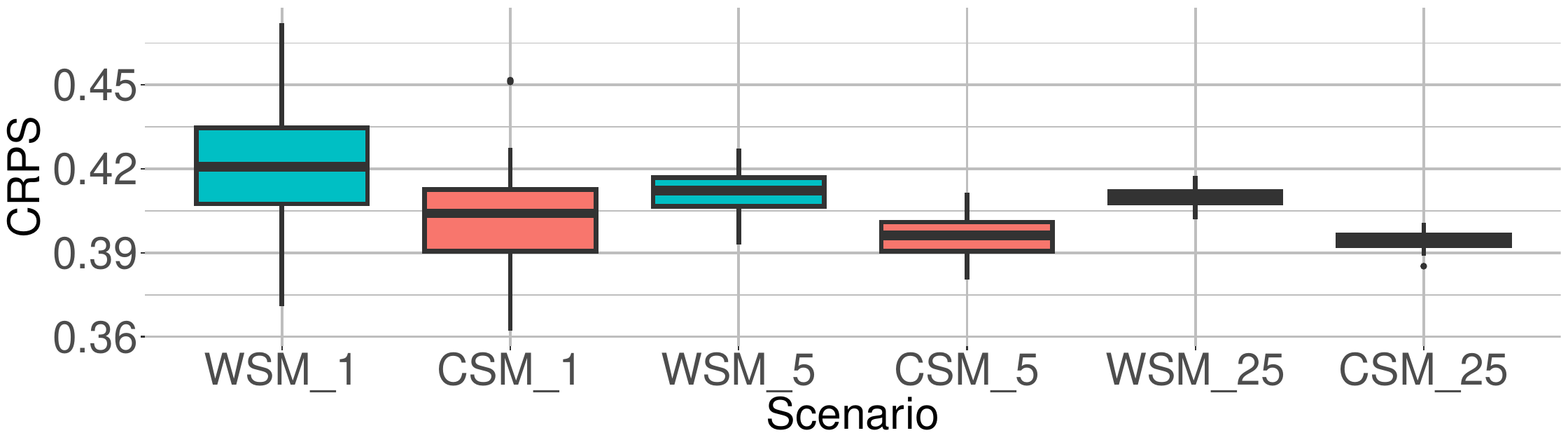}
        \label{fig:CRPS_long}
    \end{subfigure}%
    ~
    \newline
    \begin{subfigure}[b]{0.45\textwidth}
        \centering
        \includegraphics[width=\linewidth]{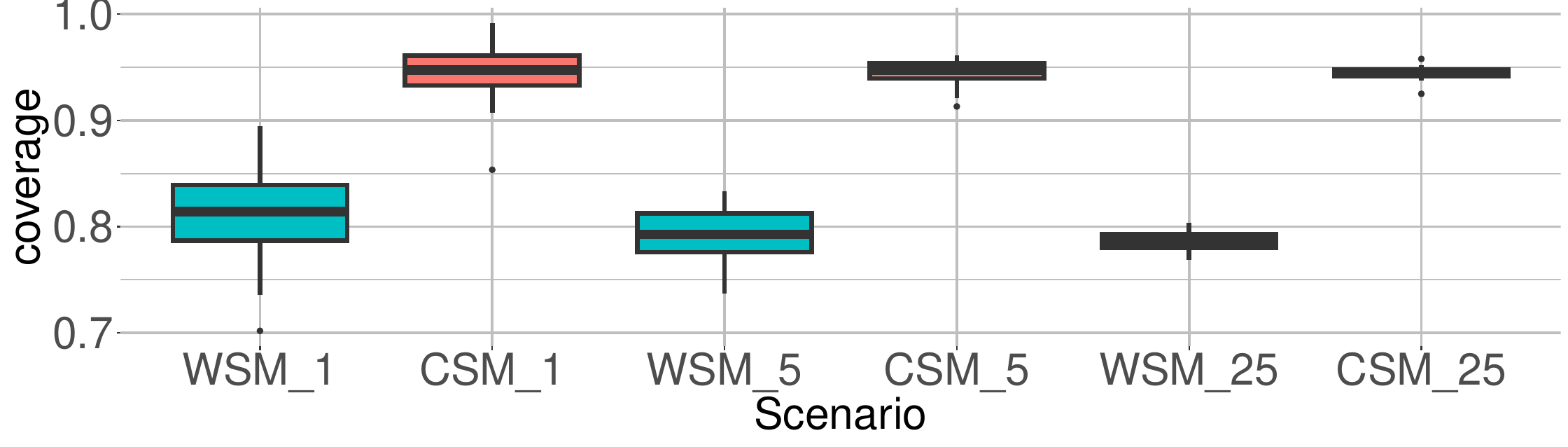}
        \caption{Medium range}
        \label{fig:coverage_short}
    \end{subfigure}
    ~
    \begin{subfigure}[b]{0.45\textwidth}
        \centering
        \includegraphics[width=\linewidth]{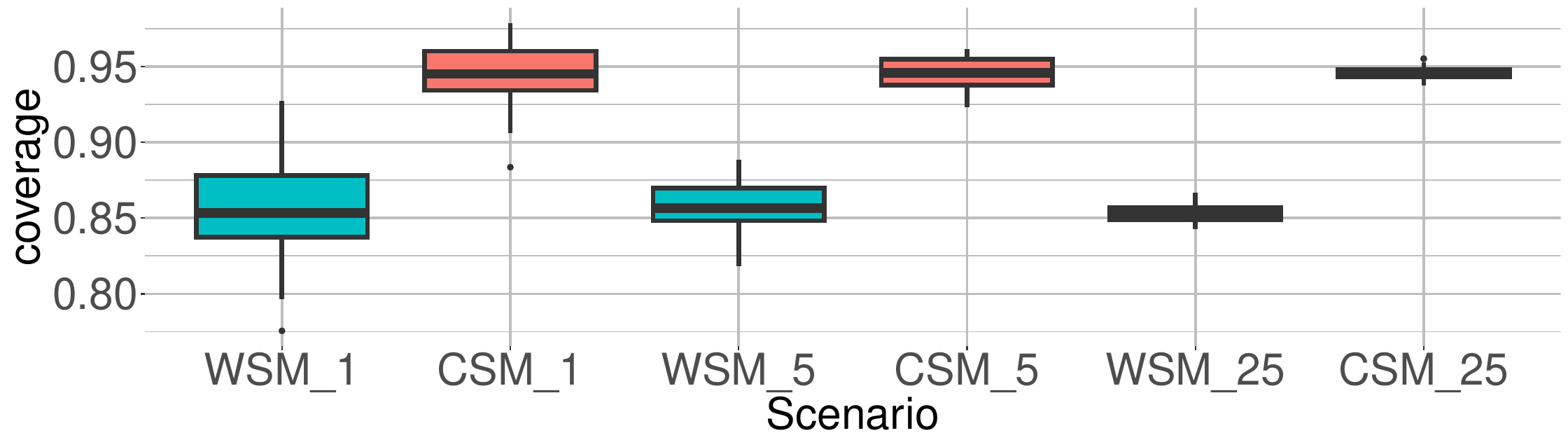}
        \caption{Long range}
        \label{fig:coverage_long}
    \end{subfigure}
    \caption{Scoring rules RMSE (first row), CRPS (second row) and coverage (third row), for the scenarios with (a) medium range (\SI{0.35}{\kilo\meter}) and (b) long range (\SI{1.0}{\kilo\meter}) with both models \SMs and \IMs and 1, 5 and 25 replicates, which are denoted by model name and the number of replicates, eg. WSM\_5 denotes model \SMs with 5 replicates.}
    \label{fig:scores}
\end{figure}

As shown in Figure \ref{fig:scores}, \IMs performs consistently better than \SMs across all scenarios in all scores. The most obvious difference is that $\IM$ is always close to nominal coverage, whereas $\SM$ has only around 80\% coverage in medium range scenarios and around 85\% coverage in long range scenarios. RMSE and CRPS for \IMs indicate that $R=5$ is better than $R=1$, and that there is less improvement from $R=5$ to $R=25$.

\begin{figure}[htb]
    \centering
    \begin{subfigure}[b]{0.45\textwidth}
        \centering
        \includegraphics[width=\linewidth]{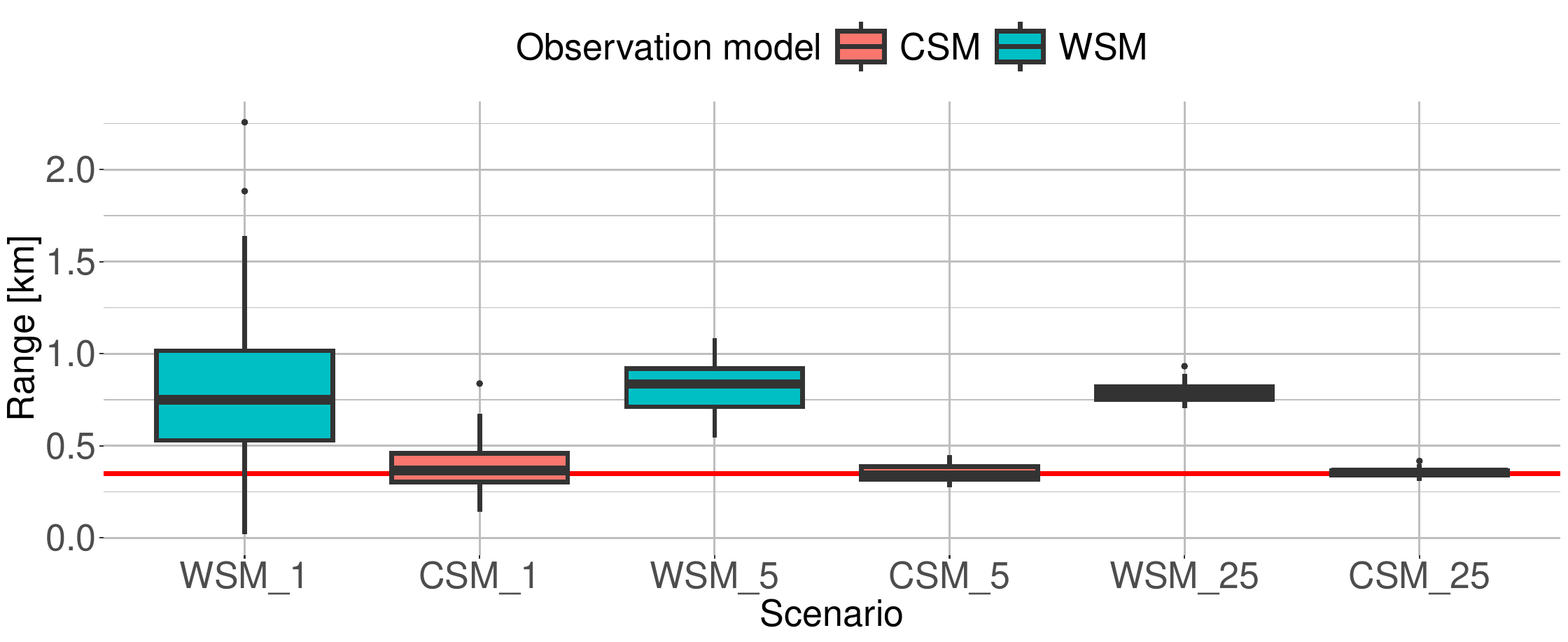}
    \end{subfigure}%
    ~
    \begin{subfigure}[b]{0.45\textwidth}
        \centering
        \includegraphics[width=\linewidth]{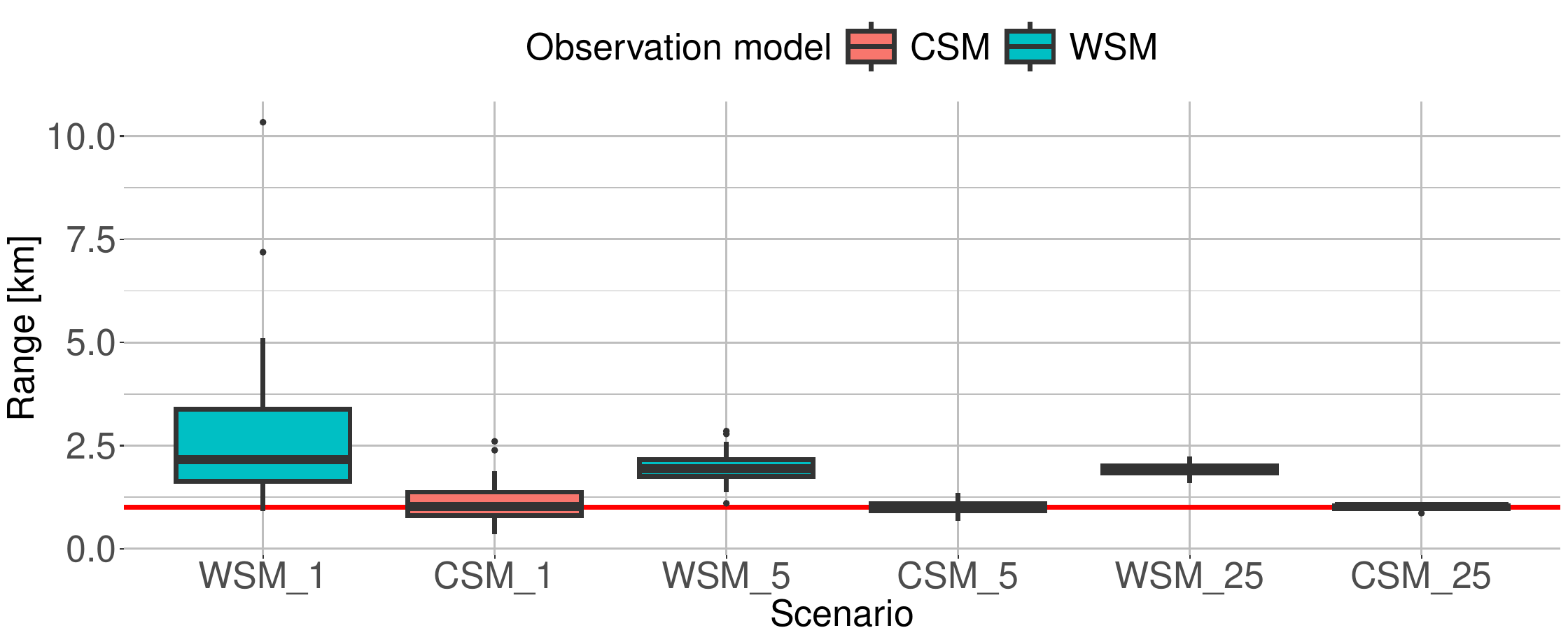}
    \end{subfigure}
    ~
    \newline
    \begin{subfigure}[b]{0.45\textwidth}
        \centering
        \includegraphics[width=\linewidth]{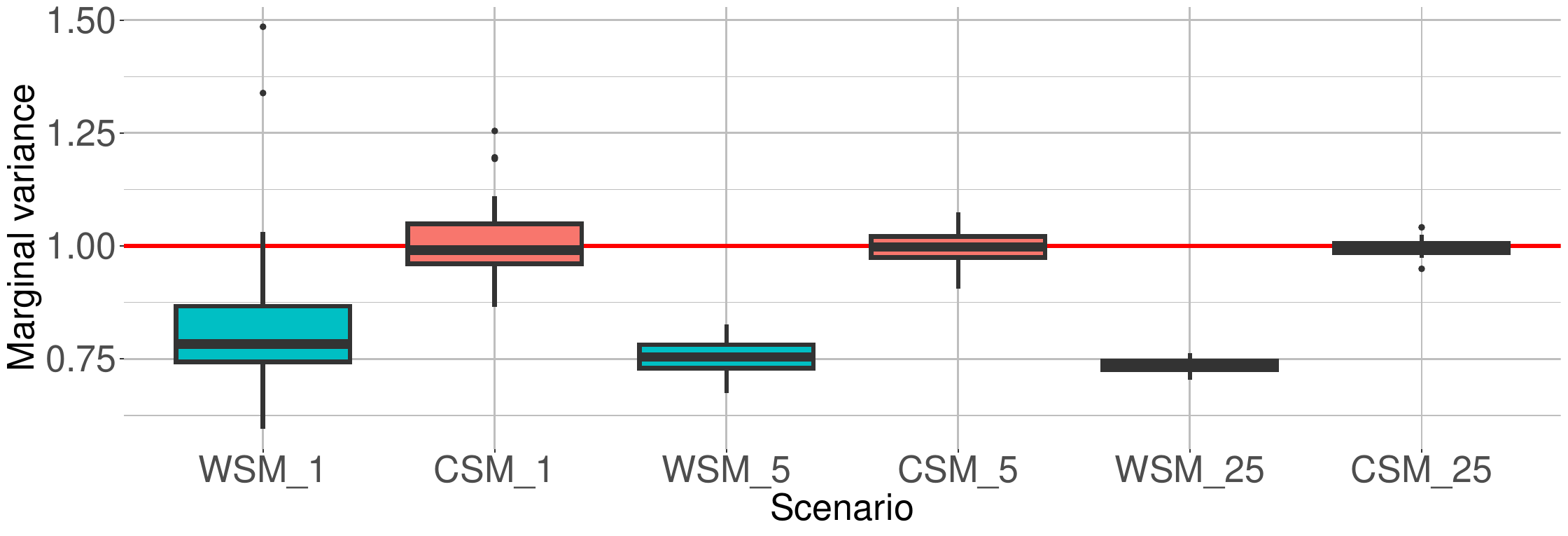}
        \caption{Medium range}
        \label{fig:range_margvar_short}
    \end{subfigure}
    ~
    \begin{subfigure}[b]{0.45\textwidth}
        \centering
        \includegraphics[width=\linewidth]{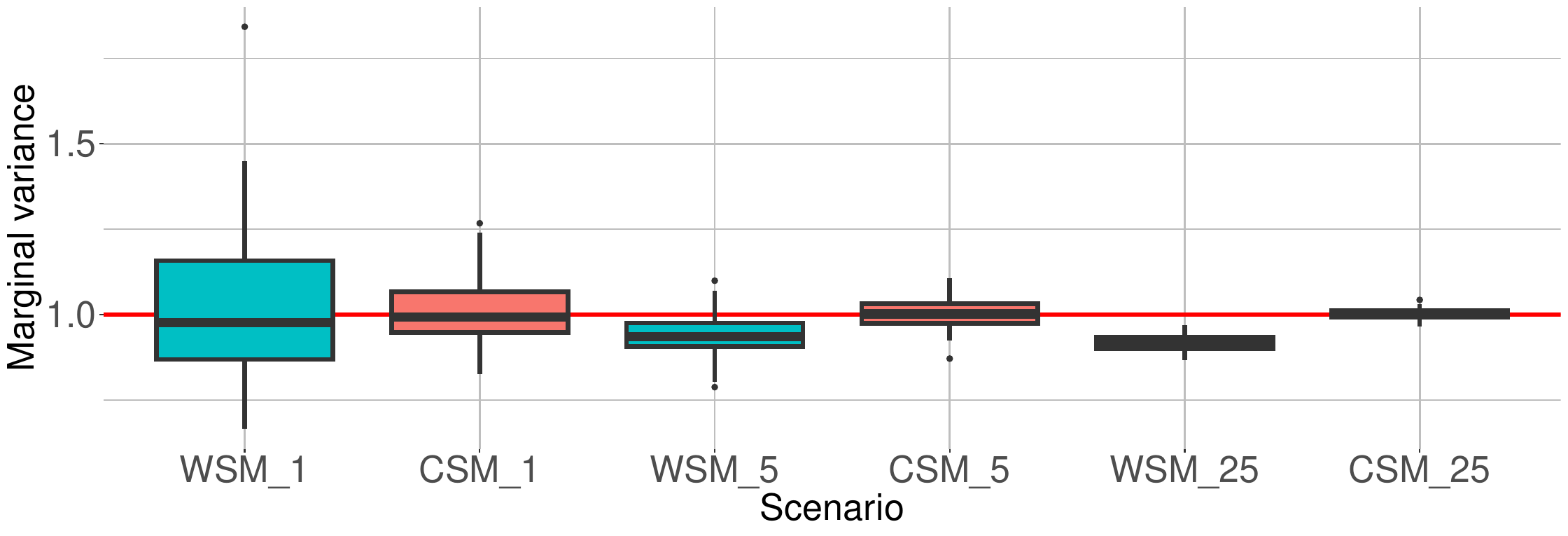}
        \caption{Long range}
        \label{fig:range_margvar_long}
    \end{subfigure}
    \caption{Parameter estimates (median) for the spatial field, $u(s)$, for range (first row) and marginal variance (second row) with (a) medium range (\SI{0.35}{\kilo\meter}) and (b) long range (\SI{1.0}{\kilo\meter}) with both models \SMs and \IMs and 1, 5 and 25 replicates, which are denoted by model name and the number of replicates, eg. WSM\_5 denotes model \SMs with 5 replicates.}
    \label{fig:spatial_param_est}
\end{figure}

From Figure \ref{fig:spatial_param_est}, we see that \SMs consistently overestimates the range, but
that \IMs estimates the range close to the true value. 
Further, \SMs strongly underestimates marginal variance for medium range and slightly underestimates marginal variance for long range.
On the other hand, \IMs estimates the marginal variance well in all scenarios. For a single replicate ($R = 1$), there appears to be some instability as can be seen from outliers in the second row of Figure \ref{fig:range_margvar_short}.
For the fixed effects $\beta_0$ and $\beta_1$, \IMs and \SMs performs comparable, and both show improved estimates with increasing number of replicates. These results are in Figure \ref{fig:fixed_parameter_est} in \ref{sup:sim_study}.
We find that estimating noise $\sigma^2_\mathrm{L}$ and $\sigma^2_\mathrm{P}$ is not feasible with as few observations as considered here. Especially $n_\mathrm{P}=6$ is too low to accurately estimate the noise. Overall \IMs performs better than \SMs when estimating these two parameters. The result for these parameters can be found in \ref{sup:sim_study} in Figure \ref{fig:line_noise_est} and Figure \ref{fig:point_noise_est}. 

Overall, we find that it is important to use the correct observation model \IM, and that we should use more than one replicate.

\FloatBarrier

\section{Traffic modelling}\label{sec:case}
\subsection{Motivation and goal}
We aim to compare traffic in rush hour and non-rush-hour time periods.
Based on the findings in the simulation study, we use $R = 5$ weeks in the DTWs introduced in Section \ref{sec:motivation}. 
The dataset DataMon consists of the observed average speeds between bus stops from five Mondays in October 2023, all with departures from the origin stations 07:00a.m.--08:00a.m., and average speeds of passing vehicles within the same hour from the six traffic stations that continuously measure traffic through induction loops. These average speeds are based on an average of 2912 vehicles each Monday. The dataset DataWed consists of observed average speeds between bus stops on  Wednesdays 07:00p.m.--08:00p.m., from November 2023, and measurements of average speed of passing cars at the six traffic stations based of on average 1389 passing vehicles each Wednesday.  For each week $r = 1, \ldots, R$, both DataMon and DataWed have $n_\mathrm{P} = 6$ point observations and 
$n_\mathrm{L}=180$ line observations. The 180 line observations arise from: 1) the 92 line segments used in Section \ref{sec:sim_study}, combined with 2) the 88 line segments from the same bus routes in the opposite direction. In this work, we do not consider direction in the modeling. A further discussion of implications and limitations of this choice is given in Section \ref{sec:discussion}. Our interest is in the differences between rush hour and non-rush hour traffic state. That is, we assume otherwise that the two data sets, DataMon and DataWed, are illustrating the general traffic in rush hour and non-rush hour, respectively.

The spatial covariate is constructed from attributes from the physical road system, obtained in the spatial objects from OpenStreetMap \citep{osm}. For this analysis, we use the speed limit as the spatial covariate, and convert it from the original \si{\kilo\meter/\hour} (speed) to \si{\sec/\meter} (pace). The covariate in all mesh vertices is displayed in Figure \ref{fig:case_covariate}. Missing values are set to 40 \si{\kilo\meter/\hour} since this is a common speed limit in the study area.

\subsection{Statistical model}
\begin{figure}[htb]
	\centering
	\includegraphics[width=0.65\linewidth]{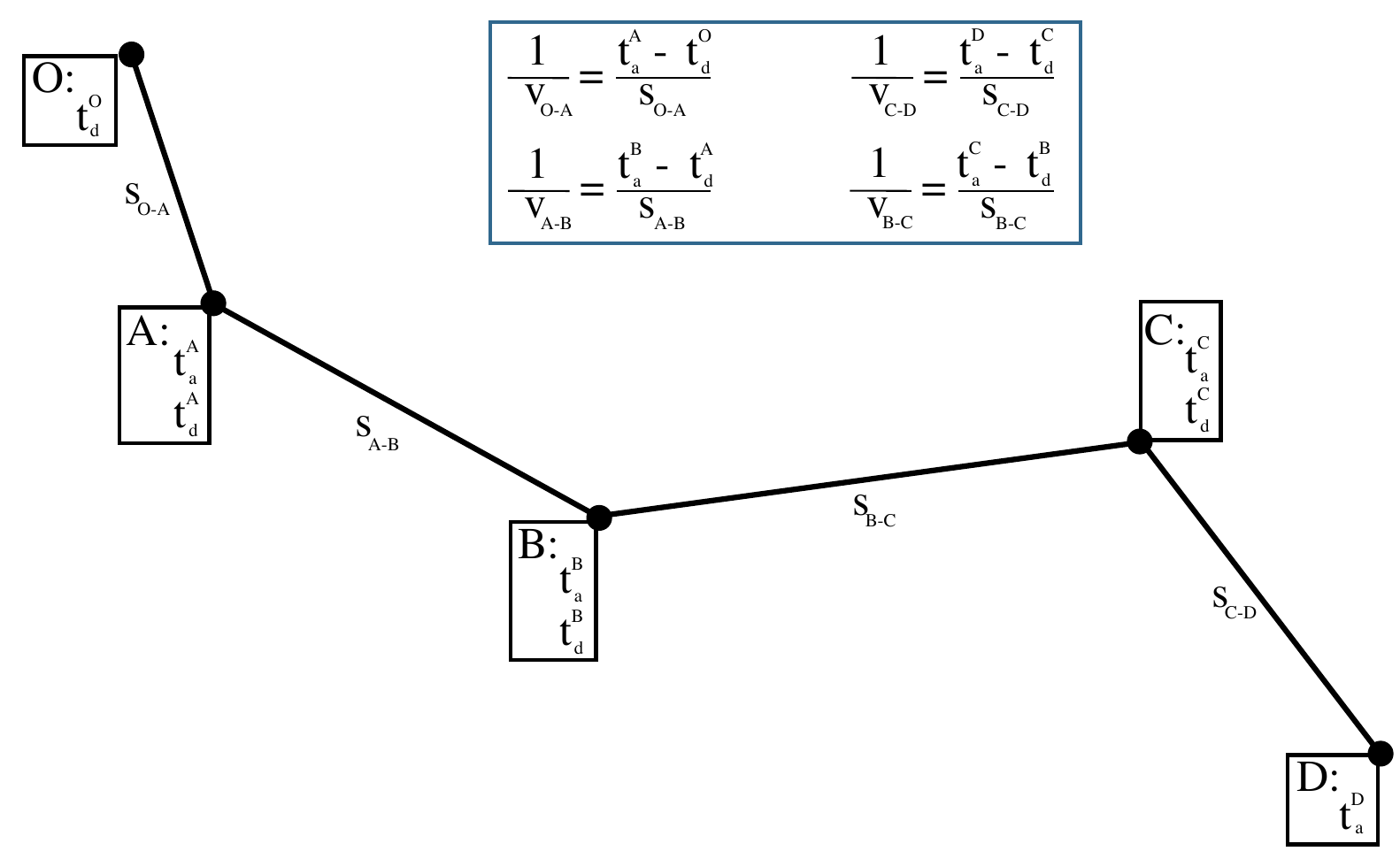}
	\caption{Simplified path from O to D with three internal stops: A, B and C. The distances between stops, e.g., $s_{\mathrm A-\mathrm B}$, are known, and for each stop arrival and departure times are collected. E.g., for stop A, $t_\mathrm{a}^{\mathrm{A}}$ and $t_\mathrm{d}^\mathrm{A}$, respectively. These quantities can be used to compute paces as illustrated in the box. 
    }
	\label{fig:od-illustration}
\end{figure}

Based on the findings in the simulation study, we use the observation model \IM. 
In what follows, we consider a generic day $r = 1,\ldots, 5$.
Assume a bus follows the average speed $v_r(\vect s)$, $\vect s \in \mathrm L_i\subset \Gamma$, for the segment $\mathrm L_i$ between two bus stops. 
Then the time spent traversing $\mathrm L_i$ is
\begin{equation*}
	t_{r,i} = \int_{\mathrm L_i} v_r(\vect s)^{-1}\dif\vect s, \quad i = 1,\ldots, n_\mathrm{L}.
\end{equation*}
The integrand, $v(\cdot)^{-1}$, is the \textit{inverse speed} and is often referred to as a \textit{pace}. 
Figure \ref{fig:od-illustration} illustrate the data collection process of a bus route from point O to point D with three stops: A, B and C. Buses move along their route and experience traffic, and adjust their behavior according to the local traffic. For each bus stop, arrival and departure times are collected. These are used to compute the corresponding paces between bus stops, which is the quantity we model. In this study we consider time spent in traffic, i.e. the dwell times at bus stops are not included.

The line observations are modelled as
\begin{equation*}
	y_{r,i}^\mathrm{L} = \frac{t_{r,i}}{\lvert \mathrm L_i\rvert } +\varepsilon_i^\mathrm{L}= \int_{\mathrm L_i} v_r(\vect s)^{-1} \frac{\dif\vect s}{\lvert \mathrm L_i\rvert} +\varepsilon_{r,i}^\mathrm{L}, \quad i = 1, \ldots, n_\mathrm{L},
\end{equation*}
where $\lvert \mathrm L_i\rvert $ is the known length of $\mathrm L_i$, and $\varepsilon_{r,1}^\mathrm{L}, \ldots, \varepsilon_{r,n_\mathrm{L}}^\mathrm{L}|\sigma_\mathrm{L}^2 \overset{\text{iid}}{\sim} \mathcal{N}(0, \sigma_\mathrm{L}^2/|\mathrm L_i|^2)$. The point observations
are modelled as 
\[
    y_{r,j}^\mathrm{P} = v_r(\vect s_j)^{-1}+\varepsilon_j^\mathrm{P}, \quad j = 1, \ldots, n_\mathrm{P},
\]
where $\vect s_j$ is location of station $j$, and $\varepsilon_{r,1}^\mathrm{P},\ldots, \varepsilon_{r,n_\mathrm{P}}^\mathrm{P}|\sigma_\mathrm{P}^2\overset{\text{iid}}{\sim}\mathcal{N}(0, \sigma_\mathrm{P}^2)$.
Pace needs to be positive, and we use a log-link function, 
\begin{equation*}
    \log(v_r(\vect s)^{-1}) = \eta_r(\vect s) = \beta_0 + x(\vect s)\beta_1 + u_r(\vect s), \quad \vect s \in \Gamma,
\end{equation*}
where the model components and priors are as described in Section \ref{ssec:hierarchicalmodel}, but we have $R$ independent realizations of the GRF sharing marginal variance and range.
As before, we can fit the model in \texttt{inlabru} with the same approach as described in Section \ref{sec:sim_study}. The only difference is the non-linear transformation of the linear predictor, $g(\cdot)=\exp(\cdot)$, that is, we specify in the formula that we consider the exponential of the linear predictor. The metric graph and mesh is the same as used in Section \ref{sec:sim_study} with maximum distance between mesh vertices $h=70\si{\meter}$.  

We compare properties of traffic in the two time windows by: 1)  comparing the parameter estimates, and 2) predictions of average speed across the road network. 
All computations are performed on a MacBook Pro with Apple M1 Pro chip with 32 GB memory, and the macOS is Sonoma 14.7.1.

\subsection{Results}
\begin{figure}
    \centering
    \begin{subfigure}[c]{0.48\textwidth}
        \centering
        \includegraphics[width=\linewidth]{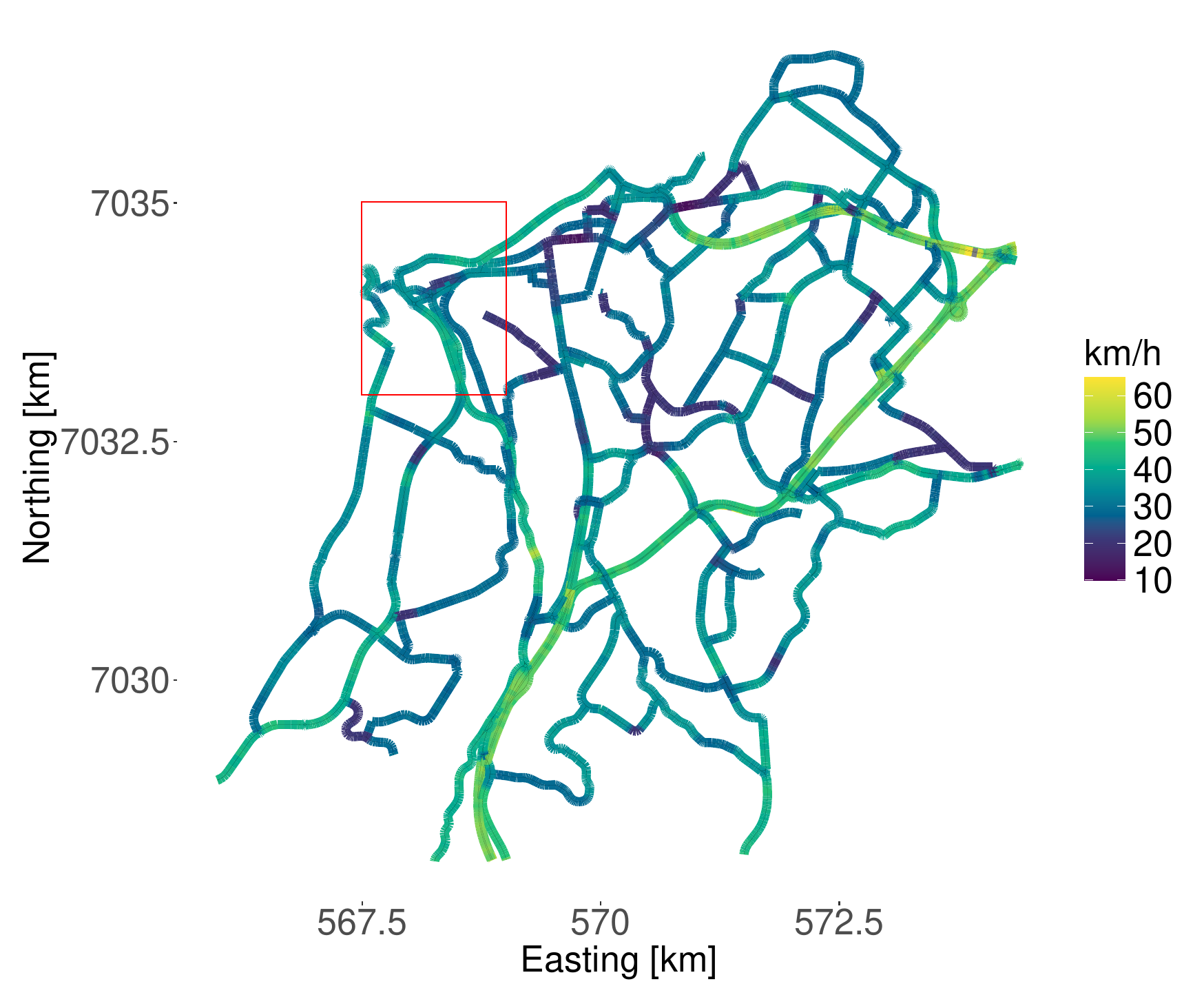}
        \caption{$\bar{v}_\mathrm{M}(\cdot)$}
        \label{fig:case_mean_mon}
    \end{subfigure}
    ~
    \begin{subfigure}[c]{0.48\textwidth}
        \centering
        \includegraphics[width=\linewidth]{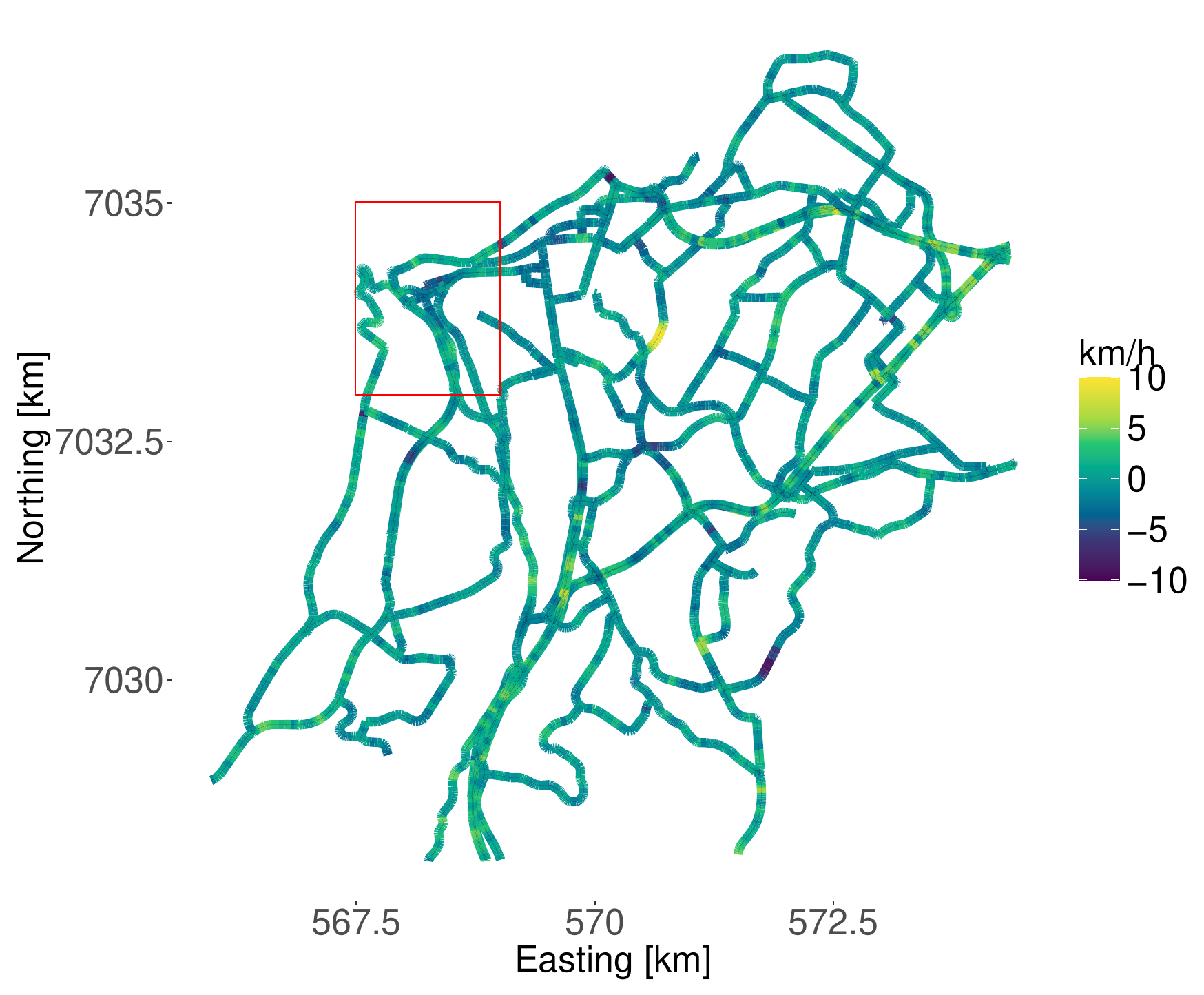}
        \caption{$\bar{v}_\mathrm{M}(\cdot) -v_\mathrm{W}(\cdot)$}
        \label{fig:case_mean_diff}
    \end{subfigure}
    \newline
    \begin{subfigure}[c]{0.48\textwidth}
        \centering
        \includegraphics[width=\linewidth]{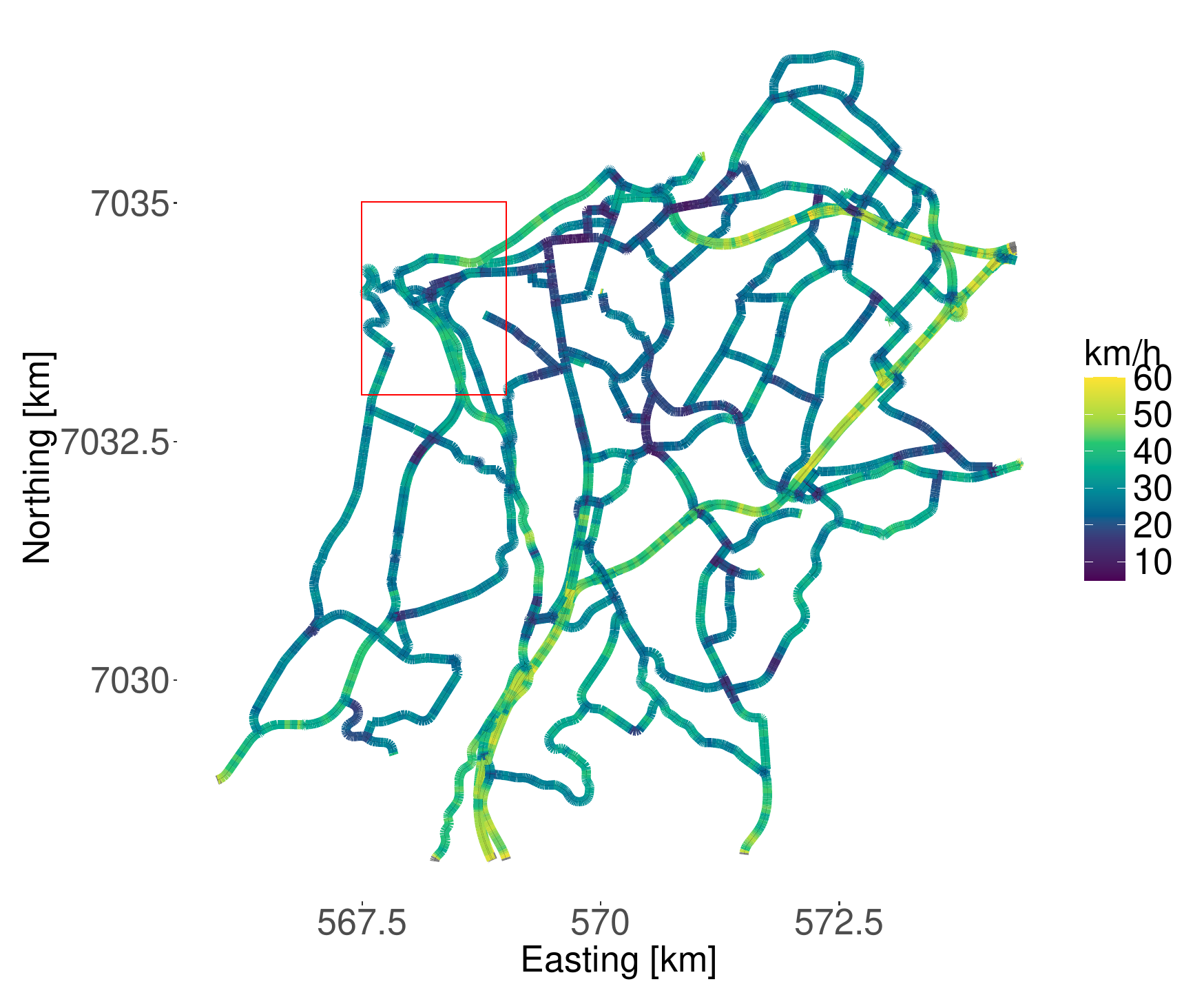}
        \caption{Width of approximate 95\% PI width $\bar{v}_\mathrm{M}(\cdot)$}
        \label{fig:case_width_mon}
    \end{subfigure}
    ~
    \begin{subfigure}[c]{0.48\textwidth}
        \centering
        \includegraphics[width=\linewidth]{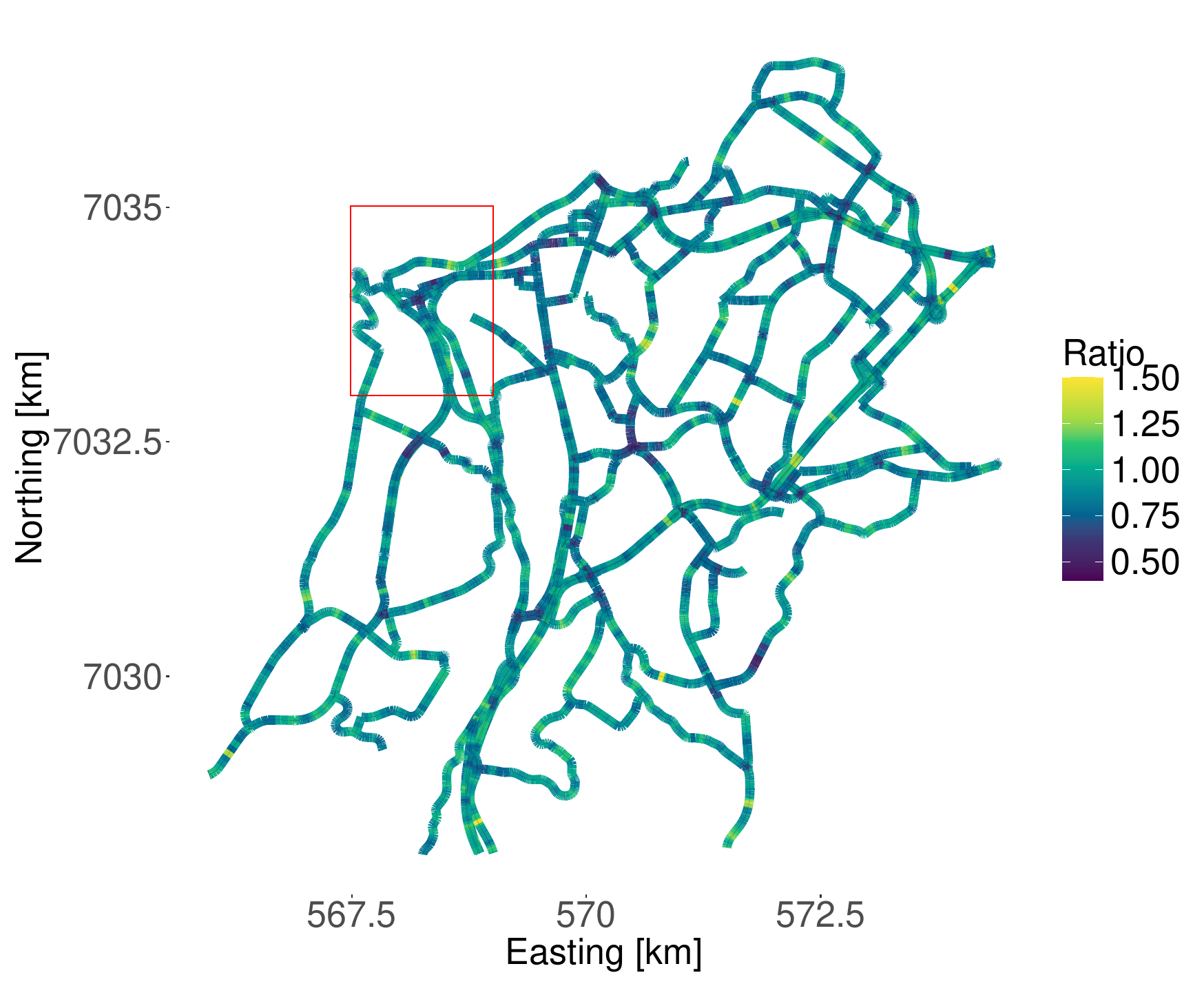}
        \caption{Ratio of widths of approximate 95\% PIs for $\bar{v}_\mathrm{M}(\cdot)/\bar{v}_\mathrm{W}(\cdot)$}
        \label{fig:case_width_ratio}
    \end{subfigure}
    \caption{Visual representation of (a) mean speed field Monday and the (b) difference in mean speed fields for Monday and Wednesday. The prediction uncertainty is shown through the (c) width of an approximated 95\% prediction interval (PI) for Monday and a comparison shown through (d) ratio between PIs for Monday and Wednesday. UTM zone is 32N.}
    \label{fig:case_prediction_full}
\end{figure}

\begin{figure}
    \centering
    \begin{subfigure}[c]{0.48\textwidth}
        \centering
        \includegraphics[width=\linewidth]{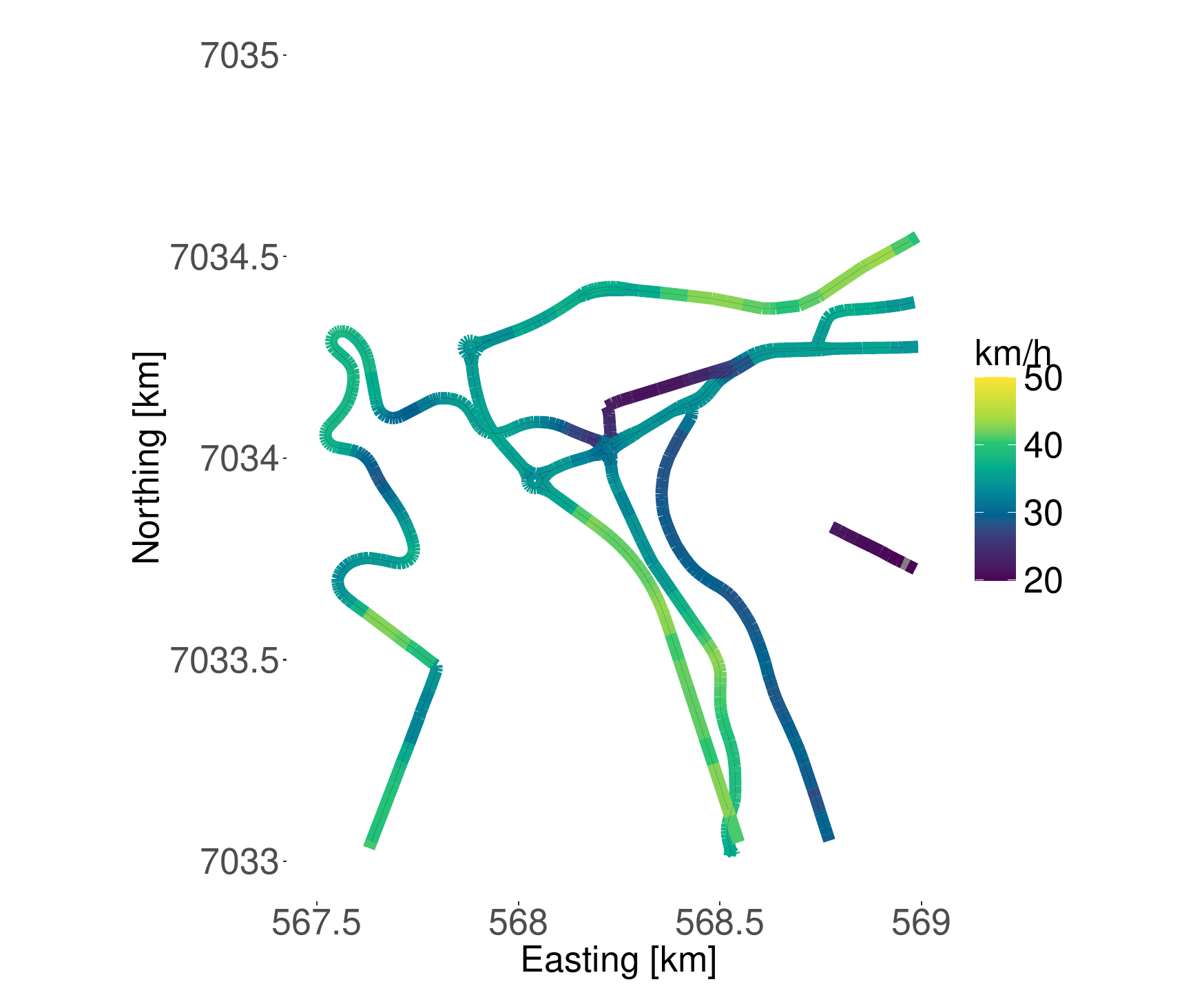}
        \caption{$\bar{v}_\mathrm{M}(\vect s)$}
        \label{fig:case_mean_monday_zoom}
    \end{subfigure}
    ~
    \begin{subfigure}[c]{0.48\textwidth}
        \centering
        \includegraphics[width=\linewidth]{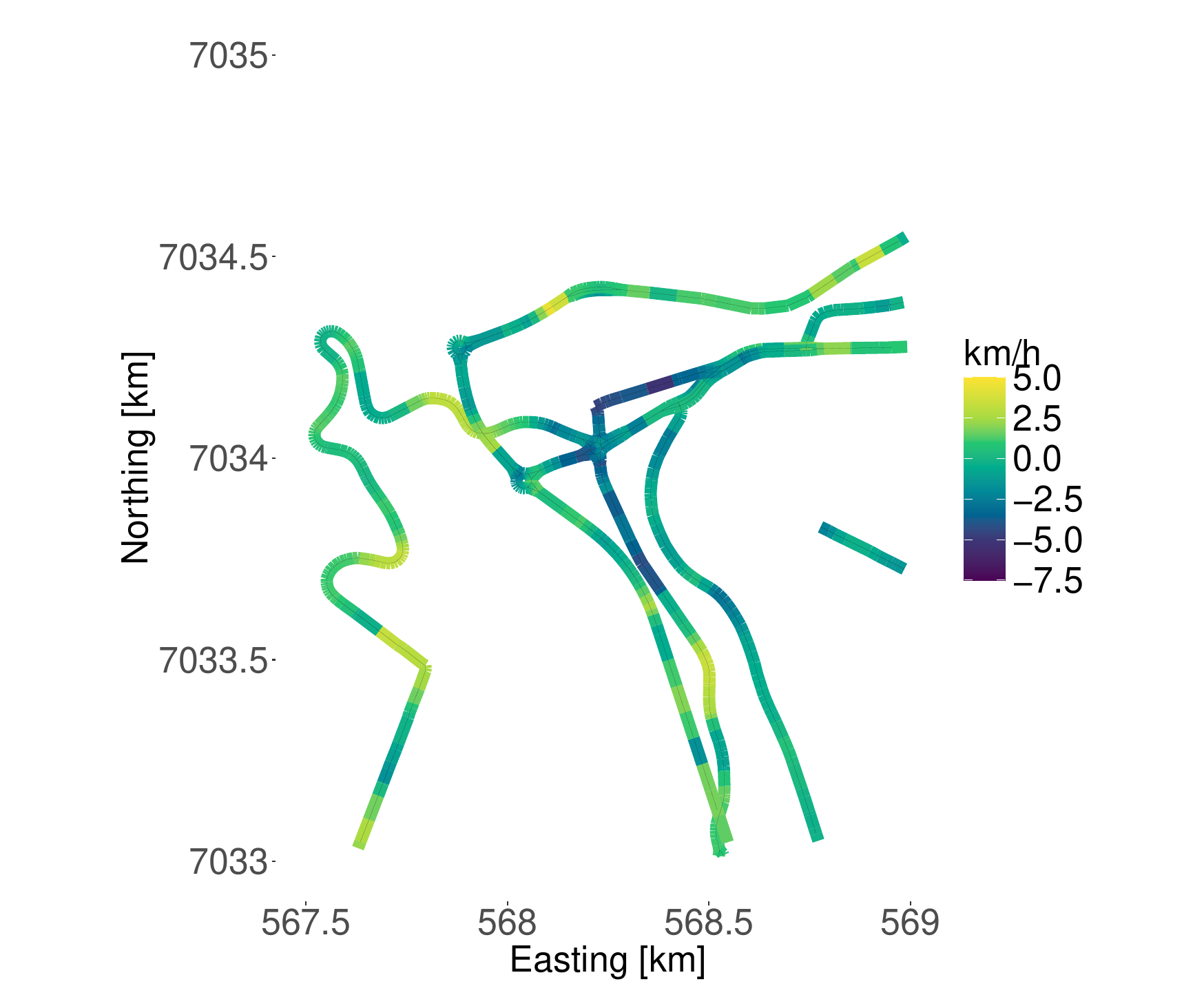}
        \caption{$\bar{v}_\mathrm{M}(\vect s) -\bar{v}_\mathrm{W}(\vect s)$}
        \label{fig:case_mean_diff_zoom}
    \end{subfigure}
    \newline
    \begin{subfigure}[c]{0.48\textwidth}
        \centering
        \includegraphics[width=\linewidth]{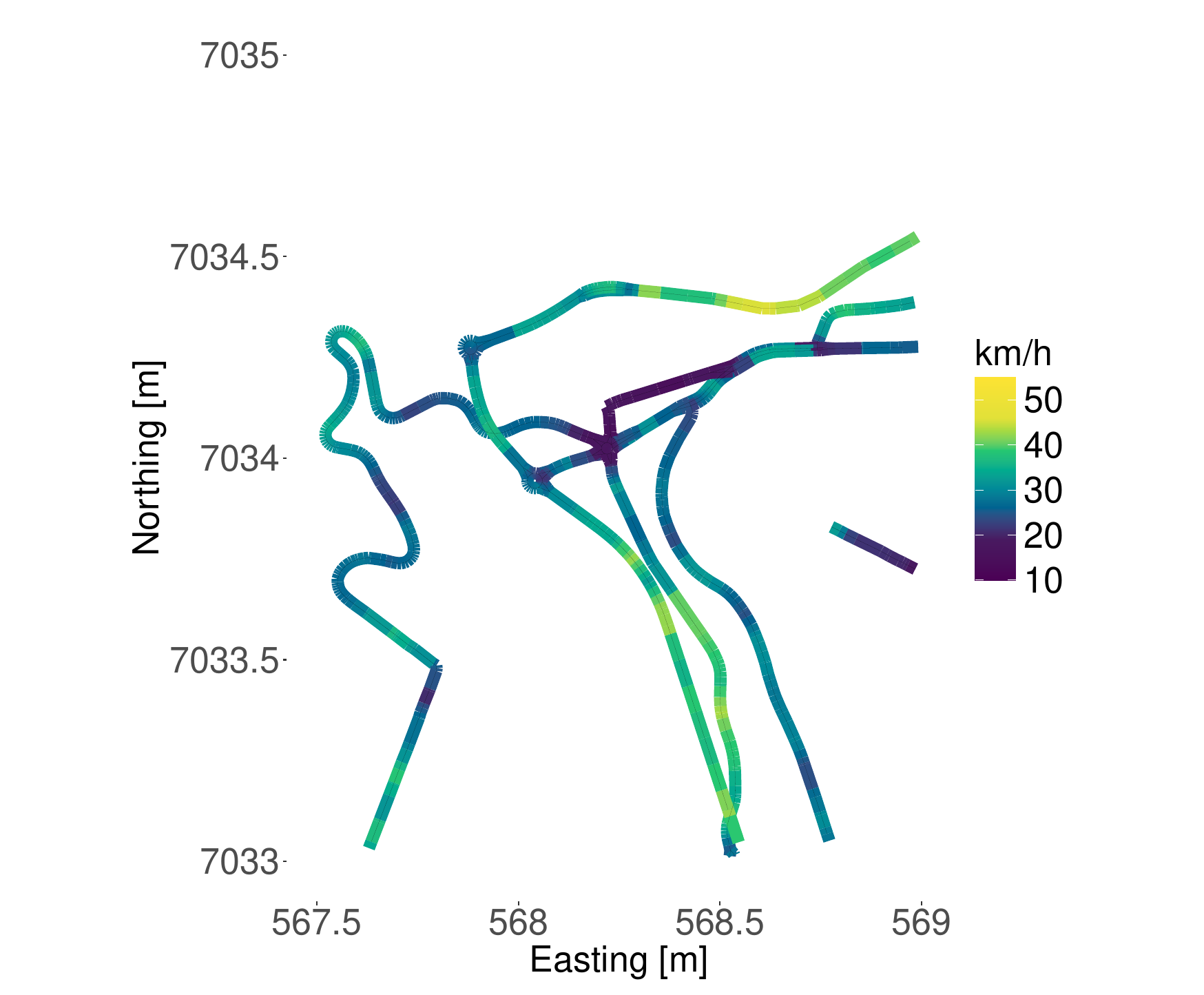}
        \caption{95\% PI width for $\bar{v}_\mathrm{M}(\vect s)$}
        \label{fig:case_width_mon_zoom}
    \end{subfigure}
    \begin{subfigure}[c]{0.48\textwidth}
        \centering
        \includegraphics[width=\linewidth]{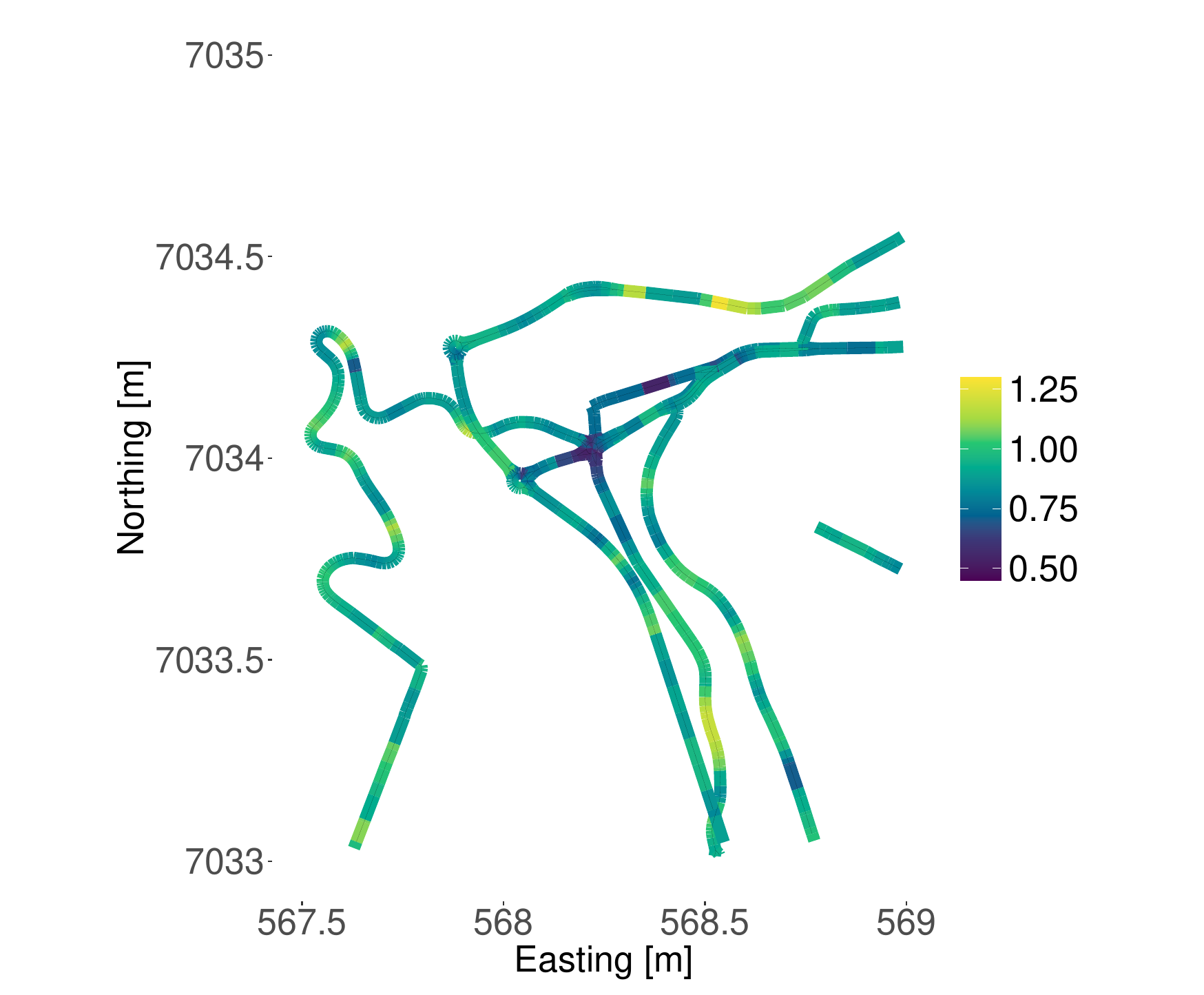}
        \caption{Ratio of widths of approximate 95\% PIs for $\bar{v}_\mathrm{M}(\cdot)/\bar{v}_\mathrm{W}(\cdot)$}
        \label{fig:case_width_ratio_zoom}
    \end{subfigure}
    \caption{Same field as in Figure \ref{fig:case_prediction_full} but zoomed in on the area inside the red square. UTM zone is 32N.}
    \label{fig:case_prediction_zoom}
\end{figure}

We are interested in typical spatial patterns of the average speed for Monday rush-hour and for Wednesday evening. We consider the estimand
\begin{equation}\label{eq:avg_pace}
\bar{v}(\vect s)=\frac{1}{5}\sum_{r = 1}^5 v_r(\vect s) = \frac{1}{5}\sum_{r = 1}^5 \exp\{-\eta_r(\vect s)\},  
\end{equation}
where $\bar v(\cdot)$ describes speed for a given DTW. We get  $\bar{v}_\mathrm{M}(\cdot)$ for Monday morning traffic and  $\bar{v}_\mathrm{W}(\cdot)$ for Wednesday evening. Posterior median and posterior $95\%$ credible intervals for these spatially varying quantities are obtained based on joint posterior samples of $\eta_1(\cdot), \ldots, \eta_R(\cdot)$. For each $\vect s \in \Gamma$, we assume that the posterior of  $\log(\bar{v}_\mathrm{M}(\vect s))$ is approximately Gaussian and compute the average $\hat{\mu}_\mathrm{M}(\vect s)$ and empirical standard deviation $\hat{\mathrm{sd}}_\mathrm{M}(\vect s)$ based on $B = 100$ samples, and similarly compute $\hat{\mu}_\mathrm{W}(\vect s)$ and $\hat{\mathrm{sd}}_\mathrm{W}(\vect s)$ for $\log(\bar{v}_\mathrm{M}(\vect s))$ based on $B = 100$ samples. For each mesh vertex, median, $2.5$ percentile, and $97.5$ percentiles are computed using the Gaussian assumption, and transformed to $\bar{v}_\mathrm{M}(\cdot)$ and $\bar{v}_\mathrm{W}(\cdot)$ using the inverse transformation $\exp(\cdot)$. We visualize the resulting field, $\bar{v}_\mathrm{M}(\cdot)$ in Figure \ref{fig:case_mean_mon} and Figure \ref{fig:case_mean_diff} show the difference, $\bar{v}_\mathrm{M}(\cdot)-\bar{v}_\mathrm{W(\cdot)}$. We can also obtain a measure of uncertainty related to the spatial prediction. Figure \ref{fig:case_width_mon} shows the width of an approximate 95\% confidence interval of the transformed field $\bar{v}_\mathrm{M}$. For comparison we show the ratio of the width of $\bar{v}_\mathrm{M}(\cdot)$ to $\bar{v}_\mathrm{W}(\cdot)$. Visual inspection of the fields on the full graph is difficult, and panning in on areas of interest can make it easier. Figure \ref{fig:case_prediction_zoom} shows how this looks for an area where the difference is non-zero and the ratio of uncertainties are not all equal to 1. We note that the mean speed on Wednesdays, in non-rush how is higher in the roundabout with five arms compared to the mean speed on Monday rush hour. Similarly, the ratio is less than one in this area, so the uncertainty of $\bar{v}_\mathrm{M}$ is smaller than for $\bar{v}_\mathrm{W}$. This could be an effect of traffic being more consistent in the rush hour as vehicles are more dependent on each other while in non-rush hour individual vehicles can choose their speeds more freely.

Note that the interpretation of a higher covariate effect e.g.\ increasing the covariate by 1, we get a factor of $\exp(-\beta_1)$ in our speed field due to the inverse. Note that the covariate is converted from \si{\kilo\meter/\hour} to \si{\meter/\sec} in this analysis, which means that increasing the covariate is equivalent to lowering the maximum speed limit. That is, we expect the sign of $\beta_1$ to be positive.
Similarly, a higher value for the global intercept, $\beta_0$, means slower speed when all other components are kept constant. For marginal variance of the GRFs, a higher value results in larger variation in pace and speed, and a larger range means longer dependence. 

From Table \ref{tab:case_study_params2} we observe that the range parameter is estimated to be longer for the morning rush hour on Monday compared to the evening non-rush hour for Wednesday. In both datasets, we find that the range is shorter than what we considered in the previous section as a medium range. The range is only 1\% and 0.4\% of the graph's diameter of 13.2 \si{\kilo\meter} for dataset DataMon and DataWed respectively. Additionally, we find that the point estimate for the marginal variance is higher for Wednesday evening than Monday morning. In total, we find that the spatial field is flatter for DataMon, while for DataWed we get more spatial variation which is not explained by the covariate. A possible explanation for such spatial behavior could be that the increased traffic affects larger areas, and with this effect of high traffic the speed is varying less due to long queues, while in the evening, without a lot of traffic/queues, the individual vehicles are free to choose their own speed and any traffic flow disturbances (traffic lights, pedestrian crossings, etc.) are locally affecting the speed. 

We observe that the point estimate for the global intercept for log-pace on the network is somewhat lower for Monday mornings compared to the Wednesday evening, while the CI is very similar. This is displayed in Table \ref{tab:case_study_params2}. The effect of the speed limit is contributing more to the Monday morning pace prediction than the Wednesday evening, as the posterior is slightly higher for rush hour traffic. That is, increasing the speed limit in a certain area (lowering the lower bound for pace) will have a stronger effect on rush hour traffic compared to off-peak hours. These differences are very small, and we can say that these parameters are comparable for the two data sets.

\begin{table}[htb]
\centering
\caption{Model parameters for the two data sets DataMon and DataWed evening traffic, the median and 2.5\% and 97.5\% quantiles for the posterior is provided.} 
	\begin{tabular}{ p{2cm} p{1cm} >{\centering\arraybackslash}p{3.5cm} >{\centering\arraybackslash}p{3.5cm} }
		Parameter & Unit &  Monday, morning & Wednesday, evening  \\
		\hline
        $\beta_0$  & $10^0$ & $-3.05$, $(-3.28,-2.83)$ & $-2.98$, $(-3.18,-2.76)$\\
		$\beta_1$ & $10^1$ & $1.16$, $(0.89,1.44)$ & $1.08$, $(0.82,1.30)$ \\
		$\rho_u$ & $\text{km}$ & $0.113$, $(0.046,0.240)$  & $0.054$, $(0.013,0.147)$\\
		$\sigma_u$ & $10^{-1}$ & $5.69$, $(4.34,7.82)$ & $6.93$, $(4.54,10.2)$ \\
		$\sigma_\mathrm{L}$ & $10^{-2}$ & $2.79$, $(2.63, 2.97)$ & $2.40$, $(2.26, 2.55)$ \\
        $\sigma_\mathrm{P}$ & $10^{-3}$ & $7.48$, $(3.22, 20.3)$ & $9.44$, $(2.96, 37.2)$ 
	\end{tabular}
	
    \label{tab:case_study_params2}
\end{table}

\FloatBarrier

\section{Discussion}\label{sec:discussion}
The paper demonstrates that line observations for Mat\'ern-like SPDE-based GRFs on metric graphs are informative for parameter estimation and prediction. However, the simulation study makes it clear that using the correct observation model is important. In particular, we observed severe undercoverage when line observations were assigned to the midpoints of the lines. 
We also saw that reasonable estimates and predictions were obtained for a single realization, but that there were substantial improvements for five realizations. This aligns with the result in \citet{Rikke}, who reached the same conclusion when studying a non-stationary, SPDE-based model. 
On the other hand, 25 realizations did not lead to a substantial improvement compared to five realizations. This means that one could consider modelling traffic patterns both at weekly and monthly scales.

The type of data used in this paper are available in many places. The use of bus data allows assessment of the traffic state at network level rather than at sparse point locations, and the modelling framework presented in this paper has demonstrated joint modelling of point and line data utilizing two sources of traffic speed data. We demonstrate that there is a potential to apply spatial modelling on metric graphs with line observations to gain insight into traffic patterns from the data. 
However, some care is needed in assessing data quality. We observed instances where data suggested speeds that are the double of the speed limit between bus stops.
There is uncertainty involved in registering arrival and departure times, which is based on onboard GPS trackers and geographic zones around bus stops.

Incorporating line data using default functionality in
\texttt{MetricGraph} is challenging since it only supports point-referenced data out-of-the-box. The line observations are more complex spatial data objects than point-referenced observations, and requires elaborate code for representing and mapping to representations of line data. The additional code extends standard functionality to handle line observations with \texttt{MetricGraph} and \texttt{inlabru}. Incorporating this functionality into the relevant packages, such as \texttt{MetricGraph} and \texttt{fmesher} is on-going work, which can be found on branch \texttt{feature/MetricGraphPaper} on \citet{fmesher}. This new code allows more easy handling of line observations, and makes modelling with line observations on metric graphs more accessible. Further, the full Bayesian inference through \texttt{inlabru} is quick and can be performed on a standard laptop.

One possible enhancement to the model would be to make the observation noise related to line observations depend on the length of the segment $\mathrm L$ and if it contains vertices of degree three or more (an intersection). In practice, that means that we include a second term in the $h(\cdot)$ function in \eqref{eq:integral}, $h(\mathrm L) = \lvert\mathrm L\rvert\sigma_{\mathrm L}^2 + \mathbb I\{\exists v\in \mathrm L \text{ s.t. } \text{deg}(v)>2\}\sigma_v^2$. It is possible to implement this type of model with \texttt{R-INLA}, and would only require some methods for determining if each observation segment does contain an intersection or not.  
We fixed smoothness to $\nu=0.5$, and investigation of other smoothnesses would be interesting. In particular, it would be interesting to understand both how hard it is to estimate smoothness, and how influential smoothness is for parameter estimation and prediction for these types of data. Smoothnesses $\nu = 1, 2, \ldots$ are directly available through FEM and $\nu = 0.5, 1.5, \ldots$ are available through least squares FEM. General $\nu$ requires fractional approximations \citep{metric_graph_FEM}.
Another interesting direction, would be to investigate whether it is reasonable to assume the same SPDE, or dependence behavior, across the entire graph. 
When modeling traffic over larger regions, we expect different traffic behavior in urban areas compared to rural areas. One could also extend to spatio-temporal models, either by using separable models or taking advantages of recent developments for non-separable, spatio-temporal models based on SPDEs \citep{Lindgren2024_sort}.

Another natural extension is to model traffic in both directions jointly. This would allow for a more precise description of the flow in the network, and one would be able to separate inflow and outflow from the city center. One possible approach is to expand the existing graph by adding edges to the graph to separate the two directions of flow for each road. This increases the computational cost of the modelling, and leads to many open research questions.  This could be important if moving to spatio-temporal models, where we would expect a strong temporal effect from the rush hours, which will likely also be highly directional.

\section*{Acknowledgments}
This research was funded by The Research Council of Norway's IKTPLUSS program, project number 332237. The authors are grateful to Alexandre Simas and David Bolin for assistance with the R-package \texttt{MetricGraph}. The authors would also like to thank Mats Lien and Martin Slaastuen at AtB, who provided helpful guidance regarding the public transport data, and Snorre Hansen at the Norwegian Public Roads Administration, who kindly provided loop detector data upon request. Last but not least, we thank the reviewers for thourough comments that improved the quality of this paper.

\appendix
\section{Minimal example code}\label{sup:code_example}

The coordinate class consists of two elements; \texttt{index} (integer) and \texttt{where} (2-column matrix). The first element refers to the edge index of the point. The second element, \texttt{where}, gives barycentric coordinates for where we are, and for implementation reasons it is given as $[1-\tilde t, \tilde t]$. Please note that if one wants to extract the \texttt{PtE}-coordinates from \texttt{fm\_bary\_MGG}, one wants the \texttt{\$index} and the \texttt{\$where[,2]}.

We provide a minimal example of how these methods can be applied. The code is also available from \citet{code}.
\begin{lstlisting}[language = R]
library(MetricGraph)
library(INLA)
library(inlabru)
library(rSPDE)
library(sf)
source("metric_graph.R")

# Edges of a simple graph
edge1 <- rbind(c(0,0),c(1,0))
edge2 <- rbind(c(0,0),c(0,1))
edge3 <- rbind(c(0,1),c(-1,1))
theta <- seq(from=pi,to=3*pi/2,length.out = 20)
edge4 <- cbind(sin(theta),1+ cos(theta))
edge5 <- rbind(c(0,0), c(1,0))
edge6 <- rbind(c(1,0), c(1,1))
edge7 <- rbind(c(1,1), c(2,1))
edge8 <- rbind(c(0,1), c(1,1))
edges = list(edge1, edge2, edge3, edge4, 
            edge5, edge6, edge7, edge8)
# Graph construction
graph <- MetricGraph::metric_graph$new(edges = edges)
# Build mesh
graph$build_mesh(h = 0.01)
# Make model construction, we fix smoothness here
rspde_model <- rspde.metric_graph(graph, nu=0.5)
# Define a valid path object from line segment to graph
line1 <- st_sfc(st_linestring(
                matrix(c(0.5,1.1,1,1),ncol=2)
                ))
line2 <- st_sfc(st_linestring(
                matrix(c(1.3,2,1,1),ncol=2)
                ))
path1 <- geom_path_to_path_MGG(line1, graph)
path2 <- geom_path_to_path_MGG(line2, graph)
# Repeated measurements of the two paths path1 and path2
paths <- list()
i <- 0
for (id in unique(path1$ID)) {
  i <- i + 1
  paths[[i]] <- path1[path1$ID == id, c("paths")]
}
for (id in unique(path2$ID)) {
  i <- i + 1
  paths[[i]] <- path2[path2$ID == id, c("paths")]
}
for (id in unique(path1$ID)) {
  i <- i + 1
  paths[[i]] <- path1[path1$ID == id, c("paths")]
}
for (id in unique(path2$ID)) {
  i <- i + 1
  paths[[i]] <- path2[path2$ID == id, c("paths")]
}
# Construct sampler using path and weight
sampler <- tibble::tibble(x = paths,
                          weight = rep(1,length(paths)))
# Mapper (here we choose aggregate and 
# rescale for integral observations)
agg <- bru_mapper_aggregate(rescale =  TRUE, 
                            n_block = nrow(sampler))
# Find integration points in the mesh
ips <- fm_int(graph, sampler)
# Simulate a true field u and its observations
rspde.order <- 2
nu <- 0.5
op <- matern.operators(
  nu = nu, range = 1.5, sigma = 3,
  parameterization = "matern",
  m = rspde.order, graph = graph
)
u <- simulate(op, nsim = 1)
y <- rnorm(nrow(sampler), sd = 0.5) + 5 +
  with(ips, ibm_eval(agg,
                     input = list(block = .block, 
                                  weights = weight),
                     state = fm_evaluate(
                        graph, 
                        loc = x, 
                        field = as.vector(u)
                        )
  ))
# Formula
formula <- y ~ ibm_eval(agg,
                        input = list(block = .block, 
                                     weights = weight),
                        state = Intercept + spde)
# Observation data for inlabru
obs <- bru_obs(formula = formula,
               response_data = data.frame(y=y),
               data = ips,
               allow_combine = TRUE)
# inlabru call
bru_res <- bru(components = y ~ Intercept(1) +
                 spde(x, model = rspde_model, 
                      mapper = bru_mapper(graph)),
               obs)
# inlabru summary
summary(bru_res)
# rSPDE parameter summary
summary(rspde.result(bru_res, "spde", rspde_model))

\end{lstlisting}

\FloatBarrier
\section{Simulation study results}\label{sup:sim_study}
\begin{figure}[htb]
    \centering
    \begin{subfigure}[b]{0.45\textwidth}
        \centering
        \includegraphics[width=\linewidth]{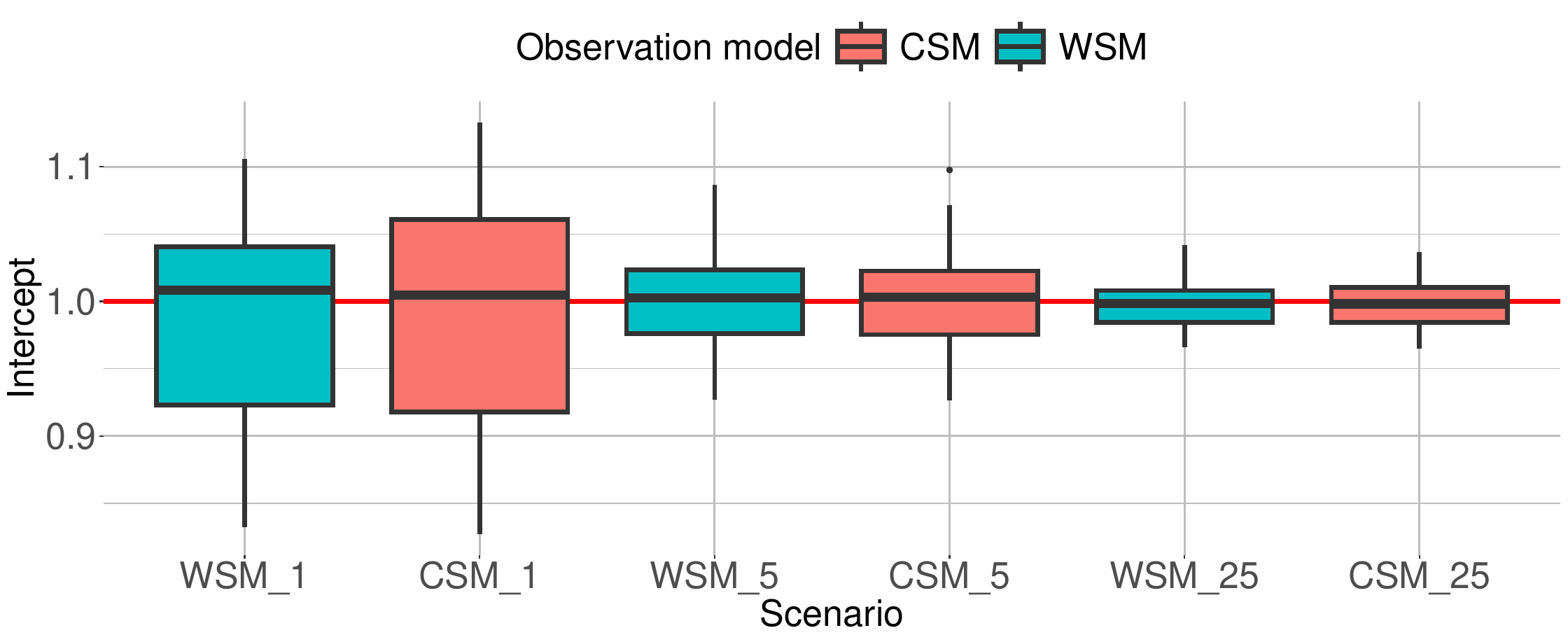}
    \end{subfigure}%
    ~
    \begin{subfigure}[b]{0.45\textwidth}
        \centering
        \includegraphics[width=\linewidth]{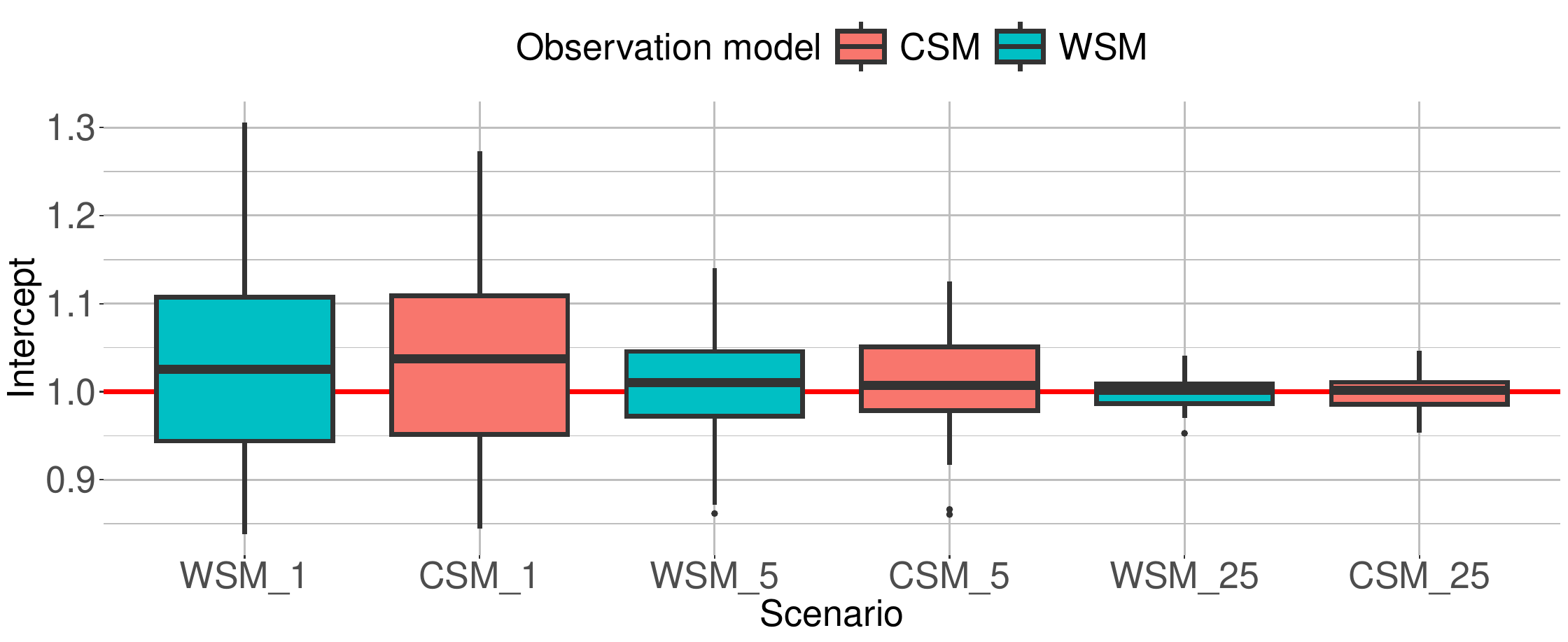}
    \end{subfigure}
    ~
    \newline
    \begin{subfigure}[b]{0.45\textwidth}
        \centering
        \includegraphics[width=\linewidth]{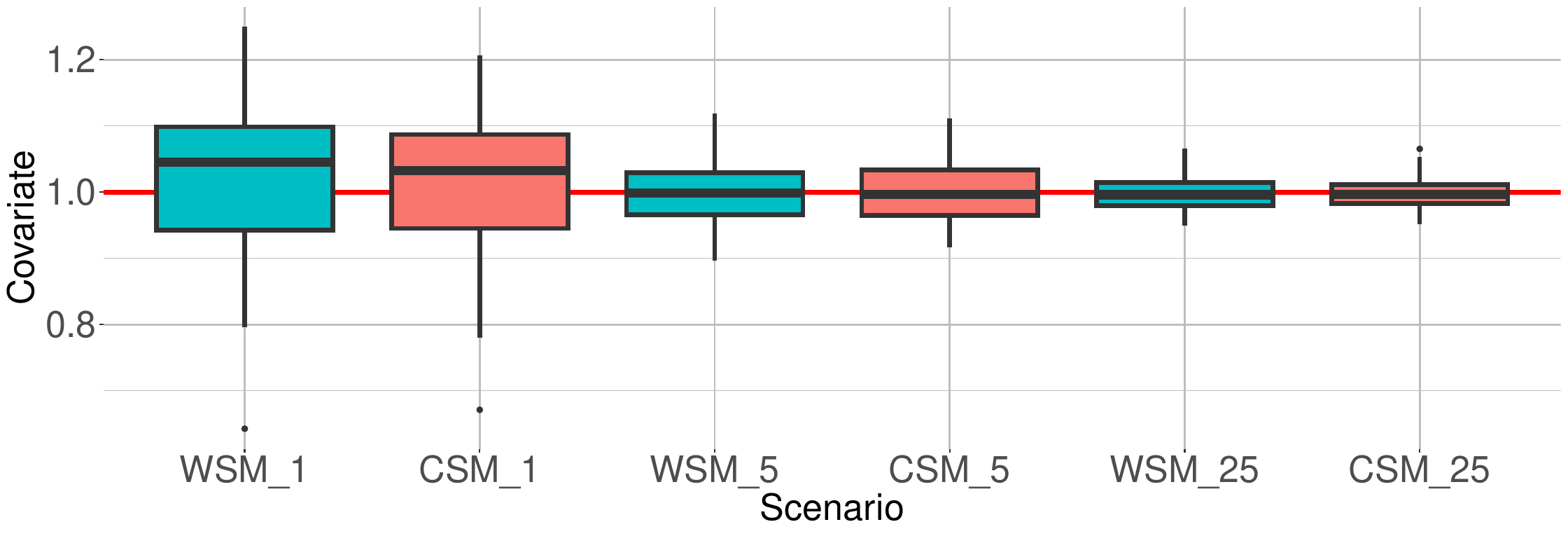}
        \caption{Medium range}
        \label{fig:beta0_beta_short}
    \end{subfigure}
    ~
    \begin{subfigure}[b]{0.45\textwidth}
        \centering
        \includegraphics[width=\linewidth]{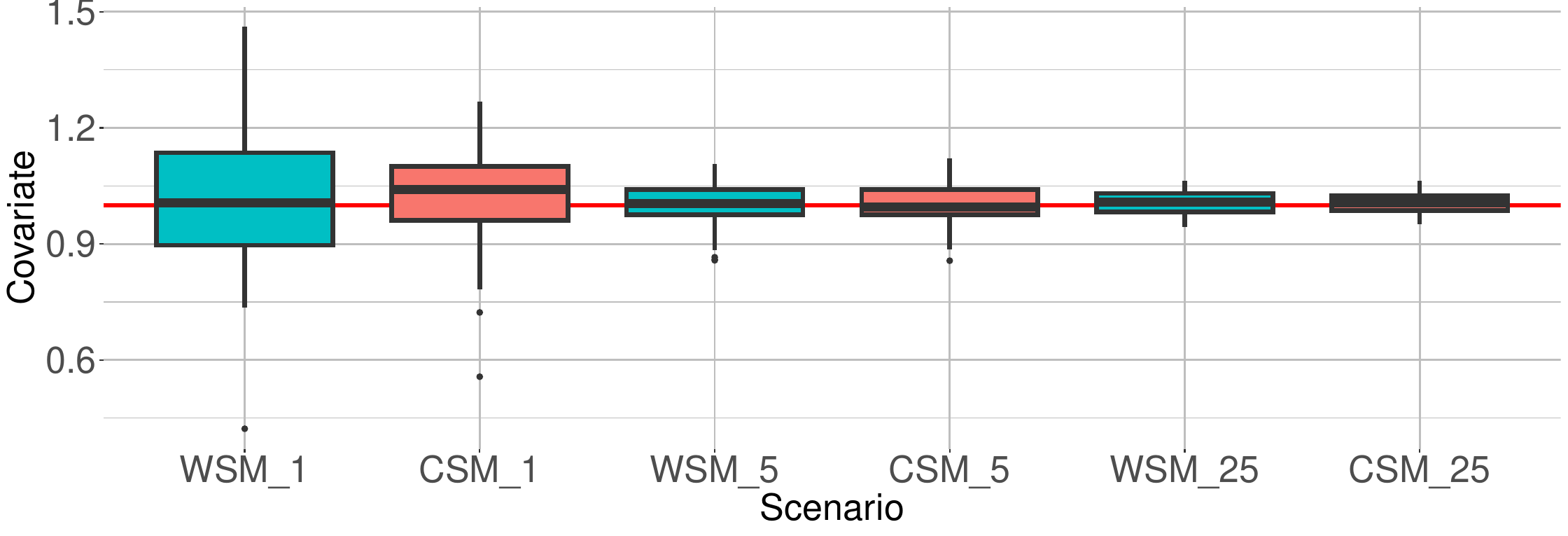}
        \caption{Long range}
        \label{fig:beta0_beta_long}
    \end{subfigure}
    \caption{Parameter estimates (median) for the fixed effects for intercept (first row) and covariate (second row) with (a) medium range and (b) long range.}
    \label{fig:fixed_parameter_est}
\end{figure}

\begin{figure}[htb]
    \centering
    \begin{subfigure}[b]{0.45\textwidth}
        \centering
        \includegraphics[width=\linewidth]{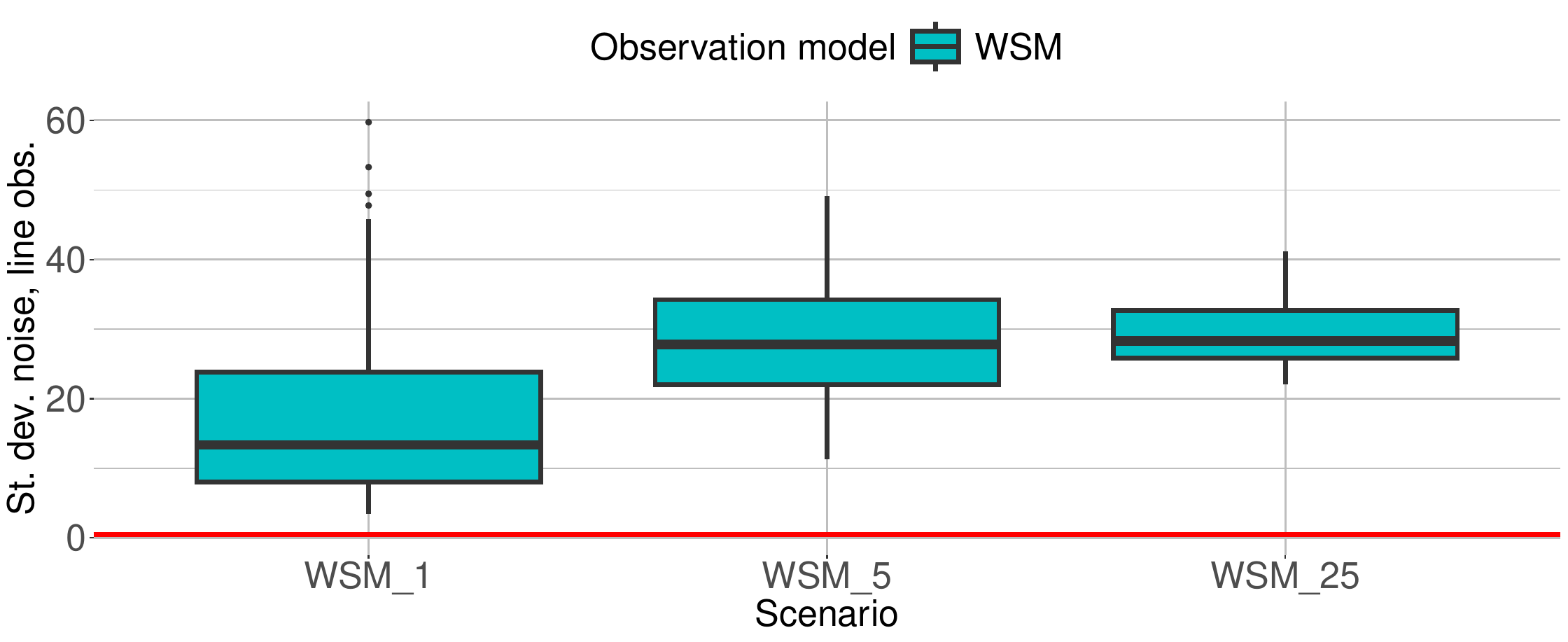}
        \caption{\SM, $\rho=\SI{350}{\meter}$}
        \label{fig:line_noise_SM_short}
    \end{subfigure}%
    ~
    \begin{subfigure}[b]{0.45\textwidth}
        \centering
        \includegraphics[width=\linewidth]{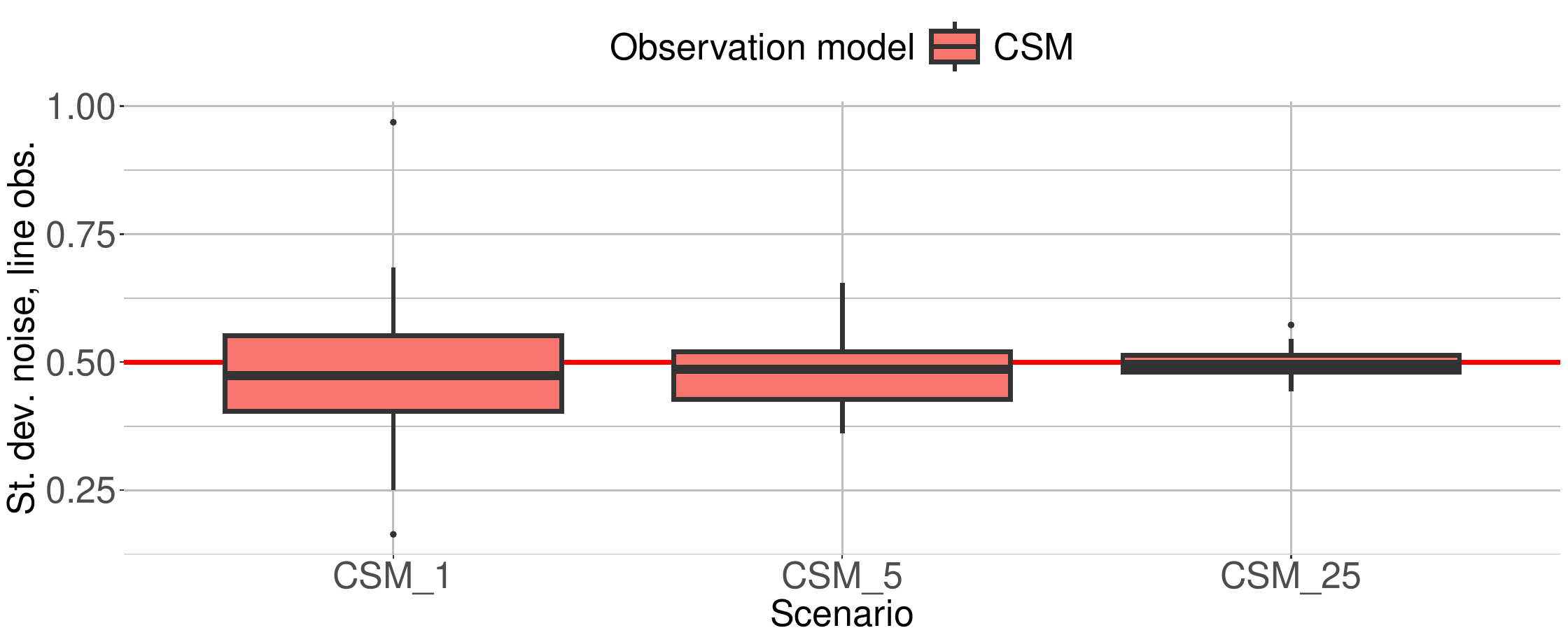}
        \caption{\IM, $\rho=\SI{350}{\meter}$}
        \label{fig:line_noise_IM_short}
    \end{subfigure}
    ~
    \newline
    \begin{subfigure}[b]{0.45\textwidth}
        \centering
        \includegraphics[width=\linewidth]{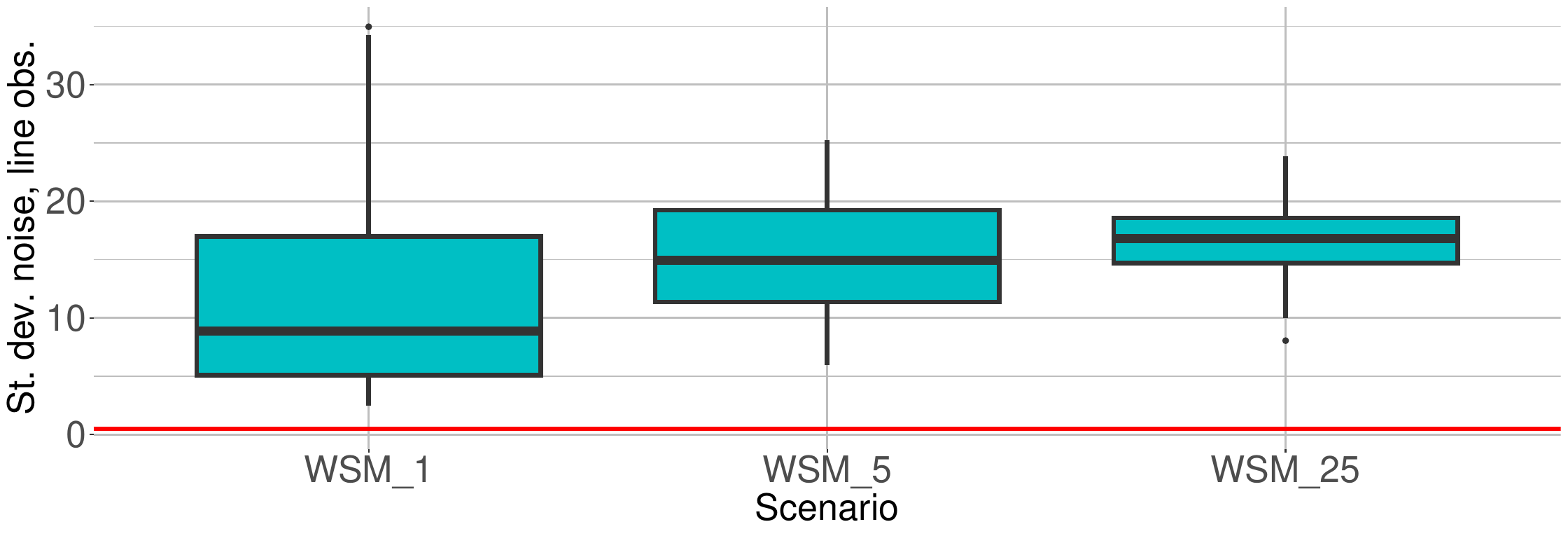}
        \caption{\SM, $\rho=\SI{1000}{\meter}$}
        \label{fig:line_noise_SM_long}
    \end{subfigure}
    ~
    \begin{subfigure}[b]{0.45\textwidth}
        \centering
        \includegraphics[width=\linewidth]{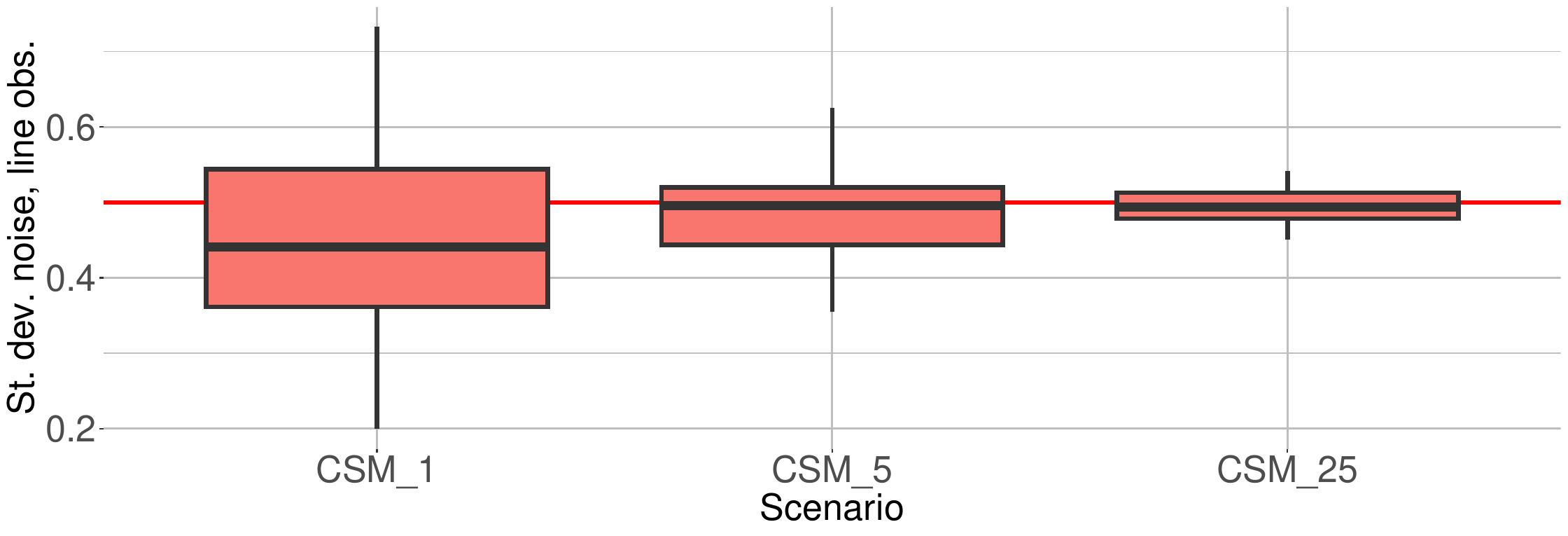}
        \caption{\IM, $\rho=\SI{1000}{\meter}$}
        \label{fig:line_noise_IM_long}
    \end{subfigure}
    \caption{Estimates of $\sigma_\mathrm{L}^2$ for both models (a,c) \SMs and (b,d) \IMs.}
    \label{fig:line_noise_est}
\end{figure}

\begin{figure}[htb]
    \centering
    \begin{subfigure}[b]{0.45\textwidth}
        \centering
        \includegraphics[width=\linewidth]{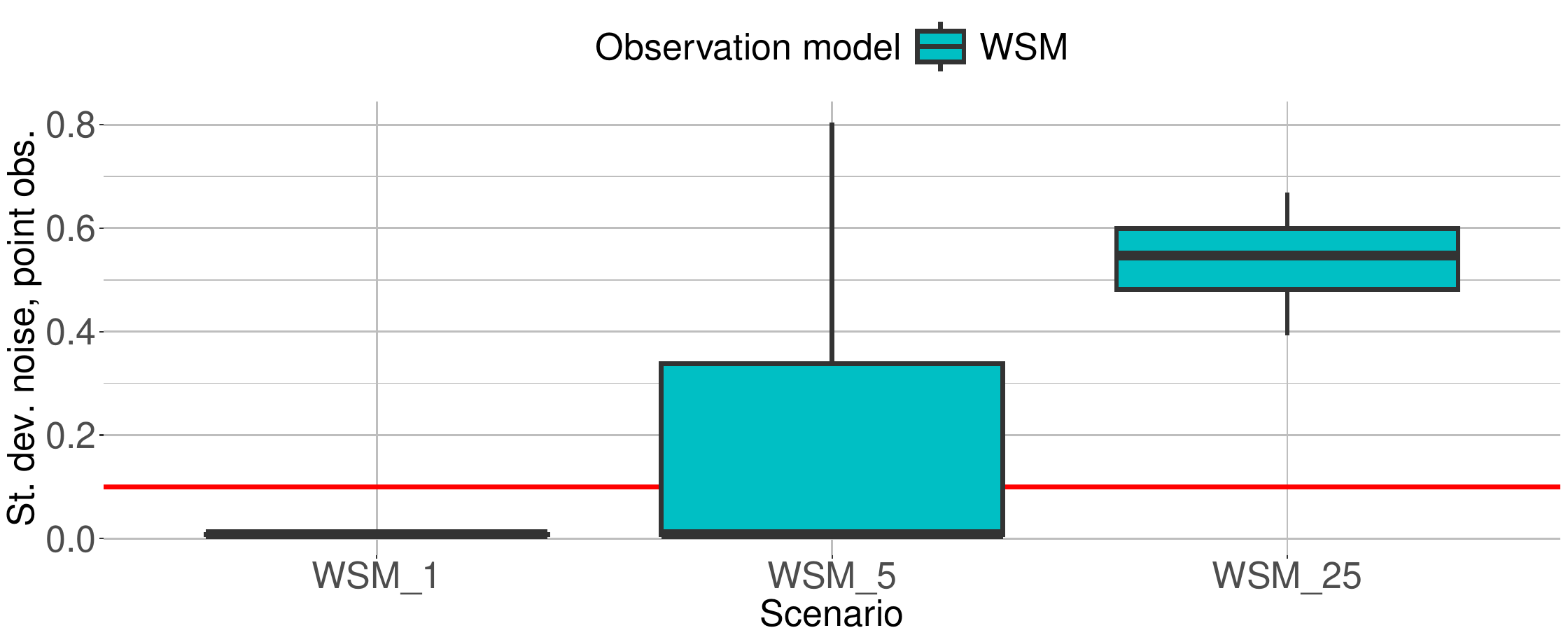}
        \caption{\SM, $\rho=\SI{350}{\meter}$}
        \label{fig:point_noise_SM_short}
    \end{subfigure}%
    ~
    \begin{subfigure}[b]{0.45\textwidth}
        \centering
        \includegraphics[width=\linewidth]{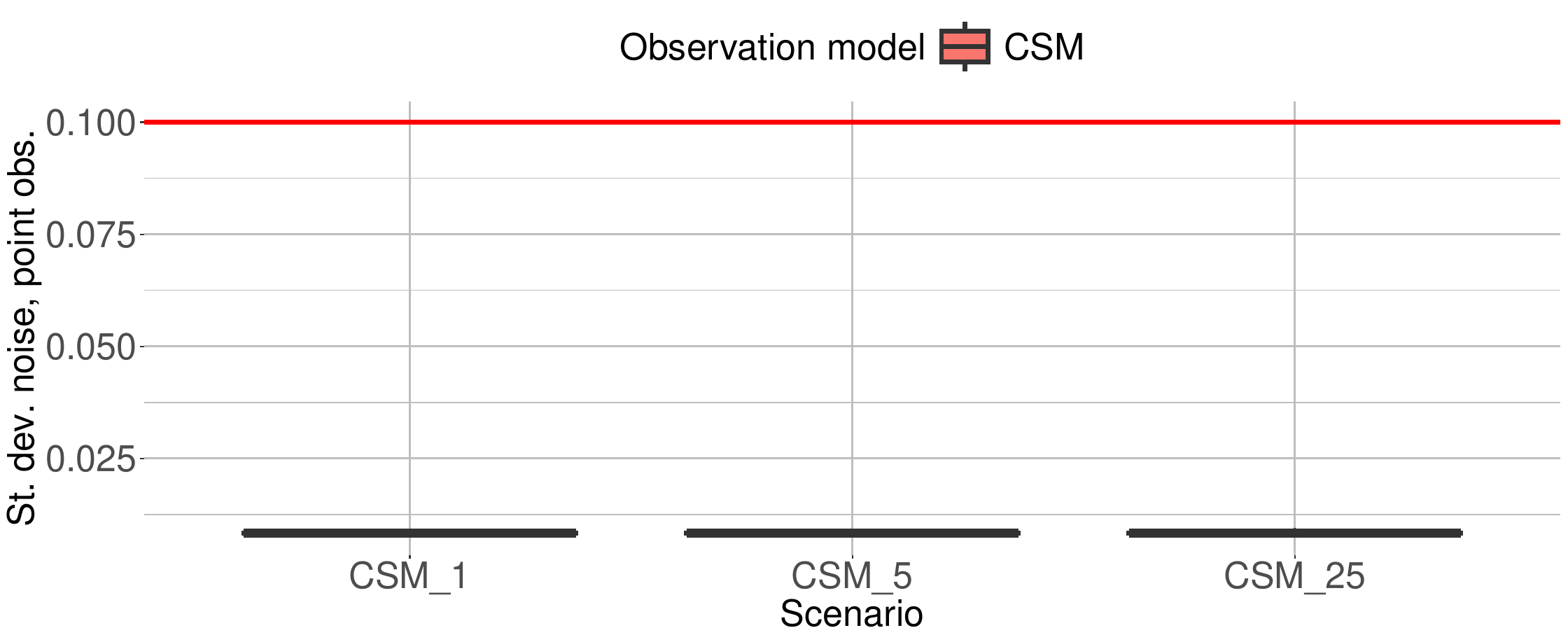}
        \caption{\IM, $\rho=\SI{350}{\meter}$}
        \label{fig:point_noise_IM_short}
    \end{subfigure}
    \newline
    \begin{subfigure}[b]{0.45\textwidth}
        \centering
        \includegraphics[width=\linewidth]{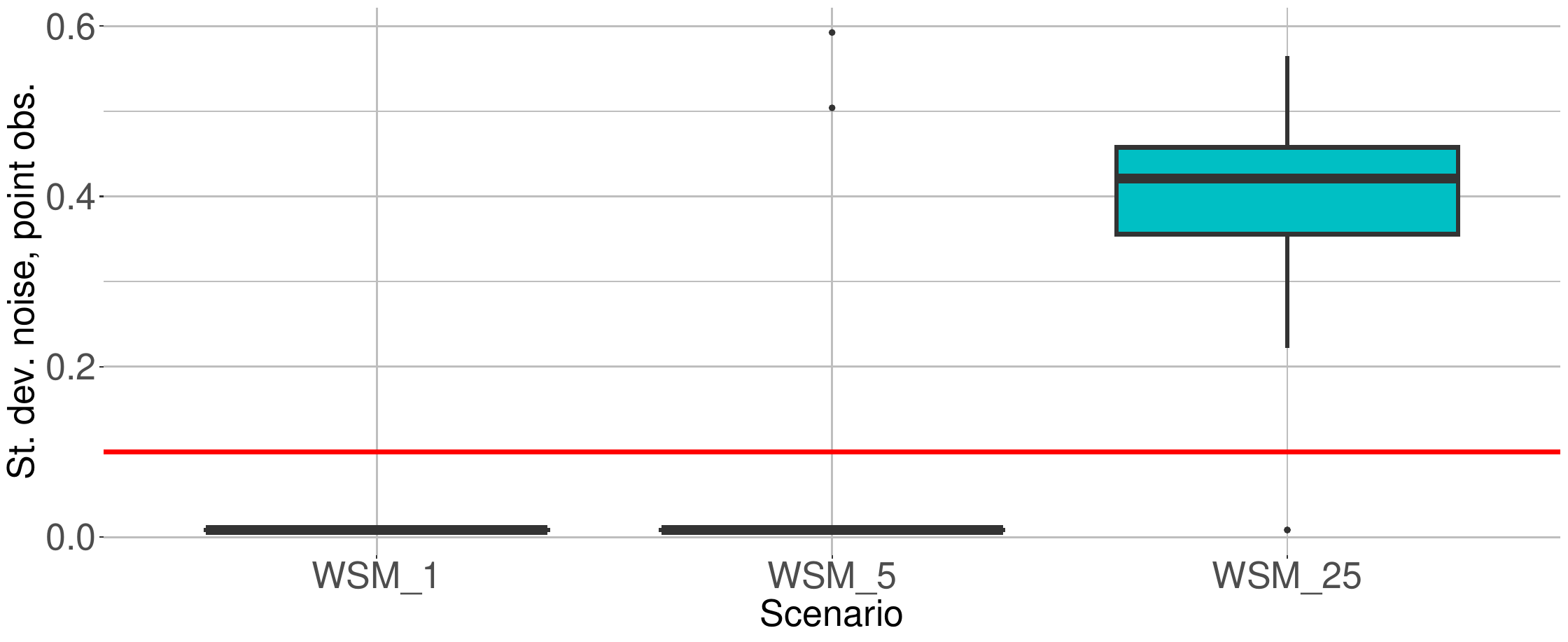}
        \caption{\SM, $\rho=\SI{1000}{\meter}$}
        \label{fig:point_noise_SM_long}
    \end{subfigure}
    ~
    \begin{subfigure}[b]{0.45\textwidth}
        \centering
        \includegraphics[width=\linewidth]{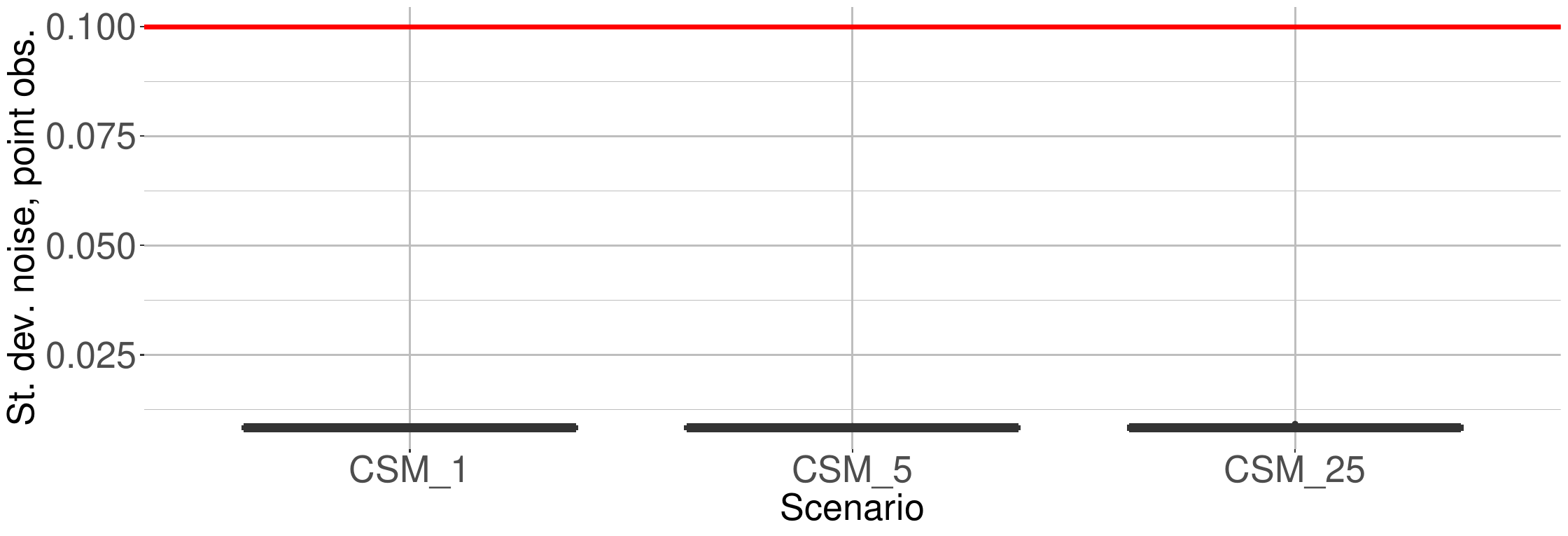}
        \caption{\IM, $\rho=\SI{1000}{\meter}$}
        \label{fig:point_noise_IM_long}
    \end{subfigure}
    \caption{Estimates of $\sigma_\mathrm{P}^2$ for both models (a,c) \SMs and (b,d) \IMs.}
    \label{fig:point_noise_est}
\end{figure}

We present the result from the simulation study in Section \ref{sec:sim_study} for parameter identifiability for the model parameters $\beta_0$, $\beta_1$, $\sigma_\mathrm L^2$ and $\sigma_\mathrm P^2$ in Figure \ref{fig:fixed_parameter_est}, Figure \ref{fig:line_noise_est} and Figure \ref{fig:point_noise_est}. Note that the number of point locations, $n_\mathrm P=6$ is too low to accurately estimate the noise for both models. Figure \ref{fig:line_noise_est} show that \IM is estimating the line noise better than \SM.

\FloatBarrier




\bibliographystyle{elsarticle-harv}
\bibliography{paper1-2}

@book{stein,
	AUTHOR = {Stein, Michael L.},
	TITLE = {Interpolation of {S}patial {D}ata},
	SERIES = {Springer Series in Statistics},
	VOLUME = {104},
	PUBLISHER = {Springer New York, NY},
	YEAR = {1999},
	PAGES = {XVII, 249},
	ISBN = {978-1-4612-1494-6},
	DOI = {10.1007/978-1-4612-1494-6},
}

@book{diggle,
title = "Model-based Geostatistics.",
author = "Diggle, {Peter J.} and Ribeiro, {Paulo J.}",
year = "2007",
month = mar,
language = "English",
isbn = "0387329072 978-0387329079",
series = "Springer Series in Statistics",
publisher = "Springer",
}

@article{gelfand,
  title = {{Bayesian Modeling and Analysis of Geostatistical Data}},
  volume = {4},
  ISSN = {2326-831X},
  DOI = {10.1146/annurev-statistics-060116-054155},
  number = {1},
  journal = {Annual Review of Statistics and Its Application},
  publisher = {Annual Reviews},
  author = {Gelfand,  Alan E. and Banerjee,  Sudipto},
  year = {2017},
  month = mar,
  pages = {245–266}
}

@article{SPDE,
  title = {{An Explicit Link between Gaussian Fields and Gaussian Markov Random Fields: The Stochastic Partial Differential Equation Approach}},
  volume = {73},
  ISSN = {1467-9868},
  DOI = {10.1111/j.1467-9868.2011.00777.x},
  number = {4},
  journal = {Journal of the Royal Statistical Society Series B: Statistical Methodology},
  publisher = {Oxford University Press (OUP)},
  author = {Lindgren,  Finn and Rue,  Håvard and Lindstr\"{o}m,  Johan},
  year = {2011},
  month = aug,
  pages = {423–498}
}

@article{SPDE_10y,
  title = {{The SPDE approach for Gaussian and non-Gaussian fields: 10 years and still running}},
  volume = {50},
  ISSN = {2211-6753},
  DOI = {10.1016/j.spasta.2022.100599},
  journal = {Spatial Statistics},
  publisher = {Elsevier BV},
  author = {Lindgren,  Finn and Bolin,  David and Rue,  Håvard},
  year = {2022},
  month = aug,
  pages = {100599}
}

@article{DTW-travel-time,
  title = {{Using GPS Data to Gain Insight into Public Transport Travel Time Variability}},
  volume = {136},
  ISSN = {1943-5436},
  DOI = {10.1061/(asce)te.1943-5436.0000126},
  number = {7},
  journal = {Journal of Transportation Engineering},
  publisher = {American Society of Civil Engineers (ASCE)},
  author = {Mazloumi,  Ehsan and Currie,  Graham and Rose,  Geoffrey},
  year = {2010},
  month = jul,
  pages = {623–631}
}

@book{transit_data_analytics,
    TITLE = {{Mobility Patterns, Big Data and Transport Analytics}},
	AUTHOR={Antoniou, Constantinos and Dimitriou, Loukas and Pereira, Francisco},
	YEAR={2019},
	PAGES = {229-261},
	PUBLISHER = {Elsevier Inc., Amsterdam, Netherlands},
	ISBN = {978-0-12-812970-8},
	DOI = {10.1016/C2016-0-03572-6}
}

@article{spat_AVL_visual,
  title = {{Identifying and Visualizing Congestion Bottlenecks with Automated Vehicle Location Systems: Application to Transantiago,  Chile}},
  volume = {2649},
  ISSN = {2169-4052},
  DOI = {10.3141/2649-07},
  number = {1},
  journal = {Transportation Research Record: Journal of the Transportation Research Board},
  publisher = {SAGE Publications},
  author = {Bucknell,  Christopher and Schmidt,  Alejandro and Cruz,  Diego and Muñoz,  Juan Carlos},
  year = {2017},
  month = jan,
  pages = {61–70}
}

@incollection{traffic_model_overview,
    title = {{Chapter Five - Modeling of Traffic and Transport Processes}},
    editor = {Andreas Schadschneider and Debashish Chowdhury and Katsuhiro Nishinari},
    booktitle = {{Stochastic Transport in Complex Systems}},
    publisher = {Elsevier},
    address = {Amsterdam},
    pages = {209-214},
    year = {2011},
    isbn = {978-0-444-52853-7},
    doi = {10.1016/B978-0-444-52853-7.00005-1},
    author = {Andreas Schadschneider and Debashish Chowdhury and Katsuhiro Nishinari}
}

@article{delay_AVL_linreg,
  title = {Assigning Bus Delay and Predicting Travel Times using Automated Vehicle Location Data},
  volume = {2673},
  ISSN = {2169-4052},
  DOI = {10.1177/0361198119832866},
  number = {3},
  journal = {Transportation Research Record: Journal of the Transportation Research Board},
  publisher = {SAGE Publications},
  author = {Coghlan,  Christy and Dabiri,  Sina and Mayer,  Brian and Wagner,  Mitch and Williamson,  Eric and Eichler,  Michael and Ramakrishnan,  Naren},
  year = {2019},
  month = mar,
  pages = {624–636}
}

@phdthesis{urban-traffic-planning,
    author = {Wendy Weijermars},
    title = {Analysis of urban traffic patterns using clustering},
    school = {TRAIL Research School},
    DOI = {10.3990/1.9789036524650},
    year = {2007}
}

@article{spatial-regression,
    author = {Fang Zhao and Soon Chung},
    title ={Contributing Factors of Annual Average Daily Traffic in a {Florida} County: Exploration with Geographic Information System and Regression Models},
    journal = {Transportation Research Record},
    volume = {1769},
    number = {1},
    pages = {113-122},
    year = {2001},
    doi = {10.3141/1769-14}
}

@article{floating-car,
  title = {{Detection of urban traffic patterns from Floating Car Data (FCD)}},
  volume = {22},
  ISSN = {2352-1465},
  DOI = {10.1016/j.trpro.2017.03.057},
  journal = {Transportation Research Procedia}, 
  publisher = {Elsevier BV},
  author = {Altintasi,  Oruc and Tuydes-Yaman,  Hediye and Tuncay,  Kagan},
  year = {2017},
  pages = {382–391}
}

@article{INLA,
    author = {Rue, Håvard and Martino, Sara and Chopin, Nicolas},
    title = {Approximate {B}ayesian inference for latent {G}aussian models by using integrated nested {L}aplace approximations},
    journal = {Journal of the Royal Statistical Society: Series B (Statistical Methodology)},
    volume = {71},
    number = {2},
    pages = {319-392},
    doi = {10.1111/j.1467-9868.2008.00700.x},
    year = {2009}
}

@Manual{INLA_web,
    author = {Håvard Rue and Finn Lindgren and Janet van Niekerk and Elias Krainski and Esmail Abdul Fattah},
    url = {https://www.r-inla.org},
    title = {R-INLA}
}

@Manual{tibble,
    title = {tibble: Simple Data Frames},
    author = {Kirill Müller and Hadley Wickham},
    year = {2023},
    note = {{R} package version 3.2.1},
    doi = {10.32614/CRAN.package.tibble}
}

@misc{inlabru,
      title={inlabru: software for fitting latent {G}aussian models with non-linear predictors}, 
      author={Finn Lindgren and Fabian Bachl and Janine Illian and Man Ho Suen and Håvard Rue and Andrew E. Seaton},
      year={2024},
      eprint={2407.00791},
      archivePrefix={arXiv},
      primaryClass={stat.ME},
}

@article{inlabru2,
  title = {{inlabru: an R package for Bayesian spatial modelling from ecological survey data}},
  volume = {10},
  ISSN = {2041-210X},
  DOI = {10.1111/2041-210x.13168},
  number = {6},
  journal = {Methods in Ecology and Evolution},
  publisher = {Wiley},
  author = {Bachl,  Fabian E. and Lindgren,  Finn and Borchers,  David L. and Illian,  Janine B.},
  editor = {Freckleton,  Robert},
  year = {2019},
  month = mar,
  pages = {760–766}
}

@Manual{rSPDE,
  title = {{rSPDE: Rational Approximations of Fractional Stochastic
      Partial Differential Equations}},
  DOI = {10.32614/cran.package.rspde},
  journal = {CRAN: Contributed Packages},
  publisher = {The R Foundation},
  author = {Bolin,  David and Simas,  Alexandre},
  year = {2019},
  note = {{R} package version 2.3.3},
  month = aug 
}

@Manual{fmesher,
  title = {{fmesher: Triangle Meshes and Related Geometry Tools}},
  author = {Finn Lindgren},
  year = {2025},
  note = {{R} package version 0.2.0, https://github.com/inlabru-org/fmesher},
  DOI = {10.32614/cran.package.fmesher},
}

@misc{metric_graph,
      title={Statistical inference for {G}aussian {W}hittle-{M}at\'ern fields on metric graphs}, 
      author={David Bolin and Alexandre Simas and Jonas Wallin},
      year={2023},
      eprint={2304.10372},
      archivePrefix={arXiv},
      primaryClass={stat.ME},
}

@article{metric_graph_FEM,
      title={Regularity and numerical approximation of fractional elliptic differential equations on compact metric graphs}, 
      author={Bolin, David and Kovács Mihály and Kumar and Vivek Alexandre B. Simas},
      year={2024},
      journal = {Mathematics of Computation},
      volume = {93},
      pages = {2439-2472},
      doi = {10.1090/mcom/3929},
}

@article{matern_metric_graph,
  title={Gaussian {W}hittle--{M}at{\'e}rn fields on metric graphs},
  author={Bolin, David and Simas, Alexandre B and Wallin, Jonas},
  journal={Bernoulli},
  volume={30},
  number={2},
  DOI = {10.3150/23-bej1647},
  pages={1611--1639},
  year={2024},
  publisher={Bernoulli Society for Mathematical Statistics and Probability}
}

@misc{nonstationary_gaussian_metric_graph,
      title={A new class of non-stationary {G}aussian fields with general smoothness on metric graphs}, 
      author={David Bolin and Lenin Riera-Segura and Alexandre B. Simas},
      year={2025},
      eprint={2501.11738},
      archivePrefix={arXiv},
      primaryClass={stat.ME},
}

@article{rspde_article,
    author = {David Bolin and Alexandre B. Simas and Zhen Xiong},
    title = {{Covariance–Based Rational Approximations of Fractional SPDEs for Computationally Efficient Bayesian Inference}},
    journal = {Journal of Computational and Graphical Statistics},
    volume = {33},
    number = {1},
    pages = {64--74},
    year = {2024},
    publisher = {ASA Website},
    doi = {10.1080/10618600.2023.2231051}
}

@misc{bolin_point,
      title={{Log-Gaussian Cox Processes on General Metric Graphs}}, 
      author={David Bolin and Damilya Saduakhas and Alexandre B. Simas},
      year={2025},
      eprint={2501.18558},
      archivePrefix={arXiv},
      primaryClass={stat.ME}, 
}

@Manual{R-language,
    title = {R: A Language and Environment for Statistical Computing},
    author = {{R Core Team}},
    organization = {R Foundation for Statistical Computing},
    address = {Vienna, Austria},
    year = {2024},
    url = {https://www.R-project.org/},
  }

@Manual{MetricGraph,
  title = {MetricGraph: Random fields on metric graphs},
  author = {David Bolin and Alexandre B. Simas and Jonas Wallin},
  year = {2023},
  note = {{R} package version 1.3.0.9000},
  doi = {10.32614/CRAN.package.MetricGraph}
}

@Manual{code,
    author = {Karina Lilleborge}, 
    year = {2025},
    title = {{Joint Modelling of Line and Point Data on Metric Graphs - code}},
    version = {1.0.0},
    publisher = {Zenodo},
    doi = {10.5281/zenodo.16979784},
}

@Manual{osm,      
	author = {{OpenStreetMap contributors}},
    title = {{Planet dump retrieved from https://planet.osm.org }},
    howpublished = "\url{ https://www.openstreetmap.org}",
    year = {2017},
    note = {Data extraction: 2025-01-17}
}

@Article{sf,
    author = {Edzer Pebesma},
    title = {{Simple Features for R: Standardized Support for Spatial
      Vector Data}},
    year = {2018},
    journal = {{The R Journal}},
    doi = {10.32614/RJ-2018-009},
    pages = {439--446},
    volume = {10},
    number = {1},
}

@book{cressie2011statistics,
  title={Statistics for spatio-temporal data},
  author={Cressie, Noel and Wikle, Christopher K},
  year={2011},
  isbn = {978‑0‑471‑69274‑4},
  publisher={John Wiley \& Sons}
}

@article{ribeiro2019mapping,
  title={Mapping 123 million neonatal, infant and child deaths between 2000 and 2017},
  author={Ribeiro, AI and others},
  journal={Nature},
  volume = {574},
  number = {7778},
  month = oct,
  pages = {353–358},
  doi = {10.1038/s41586-019-1545-0},
  year={2019},
  publisher = {Springer Science and Business Media LLC}
}

@article{anderes2020isotropic,
  title = {Isotropic covariance functions on graphs and their edges},
  volume = {48},
  ISSN = {0090-5364},
  DOI = {10.1214/19-aos1896},
  number = {4},
  journal = {The Annals of Statistics},
  publisher = {Institute of Mathematical Statistics},
  author = {Anderes,  Ethan and Møller,  Jesper and Rasmussen,  Jakob G.},
  year = {2020},
  month = aug 
}

@article{cressie2006spatial,
  title={Spatial prediction on a river network},
  author={Cressie, Noel and Frey, Jesse and Harch, Bronwyn and Smith, Mick},
  journal={{Journal of Agricultural, Biological, and Environmental Statistics}},
  volume={11},
  pages={127--150},
  doi = {10.1198/108571106X110649},
  year={2006},
  publisher={Springer}
}

@article{hoef2006spatial,
  title={Spatial statistical models that use flow and stream distance},
  author={Hoef, Jay M Ver and Peterson, Erin and Theobald, David},
  journal={Environmental and Ecological statistics},
  volume={13},
  doi = {10.1007/s10651-006-0022-8},
  pages={449--464},
  year={2006},
  publisher={Springer}
}

@article{ver2010moving,
  title={A moving average approach for spatial statistical models of stream networks},
  author={Ver Hoef, Jay M and Peterson, Erin E},
  journal={Journal of the American Statistical Association},
  volume={105},
  DOI = {10.1198/jasa.2009.ap08248},
  number={489},
  pages={6--18},
  year={2010},
  publisher={Taylor \& Francis}
}

@article{bakka2018spatial,
  title={{Spatial modeling with R-INLA: A review}},
  volume = {10},
  ISSN = {1939-0068},
  DOI = {10.1002/wics.1443},
  number = {6},
  journal = {WIREs Computational Statistics},
  publisher = {Wiley},
  author = {Bakka,  Haakon and Rue,  Håvard and Fuglstad,  Geir‐Arne and Riebler,  Andrea and Bolin,  David and Illian,  Janine and Krainski,  Elias and Simpson,  Daniel and Lindgren,  Finn},
  year = {2018},
  month = jul 
}

@article{gneiting2007strictly,
  title={Strictly proper scoring rules, prediction, and estimation},
  author={Gneiting, Tilmann and Raftery, Adrian E},
  journal={Journal of the American statistical Association},
  volume={102},
  DOI = {10.1198/016214506000001437},
  number={477},
  pages={359--378},
  year={2007},
  publisher={Taylor \& Francis}
}

@article{Lindgren2024_sort,
author = {Finn Lindgren and Haakon Bakka and David Bolin and Elias Krainski and Håvard Rue},
title={A diffusion-based spatio-temporal extension of {G}aussian {M}atérn fields: (invited article with discussion)},
journal={{SORT}-{S}tatistics and {O}perations {R}esearch {T}ransactions},
year = {2024},
DOI = {10.57645/20.8080.02.13},
volume = {48},
issue = {1},
pages={3--66}}

@article{porcu2023,
  title = {Stationary nonseparable space-time covariance functions on networks},
  volume = {85},
  ISSN = {1467-9868},
  DOI = {10.1093/jrsssb/qkad082},
  number = {5},
  journal = {Journal of the Royal Statistical Society Series B: Statistical Methodology},
  publisher = {Oxford University Press (OUP)},
  author = {Porcu,  Emilio and White,  Philip A and Genton,  Marc G},
  year = {2023},
  month = sep,
  pages = {1417–1440}
}

@article{Møller2024,
author = {Møller, Jesper and Rasmussen, Jakob G.},
title = {Cox processes driven by transformed {G}aussian processes on linear networks—A review and new contributions},
journal = {Scandinavian Journal of Statistics},
volume = {51},
number = {3},
pages = {1288-1322},
doi = {10.1111/sjos.12720},
year = {2024}
}

@article{Baddeley2021,
title = {Analysing point patterns on networks — {A} review},
journal = {Spatial Statistics},
volume = {42},
pages = {100435},
year = {2021},
issn = {2211-6753},
doi = {10.1016/j.spasta.2020.100435},
author = {Adrian Baddeley and Gopalan Nair and Suman Rakshit and Greg McSwiggan and Tilman M. Davies},
}

@article{Rikke,
title = {Estimation of a non-stationary model for annual precipitation in southern {N}orway using replicates of the spatial field},
journal = {Spatial Statistics},
volume = {14},
pages = {338-364},
year = {2015},
issn = {2211-6753},
doi = {10.1016/j.spasta.2015.07.003},
author = {Rikke Ingebrigtsen and Finn Lindgren and Ingelin Steinsland and Sara Martino}
}

\end{document}